\documentclass[3p,authoryear]{elsarticle}

\usepackage[english]{babel}

\usepackage{graphicx}

\usepackage{amsmath,amssymb}
\numberwithin{equation}{section}

\newcommand{\mymultiply}{\,}

\newcommand{\TeXmacs}{T\kern-.1667em\lower.5ex\hbox{E}\kern-.125emX\kern-.1em\lower.5ex\hbox{\textsc{m\kern-.05ema\kern-.125emc\kern-.05ems}}}
\newcommand{\cdummy}{\cdot}
\newcommand{\mathd}{\mathrm{d}}
\newcommand{\mathpi}{\pi}
\newcommand{\nobracket}{}
\newcommand{\nocomma}{}
\newcommand{\tmem}[1]{{\em #1\/}}
\newcommand{\tmmathbf}[1]{\ensuremath{\boldsymbol{#1}}}
\newcommand{\tmop}[1]{\ensuremath{\operatorname{#1}}}
\newcommand{\tmstrong}[1]{\textbf{#1}}

\newcommand{\doublecontract}{\mathbin:}
\newcommand{\nlPoisson}{p}

\usepackage[usernames,dvipsnames]{xcolor}

\begin{document}

\journal{J.~Mech.~Phys.~Solids}

\begin{frontmatter}

\title{
  Asymptotic derivation of high-order rod models\\
  from non-linear 3D elasticity
}

	\author[ba]{Basile Audoly}
	\author[cl]{Claire Lestringant}

	\address[ba]{Laboratoire de m{\'e}canique des solides, CNRS, Institut
	Polytechnique de Paris, Palaiseau, France}
	
	\address[cl]{Structures Research Group, Department of Engineering,
	University of Cambridge, Cambridge CB2 1PZ, United Kingdom}

\begin{abstract}
  We propose a method for deriving equivalent one-dimensional models for
  slender non-linear structures. The approach is designed to be broadly
  applicable, and can handle in principle finite strains, finite rotations,
  arbitrary cross-sections shapes, inhomogeneous elastic properties across the
  cross-section, arbitrary elastic constitutive laws (possibly with low
  symmetry) and arbitrary distributions of pre-strain, including finite
  pre-strain. It is based on a kinematic parameterization of the actual
  configuration that makes use of a center-line, a frame of directors, and
  local degrees of freedom capturing the detailed shape of cross-sections. A
  relaxation method is applied that holds the framed center-line fixed while
  relaxing the local degrees of freedom; it is asymptotically valid when the
  macroscopic strain and the properties of the rod vary slowly in the
  longitudinal direction. The outcome is a one-dimensional strain energy
  depending on the apparent stretching, bending and twisting strain of the
  framed center-line; the dependence on the strain gradients is also captured,
  yielding an equivalent rod model that is asymptotically exact to higher
  order. The method is presented in a fully non-linear setting and it is
  verified against linear and weakly non-linear solutions available from the
  literature.
\end{abstract}


\end{frontmatter}

\section{Introduction}

This work aims at identifying efficient and accurate models for non-linear,
slender elastic structures. Although there exists a wide range of
one-dimensional beam and rod theories, understanding nonlinear effects arising
during the compression of wide
columns~\citep{lubbers2017nonlinear,chen2020snapping}, predicting the
emergence of shape due to heterogeneous pre-stress generated by growth or
thermal effects in slender
filaments~\citep{Liu-Huang-EtAl-Structural-Transition-from-2014,%
turcaud2020twisters}
and designing structures made of complex nonlinear materials such as nematic
elastomers or active materials
\citep{agostiniani2017shape,tomassetti2017capturing} remain challenging
tasks. Well-established, classical rod theories account for the stretching,
bending and twisting strains in a linear way and therefore do not account for
finite-strain or finite-thickness effects. Extensions have been proposed to
account for some of these effects, but their justification is often patchy or
relies on restrictive hypotheses on the kinematics or the constitutive
behavior; their range of applicability is thus often limited and sometimes
ill-defined. This leaves researchers, engineers and designers with two
alternatives: either rely on full three-dimensional finite
elasticity~\citep{scherzinger1998asymptotic,%
Goriely-Vandiver-EtAl-Nonlinear-Euler-buckling-2008,%
de2011nonlinear,%
chen2020snapping}
or build their own specific reduced model~\citep{lubbers2017nonlinear}.
Should they choose this latter option, however, a clear and rigorous
methodology for deriving such a model is lacking.

This work builds up on a dimension reduction procedure introduced by the
authors in an abstract setting~\citep{LESTRINGANT2020103730} which is
applied here to the case of a hyper-elastic prismatic solid which can stretch,
bend and twist arbitrarily in three dimensions. The present work extends our
previous work on one-dimensional structures that can just
stretch~\citep{Audoly-Hutchinson-Analysis-of-necking-based-2016,%
LESTRINGANT2020103730}.

The proposed method yields one-dimensional models that account for
stretching, bending and twisting modes in a non-linear way.  It is
asymptotically correct; a scaling estimate of the error in energy with
respect to the full three-dimensional theory is available in terms of
the slenderness parameter.  The one-dimensional model is derived based
on an assumption of slow longitudinal variations, implemented by a
two-scale expansion.  Effectively, this approach splits the original
three-dimensional problem into a set of relaxation problems formulated
in the two-dimensional cross-section, and a one-dimensional
variational problem at the scale of the structure, as noted in
previous work, {\tmem{e.g.}}, by~\citet{berdichevskii1981energy},
\citet{bermudez1984justification}, \citet{trabucho1989existence} and
\citet{sanchez1999statics}, among others.

We improve on existing approaches to asymptotic dimension reduction in three
key aspects.
\begin{itemize}
  \item Our method is variational. While most of the existing work has started
  from the three-dimensional equilibrium
  equations~\citep{bermudez1984justification,trabucho1996mathematical}, we
  base our reduction on the energy formulation of the three-dimensional
  problem. This helps keeping the derivation as simple as possible, and makes
  the variational structure of the one-dimensional model stand out without any
  effort.
  
  \item We start from finite elasticity. Most of the existing work has been
  limited to linear
  strains~\citep{trabucho1996mathematical,yu2004elasticity,hodges2006nonlinear}
  but the one-dimensional models derived using the proposed method can retain
  nonlinearities coming from both the geometry and from the constitutive
  behavior.
  
  \item Our one-dimensional model is high-order and asymptotically correct,
  {\tmem{i.e.}}, it captures the energy cost arising from the longitudinal
  gradients of the stretching, bending and twisting strains. Besides
  increasing the accuracy and expanding the range of validity of the model,
  gradient terms have been found to help capture localization phenomena very
  accurately~\citep{Lestringant-Audoly-A-diffuse-interface-model-2018,%
  lestringant2020one}.
\end{itemize}

Some of the models from the literature include one or two of these
ingredients.  \citet{berdichevskii1981energy},
\citet{hodges2006nonlinear} and \citet{yu2012variational} use a
variational approach,
\citet{trabucho1989existence} and \citet{nolde2018asymptotic} introduce
higher-order terms, \citet{jiang2016nonlinear} and \citet{cimetiere1988asymptotic}
handle finite strains, \citet{moulton2020morphoelastic} work with
finite elasticity in a variational setting.  Yet this paper is the
first attempt to combine these three aspects in a unified procedure.

The proposed approach has been designed to be as general as possible. It does
not make any specific assumptions regarding the symmetry of the constitutive
law, such as isotropy~\citep{cimetiere1988asymptotic}. It is not limited to
small rotations, or to specific shapes of the cross-section. It can readily be
applied to a variety of constitutive behaviors, and in particular it can
handle inhomogeneous pre-strain as well as inhomogeneous elastic properties
across the sections. By lack of space, we cannot provide detailed
illustrations for all these capabilities but we shortly discuss
in~{\textsection}\ref{sec:nonlinear-energy-formulation} how these cases can be
covered. Besides, the approach is systematic: it is carried out by applying a
sequence of steps, much like a cooking recipe, and it lends itself naturally
to a numerical implementation.

The manuscript is organized as follows. In section~\ref{sec:full-model}, we
introduce the center-line based representation of a prismatic hyper-elastic
solid and derive the energy functional governing its elastic equilibrium in a
non-linear, three-dimensional setting. In section~\ref{sec:ideal-model}, we
introduce a relaxation method which achieves the one-dimensional reduction
formally. In section~\ref{sec:asymptotic-1d-reduction}, we combine this
relaxation method with a two-scale expansion and derive a concrete recipe for
obtaining one-dimensional models. This method is applied to the linear
analysis of the twisting of a prismatic bar in section~\ref{sec:twisting}, and
to the weakly non-linear analysis of the Euler buckling of a circular cylinder
in section~\ref{sec:Euler-buckling}.

Our mathematical notations are as follows. We use boldface for vectors such as
$\tmmathbf{e}_1$ and tensors $\tmmathbf{F}$. The longitudinal and transverse
coordinates in the prismatic body are denoted as $S$ and $\tmmathbf{T}= (T_1,
T_2)$, respectively. Einstein's implicit summation rules are used throughout,
whereby repeated indices appearing on the same side of an equal side are
implicitly summed; any index appearing once on each side of an equation is
considered to be a dummy index, {\tmem{i.e.}}, the equation holds implicitly
for any value of the index. In addition, the range of Greek indices such as
$\alpha$ is implicitly restricted to the cross-sectional directions, $\alpha
\in \{ 1, 2 \}$ although Latin indices such as $i$ run over the three
directions of the Cartesian space, $i \in \{ 1, 2, 3 \}$; as a result,
$T_{\alpha} \mymultiply \tmmathbf{d}_{\alpha} (S)$ stands for $\sum_{\alpha =
1}^2 T_{\alpha} \mymultiply \tmmathbf{d}_{\alpha} (S)$. The prime notation is
reserved for derivatives with respect to the longitudinal coordinate $S$ and
we use the notation $\partial_{\alpha}$ for partial derivatives along the
cross-sectional directions,
\[ \begin{array}{ll}
     f' (S, \tmmathbf{T}) = \frac{\partial f}{\partial S} (S, \tmmathbf{T}) &
     \partial_{\alpha} f (S, \tmmathbf{T}) = \frac{\partial f}{\partial
     T_{\alpha}} (S, \tmmathbf{T}) .
   \end{array} \]
The $\nabla$ notation is reserved for a differentiation with respect to the
macroscopic strain $\tmmathbf{h}$, see equation~(\ref{eq:nabla-notation}). The
notation $\tmmathbf{a} \odot \tmmathbf{b}= \frac{1}{2}\,(\tmmathbf{a} \otimes
\tmmathbf{b}+\tmmathbf{b} \otimes \tmmathbf{a})$ denotes the symmetrized
outer product of two vectors. The restriction of a function $f (S,
\tmmathbf{T})$ to a cross-section with coordinate $S$ is denoted as
$\nobracket f |_S$: this object is a function of the transverse coordinates
$\tmmathbf{T}$ only, such that $\nobracket f |_S (\tmmathbf{T}) = f (S,
\tmmathbf{T})$. Finally, functionals have their arguments inside square
brackets: the notation $\Psi [\tmmathbf{r}, \tmmathbf{d}_i]$ for the energy
functional implies that the arguments of $\Psi$ are the entire functions
$\tmmathbf{r}$ and $\tmmathbf{d}_i$ and not just their local values.

\section{3d model in center-line based representation}\label{sec:full-model}

In this section, the non-linear equilibrium of a finitely-strained prismatic
hyper-elastic solid is formulated without approximation. Attention is limited
to the formulation of the elasticity problem and no attempt to solve it is
made until the next sections. The formulation makes use of a center-line based
representation, which sets the stage for the forthcoming dimension reduction.
A similar parameterization was introduced in earlier work by
\citet{hodges2006nonlinear} in the framework of linear elasticity, and
extended to finite elasticity by
\citet{jiang2016nonlinear} and \citet{jiang2016nonlinear2}, where it is used as a basis
for a numerical approach to dimension reduction, without any account for the
gradient effect.

\subsection{Center-line based
representation}\label{ssec:centerline-based-parameterization}

\begin{figure}
  \centerline{\includegraphics{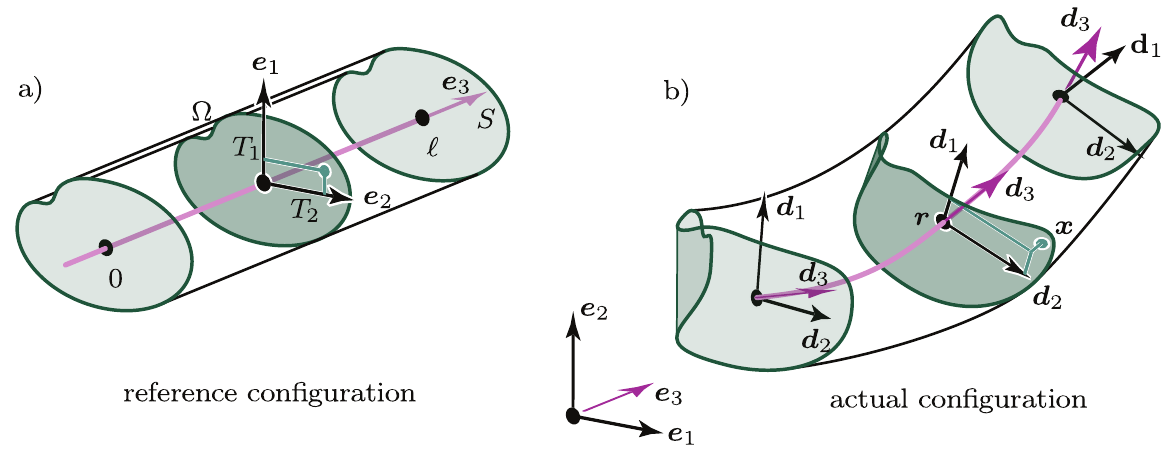}}
  \caption{Center-line based representation of a prismatic solid in
  (a)~reference and (b)~actual configurations.\label{fig:geom}}
\end{figure}

We consider a prismatic solid in reference configuration, see
figure~\ref{fig:geom}a. We denote by $\ell$ its initial length, by $S$ the
arc-length coordinate along its axis, such that $0 \leqslant S \leqslant
\ell$, by $\tmmathbf{T}= (T_1, T_2)$ the transverse coordinates and by
$(\tmmathbf{e}_1, \tmmathbf{e}_2, \tmmathbf{e}_3)$ an orthonormal frame
initially aligned with the axes $T_1$, $T_2$ and $S$, respectively. The
cross-section domain is denoted as $\Omega \subset \mathbb{R}^2$. Let $\mathd
A = \mathd T_1 \mymultiply \mathd T_2$ be the area element in the domain
$\Omega$ and $| \Omega | = \iint_{\Omega} \mathd A$ the cross-section area.
The average of a function $f (\tmmathbf{T})$ over a cross-section is denoted
as
\[ \langle f (\tmmathbf{T}) \rangle = \frac{1}{| \Omega |} \mymultiply
   \iint_{\Omega} f (\tmmathbf{T}) \mymultiply \mathd A. \]

The coordinates $(S, \tmmathbf{T}) \in (0, \ell) \times \Omega$ of a material
point in reference configuration are used as Lagrangian variables in the
elasticity problem. The position of this material point in the actual
configuration is denoted as $\tmmathbf{x} (S, \tmmathbf{T})$, see
figure~\ref{fig:geom}b. We do not assume that the internal stress is zero in
the reference configuration, i.e., pre-stress is allowed.

In terms of the mapping $\tmmathbf{x}$ from the reference to the actual
configuration, we define an apparent center-line $\tmmathbf{r} (S)$ passing
through the centroids of the cross-sections,
\begin{equation}
  \tmmathbf{r} (S) = \langle \tmmathbf{x} (S, \tmmathbf{T}) \rangle,
  \label{eq:ctr-of-mass-constraint-x}
\end{equation}
and a unit tangent to the center-line $\tmmathbf{d}_3 (S)$,
\begin{equation}
  \tmmathbf{d}_3 (S) = \frac{\frac{\mathd \tmmathbf{r}}{\mathd S} (S)}{\left|
  \frac{\mathd \tmmathbf{r}}{\mathd S} (S) \right|} .
  \label{eq:d3-from-rPrime}
\end{equation}
The unit vector $\tmmathbf{d}_3 (S)$ can be complemented by two vectors
$\tmmathbf{d}_1 (S)$ and $\tmmathbf{d}_2 (S)$ forming an orthonormal frame,
the orientation of $\tmmathbf{d}_1 (S)$ and $\tmmathbf{d}_2 (S)$ in the plane
perpendicular to $\tmmathbf{d}_3 (S)$ being fixed by the condition
\begin{equation}
  \forall S \quad \left\langle T_{\alpha} \mymultiply \tmmathbf{d}_{\alpha}
  (S) \times (\tmmathbf{x} (S, \tmmathbf{T}) -\tmmathbf{r} (S)) \right\rangle
  \cdot \tmmathbf{d}_3 (S) = 0. \label{eq:twist-condition-x}
\end{equation}
By Einstein's implicit summation rule and by our convention that Greek indices
are restricted to cross-sectional directions, the left-hand side in the
equation above is implicitly summed over $\alpha \in \{ 1, 2 \}$. By
equation~(\ref{eq:twist-condition-x}), the orthonormal frame $\tmmathbf{d}_i
(S)$ captures the average rotation of the cross-section about the tangent
$\tmmathbf{d}_3 (S)$ at any point $S$ along the center-line. The orthonormal
vectors $\tmmathbf{d}_i (S)$ are called the {\tmem{directors}} in the theory
of rods.

The condition that the directors are orthonormal writes as
\begin{equation}
  \tmmathbf{d}_i (S) \cdot \tmmathbf{d}_j (S) = \delta_{i \nocomma j},
  \label{eq:di-orthonormal-frame}
\end{equation}
for any $S$ and any integers $i$ and $j$, where $\delta_{i \nocomma j}$ is
Kronecker's symbol, equal to 1 when $i = j$ and to 0 otherwise.

The original transformation can be recovered as
\begin{equation}
  \tmmathbf{x} (S, \tmmathbf{T}) =\tmmathbf{r} (S) + y_i (S, \tmmathbf{T})
  \mymultiply \tmmathbf{d}_i (S), \label{eq:x-centerline-based-crspondence}
\end{equation}
where $y_i (S, \tmmathbf{T}) = (\tmmathbf{x} (S, \tmmathbf{T}) -\tmmathbf{r}
(S)) \cdot \tmmathbf{d}_i (S)$ for $i = 1, 2, 3$ are the microscopic
displacement functions (`displacement' is an abuse of language, since this
quantity is non-zero in the reference configuration but we will use it
anyway). In terms of the displacement $y_i$, the constraints in
equations~(\ref{eq:ctr-of-mass-constraint-x}) and~(\ref{eq:twist-condition-x})
write
\begin{equation}
  \begin{array}{crll}
    \forall (S, i) & \langle y_i (S, \tmmathbf{T}) \rangle & = & 0\\
    \forall S & \quad \left\langle \eta_{\alpha \nocomma \beta} \mymultiply
    T_{\alpha} \mymultiply y_{\beta} (S, \tmmathbf{T}) \right\rangle & = & 0.
  \end{array} \label{eq:yibar-kinematic-conditions}
\end{equation}
By Einstein's conventions, the first equation with a non-repeated Latin index
holds for $i = 1, 2, 3$, while the second equation with repeated Greek indices
contains an implicit sum over $\alpha, \beta \in \{ 1, 2 \}$. In the equation
above, $\eta_{\alpha \nocomma \beta}$ is the skew-symmetric symbol, such that
$\eta_{1 \nocomma 1} = \eta_{2 \nocomma 2} = 0$, $\eta_{1 \nocomma 2} = 1$ and
$\eta_{2 \nocomma 1} = - 1$.

Equation~(\ref{eq:x-centerline-based-crspondence}) shows that $\tmmathbf{r}
(S)$, $\tmmathbf{d}_i (S)$ and $y_i (S, \tmmathbf{T})$ can be used to
parameterize the deformed configuration. Indeed, it can be checked easily that
there is a one-to-one correspondence between the unknown $\tmmathbf{x} (S,
\tmmathbf{T})$ on the one hand, and the unknowns $\tmmathbf{r} (S)$,
$\tmmathbf{d}_i (S)$ and $y_i (S, \tmmathbf{T})$ on the other hand, provided
$\tmmathbf{d}_i (S)$ satisfies the orthonormality
condition~(\ref{eq:di-orthonormal-frame}) and $y_i (S, \tmmathbf{T})$
satisfies the four scalar kinematic
constraints~(\ref{eq:yibar-kinematic-conditions}). We will use \ $\tmmathbf{r}
(S)$, $\tmmathbf{d}_i (S)$ and $y_i (S, \tmmathbf{T})$ as the main unknowns:
we refer to this as the {\tmem{center-line based parameterization}}. It is
natural to work with this parameterization in the context of dimension
reduction as it brings in the macroscopic variables of the one-dimensional rod
model, $\tmmathbf{r} (S)$ and $\tmmathbf{d}_i (S)$.

Note that the apparent center-line $\tmmathbf{r} (S)$ is not a material
line---in the case of a hollow cylinder for instance, the curve $\tmmathbf{r}
(S)$ does not even lie within the material domain. Similarly, the directors
$\tmmathbf{d}_i (S)$ do not provide a detailed description of the microscopic
displacement on their own: the only information conveyed by the directors
frame $\tmmathbf{d}_i (S)$ is the {\tmem{average}} rotation of the
cross-section about the center-line, see~(\ref{eq:twist-condition-x}). Neither
the fact that the material frame $\tmmathbf{d}_i (S)$ is orthonormal, nor the
fact that $\tmmathbf{d}_3 (S)$ is parallel to the center-line, see
equation~(\ref{eq:d3-from-rPrime}), implies any assumption or restriction on
the microscopic displacement field: as noted above, the center-line based
representation can represent {\tmem{any}} microscopic transformation
$\tmmathbf{x} (S, \tmmathbf{T})$.

\subsection{Apparent stretching, twisting and bending
strain}\label{ssec:apparent-strain}

Together, the center-line $\tmmathbf{r} (S)$ and the directors $\tmmathbf{d}_i
(S)$ define what is known as a framed curve. The standard kinematic analysis
of framed curves goes as follows. First, we define the axial strain
$\varepsilon (S)$ by the relation
\begin{equation}
  \tmmathbf{r}' (S) = (1 + \varepsilon (S)) \mymultiply \tmmathbf{d}_3 (S),
  \label{eq:rPrime-epsilon-d3}
\end{equation}
which implies the condition of adaptation~(\ref{eq:d3-from-rPrime}).

Second, we define the bending strain $\kappa_1 (S)$ and $\kappa_2 (S)$ and the
twisting strain $\kappa_3 (S)$ by the relation
\begin{equation}
  \tmmathbf{d}_i' (S) = - \eta_{i \nocomma j \nocomma k} \mymultiply \kappa_j
  (S) \mymultiply \tmmathbf{d}_k (S), \label{eq:kappa-i}
\end{equation}
where $\eta_{i \nocomma j \nocomma k}$ is the antisymmetric (Levi-Civita)
symbol of order 3. This equation defines the quantities $\kappa_j (S)$
uniquely since the frame of directors $\tmmathbf{d}_i (S)$ is orthonormal for
all $S$. The quantities $\kappa_i (S)$ defined in this way are the components
of the rotation gradient $\tmmathbf{\kappa} (S) = \kappa_i (S) \mymultiply
\tmmathbf{d}_i (S)$ as we have $\tmmathbf{d}_i' (S) = - \eta_{i \nocomma j
\nocomma k} \mymultiply \kappa_j (S) \mymultiply \tmmathbf{d}_k (S) = \kappa_j
(S) \mymultiply \eta_{j \nocomma i \nocomma k} \mymultiply \tmmathbf{d}_k (S)
= \kappa_j (S) \mymultiply \tmmathbf{d}_j (S) \times \tmmathbf{d}_i (S)
=\tmmathbf{\kappa} (S) \times \tmmathbf{d}_i (S)$ for any $S$ and any integer
$i$.

The strain measures are collected in a {\tmem{macroscopic strain}} vector
\[ \tmmathbf{h}= (\varepsilon, \kappa_1, \kappa_2, \kappa_3) . \label{eq:h} \]
They will be referred to as {\tmem{apparent}} strain measures as they depend
on the center-line and on the directors, which are immaterial in the following
sense. Consider for instance a thin cylindrical tube made up of a soft matrix
and inextensible fibers initially oriented parallel to the axis of the
cylinder: upon twisting, the cylinder will shorten due to the inextensibility
of the fibers, making the apparent axial strain negative, $\varepsilon < 0$,
even though the longitudinal strain along any of the material (helical) fibers
is actually zero.

\subsection{Microscopic strain}\label{ssec:full-microscopic-strain}

With a view of formulating an elasticity problem for the prismatic body, we
derive the microscopic strain based on the center-line
representation~(\ref{eq:x-centerline-based-crspondence}). The deformation
gradient $\tmmathbf{F}$ such that $\mathd \tmmathbf{x}=\tmmathbf{F} \cdot
(\mathd \tmmathbf{T}, \mathd S)$ is first introduced as
\begin{equation}
  \tmmathbf{F}= \partial_{\alpha} \tmmathbf{x} \otimes \tmmathbf{e}_{\alpha}
  +\tmmathbf{x}' \otimes \tmmathbf{e}_3 = \partial_{\alpha} y_i (S,
  \tmmathbf{T}) \mymultiply \tmmathbf{d}_i (S) \otimes \tmmathbf{e}_{\alpha} +
  t_i (S, \tmmathbf{T}) \mymultiply \tmmathbf{d}_i (S) \otimes \tmmathbf{e}_3,
  \label{eq:transformation-gradient}
\end{equation}
where $t_i =\tmmathbf{x}' \cdot \tmmathbf{d}_i$ is the deformed material
tangent that was initially oriented parallel to the axis,
\[ t_i = (1 + \varepsilon (S)) \mymultiply \delta_{i \nocomma 3} + \eta_{i
   \nocomma j \nocomma k} \mymultiply \kappa_j (S) \mymultiply y_k (S,
   \tmmathbf{T}) + y_i' (S, \tmmathbf{T}) . \]
Next, we consider the microscopic Green--St-Venant deformation tensor
$\tmmathbf{E}= \frac{1}{2} \mymultiply (\tmmathbf{F}^T \cdot
\tmmathbf{F}-\tmmathbf{I})$ where $\tmmathbf{I}$ is the $3 \times 3$ identity
matrix,
\begin{equation}
  \tmmathbf{E}= \frac{t_i^2 - 1}{2} \mymultiply \tmmathbf{e}_3 \otimes
  \tmmathbf{e}_3 + t_i \mymultiply \partial_{\alpha} y_i \mymultiply
  \tmmathbf{e}_{\alpha} \odot \tmmathbf{e}_3 + \frac{\partial_{\alpha} y_i
  \mymultiply \partial_{\beta} y_i - \delta_{\alpha \nocomma \beta}}{2}
  \mymultiply \tmmathbf{e}_{\alpha} \otimes \tmmathbf{e}_{\beta} .
  \label{eq:E-tmp}
\end{equation}

We denote as $\tmmathbf{Y}= (Y_1, Y_2, Y_3)$ and $\tmmathbf{Y}^{\dag} =
(Y^{\dag}_1, Y^{\dag}_2, Y^{\dag}_3)$ the collections of functions $Y_i$ and
$Y_i^{\dag}$ obtained by {\tmem{restricting}} the microscopic displacement and
its longitudinal gradient {\tmem{to a cross-section}}, {\tmem{i.e.}},
\[ \begin{array}{ll}
     Y_i = \nobracket y_i |_S & Y_i^{\dag} = \nobracket y_i' |_S .
   \end{array} \]
By convention, the dagger in $\tmmathbf{Y}^{\dag}$ means that this
cross-sectional function evaluates to a longitudinal gradient of strain,
{\tmem{i.e.}}, $Y_i^{\dag} (\tmmathbf{T}) = y_i' (S, \tmmathbf{T})$; daggers
are roughly equivalent to primes but strictly speaking the quantity
$\tmmathbf{Y}$ cannot bear a prime $' = \frac{\mathd}{\mathd S}$ as it is a
function of $\tmmathbf{T}$ only and not of $S$.

With this notation, the strain $\tmmathbf{E}$ from equation~(\ref{eq:E-tmp})
can be written as $\tmmathbf{E}=\tmmathbf{E} (\tmmathbf{T}; \tmmathbf{h} (S) ;
\nobracket \tmmathbf{y} |_S, \nobracket \tmmathbf{y}' |_S)$ where
\begin{equation}
  \begin{array}{r}
    \tmmathbf{E} (\tmmathbf{T}; \tmmathbf{h}; \tmmathbf{Y},
    \tmmathbf{Y}^{\dag}) = \frac{t_i^2 - 1}{2} \mymultiply \tmmathbf{e}_3
    \otimes \tmmathbf{e}_3 + t_i \mymultiply \partial_{\alpha} Y_i
    (\tmmathbf{T}) \mymultiply \tmmathbf{e}_{\alpha} \odot \tmmathbf{e}_3 +
    \frac{\partial_{\alpha} Y_i (\tmmathbf{T}) \mymultiply \partial_{\beta}
    Y_i (\tmmathbf{T}) - \delta_{\alpha \nocomma \beta}}{2} \mymultiply
    \tmmathbf{e}_{\alpha} \otimes \tmmathbf{e}_{\beta}\\[.4em]
    \text{where $t_i = (1 + \varepsilon) \mymultiply \delta_{i \nocomma 3} +
    \eta_{i \nocomma j \nocomma k} \mymultiply \kappa_j \mymultiply Y_k
    (\tmmathbf{T})$} + Y_i^{\dag} (\tmmathbf{T})
  \end{array} \label{eq:E-function}
\end{equation}
The dependence of $\tmmathbf{E}$ on $\tmmathbf{h}= (\varepsilon, \kappa_1,
\kappa_2, \kappa_3)$ arises through the auxiliary quantity $t_i$.

A couple comments on the notation $\tmmathbf{E} (\tmmathbf{T}; \tmmathbf{h}
(S) ; \tmmathbf{Y}= \nobracket \tmmathbf{y} |_S, \tmmathbf{Y}^{\dag} =
\nobracket \tmmathbf{y}' |_S)$ used in equation~(\ref{eq:E-function}) are in
order. The notation implies that the strain at any point $\tmmathbf{T}$ of the
cross-section can be calculated in terms of the local macroscopic strain
$\tmmathbf{h} (S)$, and of the {\tmem{restrictions}} of the displacement
$\nobracket \tmmathbf{y} |_S$ and of its longitudinal gradient $\nobracket
\tmmathbf{y}' |_S$ to the cross-section of interest. In particular, the
notation captures the fact that the strain does not depend on the higher-order
longitudinal gradients of displacement, such as $\tmmathbf{y}''_S$. Besides,
the gradients of the displacement along the cross-section directions, namely
$\partial_1 y_i = \frac{\partial y_i}{\partial T_1}$ and $\partial_2 y_i =
\frac{\partial y_i}{\partial T_2}$, are not listed as a argument to
$\tmmathbf{E} (\tmmathbf{T}; \tmmathbf{h}; \tmmathbf{Y}, \tmmathbf{Y}^{\dag})$
as they are reconstructed `internally' from the cross-sectional restriction
$\tmmathbf{Y}$ as $\partial_{\alpha} y_i (S, \tmmathbf{T}) = \partial_{\alpha}
Y_i (\tmmathbf{T})$. As a result, the dependence of the strain on longitudinal
gradients of the displacement is explicit in this notation, but that on
transverse gradients is not.

\subsection{Energy formulation}\label{sec:nonlinear-energy-formulation}

In the classical elasticity theory, the strain energy $\Phi$ is obtained by
integration of a strain energy density $w$,
\begin{equation}
  \Phi [\tmmathbf{h}, \tmmathbf{y}] = \int_0^{\ell} \iint_{\Omega} w
  (\tmmathbf{T}, \tmmathbf{E}) \mymultiply \mathd A \mymultiply \mathd S,
  \label{eq:canonicalForm}
\end{equation}
where the microscopic strain $\tmmathbf{E}$ appearing as an argument to $w$ is
given by equation~(\ref{eq:E-function}) as
\begin{equation}
  \tmmathbf{E}=\tmmathbf{E} (\tmmathbf{T}; \tmmathbf{h} (S) ; \nobracket
  \tmmathbf{y} |_S, \nobracket \tmmathbf{y}' |_S) .
  \label{eq:E-in-canonical-form}
\end{equation}
The bracket notation in $\Phi [\tmmathbf{h}, \tmmathbf{y}]$ indicates that
$\Phi$ is a functional of its arguments.

The form of the elastic potential in equation~(\ref{eq:canonicalForm}), which
serves as a starting point for our dimension reduction method, is completely
general. In particular, the following situations can be handled (not all of
which can be illustrated in this paper, by lack of space). The elastic
properties of the body can be inhomogeneous across the section, as indicated
by the explicit dependence of the density of strain energy $w (\tmmathbf{T},
\tmmathbf{E})$ on the transverse coordinate $\tmmathbf{T}$ in
equation~(\ref{eq:canonicalForm}). Arbitrary hyper-elastic constitutive laws
can be specified through the choice of the energy density $w$; in particular,
no assumption is made on the symmetries of the material. Arbitrary pre-stress
distributions can be taken into account by an appropriate choice of $w$, the
pre-stress being the quantity $\frac{\partial w}{\partial \tmmathbf{E}}
(\tmmathbf{T}, \tmmathbf{0})$. It is also possible to treat the case where the
elastic or geometric properties of the body vary slowly in the longitudinal
direction, as discussed in the conclusion.

We assume that the prismatic solid is subjected to conservative forces,
represented by a density of external potential $V (\tmmathbf{r},
\tmmathbf{d}_i)$. At equilibrium, the total potential energy
\begin{equation}
  \Psi [\tmmathbf{r}, \tmmathbf{d}_i, \tmmathbf{y}] = \Phi [\tmmathbf{h},
  \tmmathbf{y}] + \int_0^{\ell} V (\tmmathbf{r} (S), \tmmathbf{d}_i (S))
  \mymultiply \mathd S, \label{eq:full-problem-total-potential-energy}
\end{equation}
is stationary with respect to the unknowns $\tmmathbf{r}$, $\tmmathbf{d}_i$
and $\tmmathbf{y}$. The macroscopic strain $\tmmathbf{h} (S) = (\varepsilon
(S), \kappa_1 (S), \ldots, \kappa_3 (S))$ is a dependent variable which can be
obtained in terms of the main unknowns $\tmmathbf{r}$ and $\tmmathbf{d}_i$
using equations~(\ref{eq:rPrime-epsilon-d3}) and~(\ref{eq:kappa-i}).

The stationarity of the total potential
energy~(\ref{eq:full-problem-total-potential-energy}) is subject to the
condition~(\ref{eq:di-orthonormal-frame}) that the directors are orthonormal,
to the constraint of adaptation $\tmmathbf{r}' - (1 + \varepsilon) \mymultiply
\tmmathbf{d}_3 =\tmmathbf{0}$ in equation~(\ref{eq:rPrime-epsilon-d3}), and to
the kinematic constraints~(\ref{eq:yibar-kinematic-conditions}) on the
displacement. We rewrite the latter as
\begin{equation}
  \forall S \quad \tmmathbf{q} (\nobracket \tmmathbf{y} |_S) =\tmmathbf{0},
  \label{eq:constraint-q}
\end{equation}
where $\tmmathbf{q} (\tmmathbf{Y})$ lists the constraints applicable to the
cross-sectional restriction of the displacement $\tmmathbf{Y}= \nobracket
\tmmathbf{y} |_S$,
\begin{equation}
  \tmmathbf{q} (\tmmathbf{Y}) = \left( \langle Y_1 (\tmmathbf{T}) \rangle,
  \langle Y_2 (\tmmathbf{T}) \rangle, \langle Y_3 (\tmmathbf{T}) \rangle,
  \left\langle \eta_{\alpha \nocomma \beta} \mymultiply T_{\alpha} \mymultiply
  Y_{\beta} (\tmmathbf{T}) \right\rangle \right) . \label{eq:q-vector}
\end{equation}
The first three constraints prevent the center-line from drifting away from
the real material cross-sections---we use a redundant formulation where only
the sum $\tmmathbf{r}+ y_i \mymultiply \tmmathbf{d}_i$ is physically
meaningful, see equation~(\ref{eq:x-centerline-based-crspondence}).

In equation~(\ref{eq:full-problem-total-potential-energy}), the potential $V
(\tmmathbf{r}, \tmmathbf{d}_i)$ of the external load (per unit length $\mathd
S$) depends on the macroscopic variables but not on the microscopic
displacement. This is an assumption in our model. It can typically be
justified by the scaling hypotheses that are introduced in the classical work
on dimension reduction---typically, if the load varies on a length-scale much
larger than the cross-section diameter, its potential can be derived by
assuming that cross-sections are rigid, which yields an expression of the
form~(\ref{eq:full-problem-total-potential-energy}). If, however, the external
load varies quickly or induces large strain, it might become necessary to
couple the potential $V$ with the microscopic displacement $\tmmathbf{y}$.
This requires an extension of our work, which entails appending the
microscopic variables coupled to the external load as additional entries
inside the vector $\tmmathbf{h}$. This is however beyond the scope of the
present paper, where attention is limited to an external loading of the
form~(\ref{eq:full-problem-total-potential-energy}).

\subsection{Summary}

We have completed the energy formulation of the elasticity problem. In the
center-line based parameterization, the unknowns are the center-line
$\tmmathbf{r} (S)$, the directors $\tmmathbf{d}_i (S)$ and the microscopic
displacement $y_i (S, \tmmathbf{T})$. The center-line $\tmmathbf{r} (S)$ and
the directors $\tmmathbf{d}_i (S)$ define a framed curve which is associated
with a macroscopic strain $\tmmathbf{h} (S)$, where $\tmmathbf{h}=
(\varepsilon, \kappa_1, \kappa_2, \kappa_3)$, $\varepsilon$ is the axial
strain, $\kappa_1$ and $\kappa_2$ are the curvature strains and $\kappa_3$ is
the twisting strain, see section~\ref{ssec:apparent-strain}. The microscopic
strain is then given as $\tmmathbf{E}=\tmmathbf{E} (\tmmathbf{T}; \tmmathbf{h}
(S) ; \nobracket \tmmathbf{y} |_S, \nobracket \tmmathbf{y}' |_S)$ in
equation~(\ref{eq:E-function}). The total potential energy $\Psi
[\tmmathbf{r}, \tmmathbf{d}_i, \tmmathbf{y}]$ governing the elasticity problem
is given in equation~(\ref{eq:full-problem-total-potential-energy}), and in
particular the elastic strain energy $\Phi [\tmmathbf{h}, \tmmathbf{y}] =
\iiint w (\tmmathbf{T}, \tmmathbf{E}) \mymultiply \mathd A \mymultiply
\mathd S$ is given in equation~(\ref{eq:canonicalForm}). The equilibrium
equations can be derived variationally, taking into account the kinematic
constraints~(\ref{eq:constraint-q}) for the microscopic displacement, as well
as the orthonormality and the adaptation conditions in
equations~(\ref{eq:di-orthonormal-frame}) and~(\ref{eq:rPrime-epsilon-d3}).

\section{Ideal one-dimensional model}\label{sec:ideal-model}

In this section we explore a formal method for reducing the equilibrium of the
prismatic solid, which is a problem in three-dimensional elasticity, to a
one-dimensional problem. The reduction is based on the relaxation of the
microscopic displacement $\tmmathbf{y}$. The relaxation problem will be
introduced in a formal way in this section; it will not be solved explicitly
until we introduce additional assumptions in the forthcoming sections.

\subsection{Condensing out the microscopic displacement by a thought
experiment}

What we refer to as a \emph{relaxation of the microscopic displacement
$\tmmathbf{y}$} is a minimization of the strain energy functional $\Phi
[\tmmathbf{h}, \tmmathbf{y}]$ for a prescribed distribution of macroscopic
strain $\tmmathbf{h} (S)$,
\begin{equation}
  \Phi^{\star} [\tmmathbf{h}] = \min_{\tmmathbf{y} \text{ s.t. } (\forall S)
  \tmmathbf{q} (\nobracket \tmmathbf{y} |_S) =\tmmathbf{0}} \Phi
  [\tmmathbf{h}, \tmmathbf{y}] . \label{eq:relax-y}
\end{equation}
Note that the relaxation over {\tmstrong{$y$}} is subject to the kinematic
conditions $(\forall S)\; \tmmathbf{q} (\nobracket \tmmathbf{y} |_S)
=\tmmathbf{0}$ ensuring that the microscopic displacement is consistent with
the center-line deformation, prescribed through the macroscopic strain
$\tmmathbf{h}$.

We assume that the optimization problem for $\tmmathbf{y}$ is such that the
minimum is attained and denote as $\tmmathbf{y}=\tmmathbf{y}^{\star}
[\tmmathbf{h}]$ the optimum:
\begin{equation}
  \Phi^{\star} [\tmmathbf{h}] = \Phi [\tmmathbf{h}, \tmmathbf{y}^{\star}
  [\tmmathbf{h}]], \label{eq:phi-star-by-relaxation}
\end{equation}
where all quantities obtained by relaxing the microscopic displacement
$\tmmathbf{y}$ are marked with an asterisk.

We also assume that $\tmmathbf{y}^{\star} [\tmmathbf{h}]$ is the only
stationary point of $\Phi [\tmmathbf{h}, \tmmathbf{y}]$, so that
$\tmmathbf{y}^{\star} [\tmmathbf{h}]$ is characterized by the variational
problem
\begin{equation}
  \left( \forall \hat{\tmmathbf{y}} \text{ such that $\forall S\, \tmmathbf{q}
  (\nobracket \hat{\tmmathbf{y}} |_S) =\tmmathbf{0}$} \right) \quad
  \frac{\partial \Phi}{\partial \tmmathbf{y}} [\tmmathbf{h},
  \tmmathbf{y}^{\star} [\tmmathbf{h}]] \cdot \hat{\tmmathbf{y}} = 0.
  \label{eq:variational-eq-for-y-star}
\end{equation}
All these assumptions are typically satisfied under appropriate convexity and
compactness assumptions.

In equation~(\ref{eq:variational-eq-for-y-star}), the notation $\frac{\partial
f}{\partial \tmmathbf{y}} [\tmmathbf{h}, \tmmathbf{y}] \cdot
\hat{\tmmathbf{y}}$ refers to the Fr{\'e}chet derivative of the functional $f
[\cdummy]$ at point $(\tmmathbf{h}, \tmmathbf{y})$ in the direction
$\hat{\tmmathbf{y}}$. The problem for $\tmmathbf{y}^{\star} [\tmmathbf{h}]$
in~(\ref{eq:variational-eq-for-y-star}) is a non-linear elasticity problem
with pre-stress in three dimensions, and is typically impossible to solve in
closed form.

A key remark is as follows. If we were able to solve for the optimal
microscopic displacement $\tmmathbf{y}=\tmmathbf{y}^{\star} [\tmmathbf{h}]$,
we could define a one-dimensional strain energy potential $\Phi^{\star}$
simply by inserting $\tmmathbf{y}^{\star} [\tmmathbf{h}]$ into the
three-dimensional strain energy, $\Phi^{\star} [\tmmathbf{h}] = \Phi
[\tmmathbf{h}, \tmmathbf{y}^{\star} [\tmmathbf{h}]]$, see
equation~(\ref{eq:phi-star-by-relaxation}). Based on this strain energy
functional, one could then build a one-dimensional rod model governed by the
total potential energy functional
\begin{equation}
  \Psi^{\star} [\tmmathbf{r}, \tmmathbf{d}_i] = \Phi^{\star} [\tmmathbf{h}] +
  \int_0^{\ell} V (\tmmathbf{r} (S), \tmmathbf{d}_i (S)) \mymultiply \mathd S.
  \label{eq:ideal-1d-total-potential-energy}
\end{equation}
In this one-dimensional model, $\tmmathbf{r} (S)$ and $\tmmathbf{d}_i (S)$ are
the unknowns, subjected to the same kinematic conditions as earlier in
section~\ref{sec:full-model}, and the macroscopic strain $\tmmathbf{h}$ is a
dependent variable that can be calculated as earlier
({\textsection}\ref{ssec:apparent-strain}).

We refer to this model as the {\tmem{ideal one-dimensional model}}. It is
{\tmem{one-dimensional}} in the sense that it exposes the macroscopic
variables only, the microscopic displacement
$\tmmathbf{y}=\tmmathbf{y}^{\star} [\tmmathbf{h}]$ being reconstructed `under
the hood'. It is {\tmem{ideal}} in the sense that it is rigorously equivalent
to the three-dimensional elasticity problem from section~\ref{sec:full-model},
as shown in \ref{app-sec:original-ideal-same-equilibrium}. This shows that dimension reduction is really a relaxation problem.

\subsection{Equilibrium and constitutive laws}\label{ssec:ideal-equilibrium}

We derive the equilibrium equations and the constitutive laws of the ideal
one-dimensional model variationally, starting from the total energy potential
$\Psi^{\star} [\tmmathbf{r}, \tmmathbf{d}_i]$ in
equation~(\ref{eq:ideal-1d-total-potential-energy}).

The densities of external force $\tmmathbf{p} (S)$ and external moment
$\tmmathbf{m} (S)$ are first identified from the variation $\hat{V}$ of the
external potential as follows,
\begin{equation}
  \int_0^{\ell} \hat{V} \mymultiply \mathd S = - \int_0^{\ell} (\tmmathbf{p}
  (S) \cdot \hat{\tmmathbf{r}} (S) +\tmmathbf{m} (S) \cdot
  \hat{\tmmathbf{\theta}} (S)) \mymultiply \mathd S. \label{eq:V-peturb}
\end{equation}
where $\hat{\tmmathbf{r}}$ is the perturbation to the center-line and
$\hat{\tmmathbf{\theta}}$ the infinitesimal rotation of the directors
$\tmmathbf{d}_i$, such that $\hat{\tmmathbf{d}}_i (S) =
\hat{\tmmathbf{\theta}} (S) \times \tmmathbf{d}_i (S)$. As usual in the
principle of virtual work, we limit attention to perturbations
$\hat{\tmmathbf{r}}$ and $\hat{\tmmathbf{\theta}}$ such that the incremental
form of the kinematic constraint~(\ref{eq:rPrime-epsilon-d3}) is satisfied.

As shown in \ref{app-sec:equilibrium-ideal-model}, the
condition that the energy $\Psi^{\star} [\tmmathbf{r}, \tmmathbf{d}_i]$ is
stationary yields the extensible variant of the classical Kirchhoff equations
for the equilibrium of thin rods,
\begin{equation}
  \begin{gathered}
    N (S) =\tmmathbf{R} (S) \cdot \tmmathbf{d}_3 (S)\\[.2em]
    \tmmathbf{R}' (S) +\tmmathbf{p} (S) =\tmmathbf{0}\\[.2em]
    \tmmathbf{M}' (S) +\tmmathbf{r}' (S) \times \tmmathbf{R} (S) +\tmmathbf{m}
    (S) =\tmmathbf{0}
  \end{gathered} \label{eq:rod-strong-equilibrium}
\end{equation}
together with constitutive laws for the one-dimensional stress variables $N
(S)$ and $M_i (S)$. These constitutive laws can be identified from the first
variation of the strain energy (\ref{eq:phi-star-by-relaxation}) with respect
to the macroscopic strain as follows,
\begin{equation}
  N (S) \mymultiply \hat{\varepsilon} + M_i (S) \mymultiply \hat{\kappa}_i
  \equiv \iint_{\Omega} \Sigma_{i \nocomma j} (\tmmathbf{T}, \tmmathbf{E}
  (\tmmathbf{T}; \tmmathbf{h} (S) ; \nobracket \tmmathbf{y}^{\star}
  [\tmmathbf{h}] |_S, \nobracket \tmmathbf{y}^{\star} [\tmmathbf{h}]' |_S))
  \mymultiply \frac{\partial E_{i \nocomma j}}{\partial \tmmathbf{h}}
  (\tmmathbf{T}; \tmmathbf{h} (S) ; \nobracket \tmmathbf{y}^{\star}
  [\tmmathbf{h}] |_S, \nobracket \tmmathbf{y}^{\star} [\tmmathbf{h}]' |_S)
  \cdot \hat{\tmmathbf{h}} \mymultiply \mathd A.
  \label{eq:internal-stress-full-model}
\end{equation}
Here $\hat{\tmmathbf{h}} = (\hat{\varepsilon}, \hat{\kappa}_1, \hat{\kappa}_2,
\hat{\kappa}_3)$ is a perturbation to the macroscopic strain and
$\tmmathbf{\Sigma}$ is the microscopic Piola-Kirchhoff stress tensor,
\begin{equation}
  \tmmathbf{\Sigma} (\tmmathbf{T}, \tmmathbf{E}) = \frac{\partial w}{\partial
  \tmmathbf{E}} (\tmmathbf{T}, \tmmathbf{E}) . \label{eq:microscopic-stress}
\end{equation}

In
equations~(\ref{eq:rod-strong-equilibrium}--\ref{eq:internal-stress-full-model}),
$\tmmathbf{R} (S)$ is the internal force, its component $N (S)$ along
$\tmmathbf{d}_3 (S)$ is called the normal force, $\tmmathbf{M} (S)$ is the
internal moment and $M_i (S) =\tmmathbf{M} (S) \cdot \tmmathbf{d}_i (S)$ are
its components in the directors basis. A microscopic interpretation of the
internal stress $N (S)$ and $M_i (S)$ based on
equation~(\ref{eq:internal-stress-full-model}) is given in
equation~(\ref{eq:app-NM-interpretation}) from~\ref{app:microscopic-interpretation-1d-internal-stress}. The last two
lines in equation~(\ref{eq:rod-strong-equilibrium}) are the Kirchhoff
equations for the equilibrium of rods; they are a balance of forces and
moments on an infinitesimal segment, respectively. The equilibrium
equations~(\ref{eq:rod-strong-equilibrium}) must be complemented by boundary
conditions which can be derived variationally and vary from one problem to
another.

As discussed in \ref{app-sec:original-ideal-same-equilibrium},
equations~(\ref{eq:rod-strong-equilibrium}--\ref{eq:internal-stress-full-model})
governing the equilibrium of the ideal one-dimensional model are
mathematically equivalent to those governing the original three-dimensional
model from section~\ref{sec:full-model}. The one-dimensional model involves no
approximation. It achieves the ultimate in dimension reduction: it hides the
microscopic variables while preserving the solutions of the original
three-dimensional problem. Incidentally, it also makes the connection with the
classical Kirchhoff equations~(\ref{eq:rod-strong-equilibrium}) for elastic
rods. Unfortunately, the constitutive
laws~(\ref{eq:internal-stress-full-model}) are in effect useless as they
depend on the optimal microscopic displacement $\tmmathbf{y}^{\star}
[\tmmathbf{h}]$, which is not available in closed form: the one-dimensional
potential $\Phi^{\star} [\tmmathbf{h}]$ is a mathematical object that hides a
daunting problem in non-linear three-dimensional elasticity.

In the following section, we construct approximations to the ideal
one-dimensional model that are mathematically tractable.

\section{Asymptotically exact one-dimensional
models}\label{sec:asymptotic-1d-reduction}

\subsection{Strategy}

Even though it cannot be used directly, the ideal one-dimensional model from
section~\ref{sec:ideal-model} offers a natural starting point for building
one-dimensional approximations to the original three-dimensional problem. A
critical help towards this goal is furnished by our previous
work~\citep{LESTRINGANT2020103730}, in which a method to calculate the
relaxed displacement $\tmmathbf{y}^{\star} [\tmmathbf{h}]$ in powers of the
successive gradients of $\tmmathbf{h} (S)$ has been obtained. In this section,
we apply this asymptotic method and obtain approximations to the ideal strain
energy $\Phi^{\star} [\tmmathbf{h}]$. This leads us to {\tmem{concrete}} rod,
models which are accurate approximations of the original three-dimensional
problem when the gradients of the macroscopic strain $\tmmathbf{h} (S)$ are
small.

The reduction method from~\citet{LESTRINGANT2020103730} assumes that the
macroscopic strain varies on a length-scale $\sim \rho / \zeta$ much larger
than the typical dimension of the cross-section $\sim \rho$, where $\zeta \ll
1$ is a small scalar parameter that is used as an expansion parameter,
\[ \begin{array}{ll}
     \tmmathbf{h} (S) =\mathcal{O} (1) & \frac{\mathd^i \tmmathbf{h}}{\mathd
     S^i} (S) =\mathcal{O} (\zeta^i) .
   \end{array} \]
We emphasize that $\tmmathbf{h} (S)$ is allowed to vary by a finite amount
across the length $\ell$ of the structure as long as $\tmmathbf{h}' (S)
=\mathcal{O} (\zeta)$ remains small everywhere: unlike most (if not all) of
the alternate methods from the literature, ours does not require the strain
$\tmmathbf{h} (S)$ to remain {\tmem{uniformly}} close to a specific value
$\tmmathbf{h}_0$, for all values of $S$---this property is particularly useful
for the analysis of localization, as discussed
by~\citet{Audoly-Hutchinson-Analysis-of-necking-based-2016}
and~\citet{lestringant2020one}. Besides, the expansion has been shown to
give extremely accurate results even beyond its strict conditions of
mathematical validity, when the gradient $\tmmathbf{h}' (S)$ is not small.

The reduction method uses as input the expressions of the strain $\tmmathbf{E}
(\tmmathbf{T}; \tmmathbf{h} (S) ; \tmmathbf{Y}, \tmmathbf{Y}^{\dag})$ and the
constraint $\tmmathbf{q} (\tmmathbf{Y})$ relevant to our particular problem
from equations~(\ref{eq:E-function}) and~(\ref{eq:q-vector}), and furnishes an
approximation to the one-dimensional strain energy functional
\begin{equation}
  \Phi^{\star} [\tmmathbf{h}] \approx \Phi_{(2)}^{\star} [\tmmathbf{h}] + \ell
  \mymultiply \mathcal{O} (\zeta^3) \label{eq:reduced-model-energy-sketch}
\end{equation}
of the form
\begin{equation}
  \Phi_{(2)}^{\star} [\tmmathbf{h}] = \int_0^{\ell} \left[ W_{\text{hom}}
  (\tilde{\tmmathbf{h}} (S)) +\tmmathbf{A} (\tmmathbf{h} (S)) \cdot
  \tmmathbf{h}' (S) + \frac{1}{2} \mymultiply \tmmathbf{h}' (S) \cdot
  \tmmathbf{D} (\tmmathbf{h} (S) \nobracket \cdot \tmmathbf{h}' (S) \right]
  \mymultiply \mathd S, \label{eq:phi-gr}
\end{equation}
where $\tilde{h}_i = h_i (S) + \xi_i (\tmmathbf{h} (S)) \mymultiply h_i'' (S)$
(no implicit sum on $i$) is a modified strain measure, see
equation~(\ref{eq:hi-tilde}) below.

The reduction method of~\citet{LESTRINGANT2020103730} is summarized 
in \ref{app:compendium}. Explicit expressions for the potential
$W_{\text{hom}} (\tmmathbf{h})$, the coefficients $\xi_i (\tmmathbf{h})$
entering in the alternate strain measure $\tilde{\tmmathbf{h}}$, and for the
elastic moduli $\tmmathbf{A} (\tmmathbf{h})$ and $\tmmathbf{D} (\tmmathbf{h})$
are available. Both geometric and material nonlinearities are accounted for,
as reflected by the fact that the quantities $\xi_i$, $\tmmathbf{A}$ and
$\tmmathbf{D}$ all depend on $\tmmathbf{h}$, typically in a non-linear way.

A lower-order approximation $\Phi^{\star} [\tmmathbf{h}] \approx
\Phi_{(0)}^{\star} [\tmmathbf{h}] + \ell \mymultiply \mathcal{O} (\zeta)$ can
also be obtained by discarding the gradient terms \ in $\Phi_{(2)}^{\star}
[\tmmathbf{h}]$, which are of order $\zeta$ or higher,
\begin{equation}
  \Phi_{(0)}^{\star} [\tmmathbf{h}] = \int_0^{\ell} W_{\text{hom}}
  (\tmmathbf{h} (S)) \mymultiply \mathd S. \label{eq:phi-no-gradient}
\end{equation}
Unlike $\Phi_{(2)}^{\star} [\tmmathbf{h}]$, the strain potential
$\Phi_{(0)}^{\star} [\tmmathbf{h}]$ does not capture the gradient effect: the
strain energy $W_{\text{hom}} (\tmmathbf{h})$ in $\Phi_{(0)}^{\star}
[\tmmathbf{h}]$ is a function of the local strain $\tmmathbf{h} (S)$ only.

The term $\tmmathbf{A} (\tmmathbf{h} (S)) \cdot \tmmathbf{h}' (S)$ being
incompatible with the most common material symmetries, see
section~\ref{sssec:beam-symmetries}, the modulus $\tmmathbf{A} (\tmmathbf{h}
(S))$ is often zero. In this case, $\Phi_{(0)}^{\star} [\tmmathbf{h}]$ is a
better approximation than announced above, {\tmem{i.e.}}, \ $\Phi^{\star}
[\tmmathbf{h}] \approx \Phi_{(0)}^{\star} [\tmmathbf{h}] + \ell \mymultiply
\mathcal{O} (\zeta^2)$; by a similar argument the other estimate can be
improved as well in the presence of additional symmetries, {\tmem{i.e.}},
$\Phi^{\star} [\tmmathbf{h}] \approx \Phi_{(2)}^{\star} [\tmmathbf{h}] + \ell
\mymultiply \mathcal{O} (\zeta^4)$.

In the remainder of this section, we apply the reduction method from our
previous work to the prismatic solid. The potential $W_{\text{hom}}
(\tmmathbf{h})$ entering in the lower-order approximation $\Phi_{(0)}^{\star}
[\tmmathbf{h}]$ is derived in
section~\ref{ssec:non-regularized-model-outline}. The higher-order
approximation $\Phi_{(2)}^{\star} [\tmmathbf{h}]$ is derived in the subsequent
section~\ref{sec:gradient-effect}.

\subsection{Analysis of homogeneous
solutions}\label{ssec:non-regularized-model-outline}

As recalled in \ref{app:compendium-homogeneous}, the
elastic potential $W_{\text{hom}} (\tmmathbf{h})$ has to be constructed from a
catalog of homogeneous solutions. By homogeneous solutions, we refer to the
case where neither $\tmmathbf{h} (S)$ nor the microscopic displacement
$\tmmathbf{y} (S, \tmmathbf{T})$ depend on $S$. Homogeneous solutions are
analyzed in this section; accordingly, the macroscopic strain $\tmmathbf{h}=
(\varepsilon, \kappa_1, \kappa_2, \kappa_3)$ is treated as a constant. Doing
so, we are temporarily limiting attention to configurations of the center-line
that are a helix, an arc of circle or a straight line.

The optimal microscopic displacement $\tmmathbf{y}^{\star} [\tmmathbf{h}]$
being now independent of $S$, we denote by $\tmmathbf{Y}^{\tmmathbf{h}} =
\nobracket (\tmmathbf{y}^{\star} [\tmmathbf{h}]) |_S$ its restriction to any
particular cross-section $S$. Then,
\[ \tmmathbf{y}^{\star} [\tmmathbf{h}] (S, \tmmathbf{T})
   =\tmmathbf{Y}^{\tmmathbf{h}} (\tmmathbf{T}) \qquad \text{(homogeneous case,
   $\tmmathbf{h}$ is constant)} . \]
Here, $\tmmathbf{Y}^{\tmmathbf{h}} = (Y_i^{\tmmathbf{h}})_{1 \leqslant i
\leqslant 3}$ denotes a triple of functions defined over the cross-section,
each function $Y_i^{\tmmathbf{h}} (\tmmathbf{T})$ being a component of the
local displacement in the basis of directors, see
equation~(\ref{eq:x-centerline-based-crspondence}). The superscript
$\tmmathbf{h}$ in the notation $\tmmathbf{Y}^{\tmmathbf{h}}$ serves both as an
abbreviation for `homogeneous', and as a container for the macroscopic strain
values, on which $\tmmathbf{Y}^{\tmmathbf{h}} =\tmmathbf{Y}^{(\varepsilon,
\kappa_1, \kappa_2, \kappa_3)}$ depends.

From equation~(\ref{eq:E-function}), the strain $\tilde{\tmmathbf{E}}
(\tmmathbf{T}, \tmmathbf{h}, \tmmathbf{Y}) =\tmmathbf{E} (\tmmathbf{T};
\tmmathbf{h}; \tmmathbf{Y}, \tmmathbf{0})$ relevant to homogeneous solutions
writes
\begin{equation}
  \begin{array}{r}
    \tilde{\tmmathbf{E}} (\tmmathbf{T}, \tmmathbf{h}, \tmmathbf{Y}) =
    \frac{\tilde{t}_i^2 - 1}{2} \mymultiply \tmmathbf{e}_3 \otimes
    \tmmathbf{e}_3 + \tilde{t}_i \mymultiply \partial_{\alpha} Y_i
    (\tmmathbf{T}) \mymultiply \tmmathbf{e}_{\alpha} \odot \tmmathbf{e}_3 +
    \frac{\partial_{\alpha} Y_i (\tmmathbf{T}) \mymultiply \partial_{\beta}
    Y_i (\tmmathbf{T}) - \delta_{\alpha \nocomma \beta}}{2} \mymultiply
    \tmmathbf{e}_{\alpha} \otimes \tmmathbf{e}_{\beta}\\[.3em]
    \text{where $\tilde{t}_i = (1 + \varepsilon) \mymultiply \delta_{i
    \nocomma 3} + \eta_{i \nocomma j \nocomma k} \mymultiply \kappa_j
    \mymultiply Y_k (\tmmathbf{T})$} .
  \end{array} \label{eq:homogeneous-strain}
\end{equation}
This specific expression of the strain is derived from the generic one in
equation~(\ref{eq:E-function}) with the gradient term $\tmmathbf{Y}^{\dag}$
set to zero.

For any value of the macroscopic strain $\tmmathbf{h}$, the relaxed
displacement $\tmmathbf{Y}^{\tmmathbf{h}}$ of the homogeneous solution must be
found by minimizing the strain energy potential $\Phi$ with respect to
$\tmmathbf{Y}$ among all the microscopic displacements satisfying the
kinematic conditions $\tmmathbf{q} (\tmmathbf{Y}) =\tmmathbf{0}$, see
equations~(\ref{eq:relax-y}) or~(\ref{eq:variational-eq-for-y-star}). This
leads to the following variational problem, as derived in
equation~(\ref{eq:Yh-variational-abstract}) from the appendix,
\begin{equation}
	\begin{gathered}
		\iint_{\Omega} Y_i^{\tmmathbf{h}} (\tmmathbf{T}) \mymultiply \mathd A
    =\tmmathbf{0}\\
    \iint_{\Omega} \left[ T_1 \mymultiply Y_2^{\tmmathbf{h}} (\tmmathbf{T})
    - T_{2 \mymultiply} \mymultiply Y_1^{\tmmathbf{h}} (\tmmathbf{T}) \right]
    \mymultiply \mathd A =\tmmathbf{0}\\
    \forall \hat{\tmmathbf{Y}} \quad \iint_{\Omega} \left[
    \tmmathbf{\Sigma} (\tmmathbf{T}, \tmmathbf{E}^{\tmmathbf{h}}
    (\tmmathbf{T})) \doublecontract
    \widehat{\tilde{\tmmathbf{E}}}^{\tmmathbf{h}} (\tmmathbf{T}) +
    F_i^{\tmmathbf{h}} \mymultiply \hat{Y}_i (\tmmathbf{T}) + Q^{\tmmathbf{h}}
    \mymultiply \left( T_1 \mymultiply \hat{Y}_2 (\tmmathbf{T}) - T_{2
    \mymultiply} \mymultiply \hat{Y}_1 (\tmmathbf{T}) \right) \right]
    \mymultiply \mathd A = 0.
	\end{gathered}
 \label{eq:app-red-variational-pb-homogeneous}
\end{equation}
where $\widehat{\tilde{\tmmathbf{E}}}^{\tmmathbf{h}} (\tmmathbf{T}) =
\frac{\mathd \tilde{\tmmathbf{E}}}{\mathd \tmmathbf{Y}} (\tmmathbf{T},
\tmmathbf{h}, \tmmathbf{Y}^{\tmmathbf{h}}) \cdot \hat{\tmmathbf{Y}}$ is the
virtual change of strain, and the four scalars $(F_1^{\tmmathbf{h}},
F_2^{\tmmathbf{h}}, F_3^{\tmmathbf{h}}, Q^{\tmmathbf{h}})$ are Lagrange
multipliers enforcing the constraints $\tmmathbf{q}
(\tmmathbf{Y}^{\tmmathbf{h}}) = 0$ that have been spelled out in the first two
lines of equation~(\ref{eq:app-red-variational-pb-homogeneous}). This
variational problem is a two-dimensional, non-linear problem of elasticity in
the cross-section $\Omega$ with pre-strain that depends on $\tmmathbf{h}$. For
the simple examples given at the end of this paper, the solution
$\tmmathbf{Y}^{\tmmathbf{h}}$ will be obtained analytically but a numerical
solution might be required in more complex geometries.

Solving this variational problem repeatedly for all possible values of
$\tmmathbf{h}$, one obtains a catalog of homogeneous solutions
$\tmmathbf{Y}^{\tmmathbf{h}}$ indexed by the macroscopic
strain~$\tmmathbf{h}$. The elastic potential $W_{\text{hom}} (\tmmathbf{h})$
is then defined as the strain energy per unit length of the homogeneous
solution $\tmmathbf{Y}^{\tmmathbf{h}}$,
\begin{equation}
  W_{\text{hom}} (\tmmathbf{h}) = \iint_{\Omega} w (\tmmathbf{T},
  \tmmathbf{E}^{\tmmathbf{h}} (\tmmathbf{T})) \mymultiply \mathd A \text{,
  \quad$\tmop{where} \tmmathbf{E}^{\tmmathbf{h}} (\tmmathbf{T}) =
  \tilde{\tmmathbf{E}} (\tmmathbf{T}, \tmmathbf{h},
  \tmmathbf{Y}^{\tmmathbf{h}})$} . \label{eq:Wh-def}
\end{equation}
The lower-order one-dimensional strain energy potential $\Phi_{(0)}^{\star}
[\tmmathbf{h}]$ can then be readily constructed from
equation~(\ref{eq:phi-no-gradient}): most engineering models for slender
structures make use of the energy potential $\Phi_{(0)}^{\star}
[\tmmathbf{h}]$ which we have just obtained, see
equation~(\ref{eq:twisting-Phi0}) for instance.

In terms of the catalog of homogeneous solutions
$\tmmathbf{Y}^{\tmmathbf{h}}$, we introduce the following auxiliary quantities
relevant to the homogeneous solution,
\begin{equation}
  \begin{split}
    F^{\tmmathbf{h}}_{i \nocomma 3} (\tmmathbf{T}) & = (1 + \varepsilon)
    \mymultiply \delta_{i \nocomma 3} + \eta_{i \nocomma j \nocomma k}
    \mymultiply \kappa_j \mymultiply Y_k^{\tmmathbf{h}} (\tmmathbf{T})\\
    F^{\tmmathbf{h}}_{i \nocomma \alpha} (\tmmathbf{T}) & =
    \partial_{\alpha} Y_i^{\tmmathbf{h}} (\tmmathbf{T})\\
    \tmmathbf{E}^{\tmmathbf{h}} (\tmmathbf{T}) & = \tmmathbf{E}
    (\tmmathbf{T}; \tmmathbf{h}; \tmmathbf{Y}^{\tmmathbf{h}}, \tmmathbf{0})\\
    \tmmathbf{\Sigma}^{\tmmathbf{h}} (\tmmathbf{T}) & = \tmmathbf{\Sigma}
    (\tmmathbf{T}, \tmmathbf{E}^{\tmmathbf{h}} (\tmmathbf{T}))\\
    \tmmathbf{K}^{\tmmathbf{h}} (\tmmathbf{T}) & = \frac{\partial^2
    w}{\partial \tmmathbf{E}^2} (\tmmathbf{T}, \tmmathbf{E}^{\tmmathbf{h}}
    (\tmmathbf{T}))
  \end{split} \label{eq:gr-effect-homogeneous-qties}
\end{equation}
where $F^{\tmmathbf{h}}_{i \nocomma j} (\tmmathbf{T})$ are the components of
the deformation gradient $\tmmathbf{F}^{\tmmathbf{h}} (S, \tmmathbf{T}) = F_{i
\nocomma j}^{\tmmathbf{h}} (\tmmathbf{T}) \mymultiply \tmmathbf{d}_i (S)
\otimes \tmmathbf{e}_j$, $\tmmathbf{E}^{\tmmathbf{h}}$ is the microscopic
strain, $\tmmathbf{\Sigma}^{\tmmathbf{h}}$ the microscopic stress, and
$\tmmathbf{K}^{\tmmathbf{h}}$ is the matrix of tangent elastic moduli.

\subsection{Analysis of the gradient effect}\label{sec:gradient-effect}

This section aims at deriving the higher-order approximation
$\Phi_{(2)}^{\star} [\tmmathbf{h}]$ to the strain energy, that captures the
gradient effect. We do so by following the general method from
section~\ref{sec-app:general-method-gradient} in \ref{app:compendium}.

Given a distribution of macroscopic stress $\tmmathbf{h} (S)$, the idea of the
method is to seek the solution to the relaxation problem~(\ref{eq:relax-y}) in
the form
\begin{equation}
  \tmmathbf{y}^{\star} [\tmmathbf{h}] (S, \tmmathbf{T})
  =\tmmathbf{Y}^{\tmmathbf{h} (S)} (\tmmathbf{T})
  +\tmmathbf{Z}_{\text{opt}}^{\tmmathbf{h} (S)} (\tmmathbf{h}' (S),
  \tmmathbf{T}) +\mathcal{O} (\zeta^2), \label{eq:y-star-expansion}
\end{equation}
where $\tmmathbf{Y}^{\tmmathbf{h} (S)}$ is the displacement predicted by the
catalog of homogeneous solutions based on the local value $\tmmathbf{h} (S)$
of the macroscopic strain, $\tmmathbf{Z}_{\text{opt}}^{\tmmathbf{h} (S)}
(\tmmathbf{T}, \tmmathbf{h}' (S))$ is a correction proportional to the local
strain gradient $\tmmathbf{h}' (S)$, to be determined, and $\mathcal{O}
(\zeta^2)$ denotes higher-order terms which do not enter in the determination
of $\Phi_{(2)}^{\star} [\tmmathbf{h}]$. We proceed to show how the correction
$\tmmathbf{Z}_{\text{opt}}^{\tmmathbf{h} (S)} (\tmmathbf{h}' (S),
\tmmathbf{T})$ can be obtained, which is a first step towards constructing the
functional $\Phi_{(2)}^{\star} [\tmmathbf{h}]$.

\subsubsection{Optimal displacement}

As shown in previous work and summarized in \ref{app:compendium}, the
optimal correction $\tmmathbf{Z}_{\text{opt}}^{\tmmathbf{h} (S)}
(\tmmathbf{T}, \tmmathbf{h}' (S))$ can be found by solving a variational
problem on the cross-section that effectively enforces the optimality
condition~(\ref{eq:relax-y}). This variational problem makes use of an
operator $\mathcal{B}^{\tmmathbf{h}} (\tmmathbf{h}^{\dag}, \tmmathbf{Z})$ that
takes the strain $\tmmathbf{h}$, its gradient $\tmmathbf{h}^{\dag}$ and a
generic displacement field $\tmmathbf{Z}$ defined on the cross-section as
arguments. We follow the step-by-step recipe from the appendix to build the
operator $\mathcal{B}^{\tmmathbf{h}} (\tmmathbf{h}^{\dag}, \tmmathbf{Z})$,
based on the knowledge of the homogeneous solutions
$\tmmathbf{Y}^{\tmmathbf{h}}$. Auxiliary operators need to be introduced in
this process.

The first step is to identify the {\tmem{structure operators}}
$\tmmathbf{e}_{i \nocomma j}^k$, which are the gradients with respect to
$\tmmathbf{h}$, $\tmmathbf{h}^{\dag}$, $\tmmathbf{Y}$ and
$\tmmathbf{Y}^{\dag}$ of the microscopic strain function $\tmmathbf{E}
(\tmmathbf{T}; \tmmathbf{h}; \tmmathbf{Y}, \tmmathbf{Y}^{\dag})$ in
equation~(\ref{eq:E-function}) about a homogeneous solution. These structure
operators are defined in \ref{app-sec:structure-operators} and are
calculated in \ref{app:structure-operators}. They are purely
geometric quantities.

Next, the linear increment of strain $\mathcal{E}^{\tmmathbf{h}}
(\tmmathbf{T}, \tmmathbf{h}^{\dag}, \tmmathbf{Z})$ associated with a small
strain gradient $\tmmathbf{h}^{\dag} = (\varepsilon^{\dag}, \kappa_1^{\dag},
\kappa_2^{\dag}, \kappa_3^{\dag})$ and with the corrective displacement
$\tmmathbf{Z}$ are obtained from equation~(\ref{eq:perturbed-strain-Ecal}) as
\begin{equation}
  \mathcal{E}^{\tmmathbf{h}} (\tmmathbf{T}, \tmmathbf{h}^{\dag}, \tmmathbf{Z})
  = (\tmmathbf{h}^{\dag} \cdot \nabla Y^{\tmmathbf{h}}_i (\tmmathbf{T}))
  \mymultiply F^{\tmmathbf{h}}_{i \nocomma j} (\tmmathbf{T}) \mymultiply
  \tmmathbf{e}_j \odot \tmmathbf{e}_3 + \eta_{i \nocomma j \nocomma k}
  \mymultiply \kappa_k \mymultiply F^{\tmmathbf{h}}_{j \nocomma l}
  (\tmmathbf{T}) \mymultiply Z_i (\tmmathbf{T}) \mymultiply \tmmathbf{e}_l
  \odot \tmmathbf{e}_3 + F^{\tmmathbf{h}}_{i \nocomma j} (\tmmathbf{T})
  \mymultiply \partial_{\alpha} Z_i (\tmmathbf{T}) \tmmathbf{e}_j \odot
  \tmmathbf{e}_{\alpha} . \label{eq:Lh}
\end{equation}
The last argument $\tmmathbf{Z}= (Z_1, Z_2, Z_3)$ is a triple of functions
defined on the cross-section, representing the corrective displacement;
$\mathcal{E}^{\tmmathbf{h}}$ is viewed as an operator acting on the
cross-sectional functions $(Z_1, Z_2, Z_3)$ that are not yet known.

The $\nabla$ notation is systematically used to denote a gradient with respect
to the macroscopic gradient $\tmmathbf{h}$, with the convention that the
increment of $\tmmathbf{h}$ is applied by a left multiplication: in
equation~(\ref{eq:Lh}), the first term in the right-hand side must be
interpreted as
\begin{equation}
  \begin{split}
    \tmmathbf{h}^{\dag} \cdot \nabla Y^{\tmmathbf{h}}_i (\tmmathbf{T}) & = 
    \frac{\partial Y^{\tmmathbf{h}}_i (\tmmathbf{T})}{\partial \tmmathbf{h}}
    \cdot \tmmathbf{h}^{\dag}\\
    & =  \frac{\partial Y^{(\varepsilon, \kappa_1, \kappa_2, \kappa_3)}_i
    (\tmmathbf{T})}{\partial \varepsilon} \mymultiply \varepsilon^{\dag} +
    \frac{\partial Y^{(\varepsilon, \kappa_1, \kappa_2, \kappa_3)}_i
    (\tmmathbf{T})}{\partial \kappa_1} \mymultiply \kappa_1^{\dag}
  \end{split} \label{eq:nabla-notation}
\end{equation}

Following the general method, we introduce yet another operator
$\tmmathbf{C}_{\tmmathbf{h}}^{(1)}$. For a given value of the macroscopic
strain $\tmmathbf{h}$ and for a triple $\tmmathbf{Z}^{\dag}$ of scalar
functions $Z_i^{\dag}$ defined over the cross-sections (representing the
components of the longitudinal gradient of the corrective displacement, hence
the dagger notation), $\tmmathbf{C}_{\tmmathbf{h}}^{(1)}$ is defined from
equation~(\ref{eq:ACB-operators}) as
\begin{equation}
  \tmmathbf{C}_{\tmmathbf{h}}^{(1)} \cdot \tmmathbf{Z}^{\dag} =
  \iint_{\Omega} F^{\tmmathbf{h}}_{i \nocomma j} (\tmmathbf{T}) \mymultiply
  \Sigma_{j \nocomma 3}^{\tmmathbf{h}} (\tmmathbf{T}) \mymultiply Z_i^{\dag}
  (\tmmathbf{T}) \mymultiply \mathd A. \label{eq:A-C1h}
\end{equation}
A related operator $\nabla \tmmathbf{C}^{(1)}_{\tmmathbf{h}}$ is defined by a
formal integration by parts with respect to the $\dag$ symbol (formally
representing the longitudinal derivative $\mathd / \mathd S$), see
equation~(\ref{eq:minus-grad-C1h-abstract}),
\begin{equation}
  -\tmmathbf{h}^{\dag} \cdot \nabla \tmmathbf{C}^{(1)}_{\tmmathbf{h}} \cdot
  \tmmathbf{Z}= - \iint_{\Omega} \left( \frac{\mathd \left(
  F^{\tmmathbf{h}}_{i \nocomma j} (\tmmathbf{T}) \mymultiply \Sigma_{j
  \nocomma 3}^{\tmmathbf{h}} (\tmmathbf{T}) \right)}{\mathd \tmmathbf{h}}
  \cdot \tmmathbf{h}^{\dag} \right) \mymultiply Z_i (\tmmathbf{T}) \mymultiply
  \mathd A. \label{eq:minus-grad-C1h}
\end{equation}

In our previous work, we have shown that the perturbation to the strain energy
per unit length caused by a strain gradient $\tmmathbf{h}^{\dag}$ and by the
corrective displacement $\tmmathbf{Z}$ is given by the operator
$\mathcal{B}^{\tmmathbf{h}} (\tmmathbf{h}^{\dag}, \tmmathbf{Z})$ defined in
equation~(\ref{eq:ACB-operators}) as
  \begin{multline}
    \mathcal{B}^{\tmmathbf{h}} (\tmmathbf{h}^{\dag}, \tmmathbf{Z}) =
    \iint_{\Omega} \frac{1}{2} \mymultiply \mathcal{E}^{\tmmathbf{h}}
    (\tmmathbf{T}, \tmmathbf{h}^{\dag}, \tmmathbf{Z}) \doublecontract
    \tmmathbf{K}^{\tmmathbf{h}} (\tmmathbf{T}) \doublecontract
    \mathcal{E}^{\tmmathbf{h}} (\tmmathbf{T}, \tmmathbf{h}^{\dag},
    \tmmathbf{Z}) \mymultiply \mathd A \ldots\\
    \nobracket \nobracket \hspace{2em} + \iint_{\Omega} \left( \left(
    \frac{1}{2} \mymultiply \sum_i (\tmmathbf{h}^{\dag} \cdot \nabla
    Y^{\tmmathbf{h}}_i (\tmmathbf{T}))^2 \mymultiply \Sigma_{3 \nocomma
    3}^{\tmmathbf{h}} (\tmmathbf{T}) \right) +\tmmathbf{h}^{\dag} \cdot \nabla
    Y^{\tmmathbf{h}}_i (\tmmathbf{T}) \mymultiply \left( \eta_{i \nocomma j
    \nocomma k} \mymultiply \kappa_j \mymultiply \Sigma_{3 \nocomma
    3}^{\tmmathbf{h}} (\tmmathbf{T}) \mymultiply Z_k (\tmmathbf{T}) +
    \Sigma_{\beta \nocomma 3}^{\tmmathbf{h}} (\tmmathbf{T}) \mymultiply
    \partial_{\beta} Z_i (\tmmathbf{T}) \right) \right. \ldots\\
    \hspace{6em} + \frac{1}{2} \mymultiply \left( \delta_{i \nocomma j}
    \mymultiply \kappa_l^2 - \kappa_i \mymultiply \kappa_j \right) \mymultiply
    \Sigma_{3 \nocomma 3}^{\tmmathbf{h}} (\tmmathbf{T}) \mymultiply Z_i
    (\tmmathbf{T}) \mymultiply Z_j (\tmmathbf{T}) + \eta_{i \nocomma j
    \nocomma k} \mymultiply \kappa_j \mymultiply \Sigma_{\alpha \nocomma
    3}^{\tmmathbf{h}} (\tmmathbf{T}) \mymultiply Z_k (\tmmathbf{T})
    \mymultiply \partial_{\alpha} Z_i (\tmmathbf{T}) \ldots\\
    \hspace{8em} \left. + \frac{1}{2} \mymultiply \Sigma_{\alpha \nocomma
    \beta}^{\tmmathbf{h}} (\tmmathbf{T}) \mymultiply \partial_{\alpha} Z_i
    (\tmmathbf{T}) \mymultiply \partial_{\beta} Z_i (\tmmathbf{T}) \right)
    \mymultiply \mathd A -\tmmathbf{h}^{\dag} \cdot \nabla
    \tmmathbf{C}^{(1)}_{\tmmathbf{h}} \cdot \tmmathbf{Z}.
	\label{eq:B-cal}
  \end{multline} 
This operator $\mathcal{B}^{\tmmathbf{h}} (\tmmathbf{h}^{\dag}, \tmmathbf{Z})$
is a quadratic form with respect to each one of its arguments
$\tmmathbf{h}^{\dag}$ and $\tmmathbf{Z}$.

The optimal corrective displacement $\tmmathbf{Z}_{\text{opt}}^{\tmmathbf{h}}
(\tmmathbf{h}^{\dag})$ is characterized by the fact that it is the stationary
point of the quadratic functional $\mathcal{B}^{\tmmathbf{h}}
(\tmmathbf{h}^{\dag}, \tmmathbf{Z})$ over the set of cross-sectional functions
$\tmmathbf{Z}$'s satisfying the kinematic constraint $\tmmathbf{q}
(\tmmathbf{Z}) =\tmmathbf{0}$; this is fully in line with the interpretation
of $\mathcal{B}^{\tmmathbf{h}} (\tmmathbf{h}^{\dag}, \tmmathbf{Z})$ as the
increment of strain energy caused by the gradient effect. This leads to the
variational problem stated in equation~(\ref{eq:Z-variational-pb-abstract}):
given $\tmmathbf{h}$ and $\tmmathbf{h}^{\dag}$, find the corrective
cross-sectional displacement $\tmmathbf{Z}_{\text{opt}}^{\tmmathbf{h}}
(\tmmathbf{h}^{\dag})$ and the Lagrange multipliers
$\tmmathbf{F}_{\text{opt}}^{\tmmathbf{h}} (\tmmathbf{h}^{\dag})$ and
$Q_{\text{opt}}^{\tmmathbf{h}} (\tmmathbf{h}^{\dag})$ such that
\begin{equation}
  \begin{gathered}
    \iint_{\Omega} Z_{\text{opt}, i}^{\tmmathbf{h}} (\tmmathbf{h}^{\dag},
    \tmmathbf{T}) \mymultiply \mathd A =\tmmathbf{0}\\
    \iint_{\Omega} \left[ T_1 \mymultiply Z_{\text{opt}, 2}^{\tmmathbf{h}}
    (\tmmathbf{h}^{\dag}, \tmmathbf{T}) - T_{2 \mymultiply} \mymultiply
    Z_{\text{opt}, 1}^{\tmmathbf{h}} (\tmmathbf{h}^{\dag}, \tmmathbf{T})
    \right] \mymultiply \mathd A =\tmmathbf{0}\\
    \forall \hat{\tmmathbf{Z}} \quad \frac{\partial
    \mathcal{B}^{\tmmathbf{h}}}{\partial \tmmathbf{Z}} \left(
    \tmmathbf{h}^{\dag}, \tmmathbf{Z}_{\text{opt}}^{\tmmathbf{h}}
    (\tmmathbf{h}^{\dag}) \right) \cdot \hat{\tmmathbf{Z}} + \iint_{\Omega}
    \left[ F_{\text{opt}, i}^{\tmmathbf{h}} (\tmmathbf{h}^{\dag}) \mymultiply
    \hat{Z}_i (\tmmathbf{T}) + Q^{\tmmathbf{h}}_{\text{opt}}
    (\tmmathbf{h}^{\dag}) \mymultiply \left( T_1 \mymultiply \hat{Z}_2
    (\tmmathbf{T}) - T_{2 \mymultiply} \mymultiply \hat{Z}_1 (\tmmathbf{T})
    \right) \right] \mymultiply \mathd A = 0,
  \end{gathered} \label{eq:Z-variational-pb}
\end{equation}
As $\mathcal{B}^{\tmmathbf{h}} (\tmmathbf{h}^{\dag}, \tmmathbf{Z})$ is
quadratic, this is a two-dimensional problem of linear elasticity in the
cross-section, with residual strain proportional to $\tmmathbf{h}^{\dag}$: its
solution $\tmmathbf{Z}_{\text{opt}}^{\tmmathbf{h}} (\tmmathbf{T},
\tmmathbf{h}^{\dag})$ is linear with respect to the strain gradient
$\tmmathbf{h}^{\dag}$.

This completes the determination of the corrective displacement
$\tmmathbf{Z}_{\text{opt}}^{\tmmathbf{h}} (\tmmathbf{h}^{\dag})$ as a function
of the macroscopic strain $\tmmathbf{h}$ and its longitudinal gradient
$\tmmathbf{h}^{\dag}$.

\subsubsection{Definition of the one-dimensional model}

We have solved for the corrective displacement
$\tmmathbf{Z}_{\text{opt}}^{\tmmathbf{h} (S)} (\tmmathbf{h}' (S),
\tmmathbf{T})$. There remains to insert the relaxed displacement
$\tmmathbf{y}^{\star} [\tmmathbf{h}]$ from
equation~(\ref{eq:y-star-expansion}) into the original strain energy
$\Phi^{\star} [\tmmathbf{h}] = \Phi [\tmmathbf{h}, \tmmathbf{y}^{\star}
[\tmmathbf{h}]]$. This yields the reduced strain energy $\Phi_{(2)}^{\star}
[\tmmathbf{h}]$ announced earlier in equation~(\ref{eq:phi-gr}),
\[ \Phi_{(2)}^{\star} [\tmmathbf{h}] = \int_0^{\ell} \left[ W_{\text{hom}}
   (\tilde{\tmmathbf{h}} (S)) +\tmmathbf{A} (\tmmathbf{h} (S)) \cdot
   \tmmathbf{h}' (S) + \frac{1}{2} \mymultiply \tmmathbf{h}' (S) \cdot
   \tmmathbf{D} (\tmmathbf{h} (S) \nobracket \cdot \tmmathbf{h}' (S) \right]
   \mymultiply \mathd S, \]
together with explicit expressions for the elastic moduli $\tmmathbf{A}
(\tmmathbf{h})$, $\tmmathbf{D} (\tmmathbf{h})$ and for the modified strain
$\tilde{\tmmathbf{h}} (S)$, which are obtained as follows---the reader is
referred to \ref{app:compendium} for details.

The auxiliary one-dimensional moduli $\tmmathbf{B} (\tmmathbf{h})$ and
$\tmmathbf{C} (\tmmathbf{h})$ are obtained by inserting the optimal
displacement into the operators $\mathcal{B}^{\tmmathbf{h}}$ and
$\tmmathbf{C}_{\tmmathbf{h}}^{(1)}$ introduced earlier, and by identification
from the following equations,
\begin{equation}
	\begin{aligned}
    \frac{1}{2} \mymultiply \tmmathbf{h}^{\dag} \cdot \tmmathbf{B}
    (\tmmathbf{h}) \cdot \tmmathbf{h}^{\dag} & = \mathcal{B}^{\tmmathbf{h}}
    \left( \tmmathbf{h}^{\dag}, \tmmathbf{Z}_{\text{opt}}^{\tmmathbf{h}}
    (\tmmathbf{h}^{\dag}) \right)\\
    \tmmathbf{C} (\tmmathbf{h}) \cdot \tmmathbf{h}^{\dag} & =
    \tmmathbf{C}_{\tmmathbf{h}}^{(1)} \cdot
    \tmmathbf{Z}_{\text{opt}}^{\tmmathbf{h}} (\tmmathbf{h}^{\dag}) .
\end{aligned}
	\label{eq:B-C}
\end{equation} 
The one-dimensional moduli $\tmmathbf{A} (\tmmathbf{h})$ and $\tmmathbf{D}
(\tmmathbf{h})$ appearing in $\Phi_{(2)}^{\star} [\tmmathbf{h}]$ write, from
section~\ref{sec-app:elimination-of-boundary-terms},
\begin{equation}
	\begin{aligned}
    \tmmathbf{A} (\tmmathbf{h}) \cdot \tmmathbf{h}^{\dag} & =
    \iint_{\Omega} \tmmathbf{\Sigma}^{\tmmathbf{h}} (\tmmathbf{T})
    \doublecontract \mathcal{E}^{\tmmathbf{h}} (\tmmathbf{T},
    \tmmathbf{h}^{\dag}, \tmmathbf{0}) \mymultiply \mathd A\\
    \tmmathbf{D} (\tmmathbf{h}) & = \tmmathbf{B} (\tmmathbf{h}) + 2
    \mymultiply \frac{\mathd \tmmathbf{C}}{\mathd \tmmathbf{h}} (\tmmathbf{h})
    \end{aligned}
	\label{eq:D-of-h}
\end{equation}
and the modified strain $\tilde{\tmmathbf{h}} (S)$ is defined by
\begin{equation}
  \begin{aligned}
    \tilde{h}_i (S) & =  h_i (S) + \xi_i (\tmmathbf{h} (S)) \mymultiply h_i''
    (S) \text{\quad (no sum on $i$)}\\
    \xi_i (\tmmathbf{h}) & =  \frac{C_i (\tmmathbf{h})}{\frac{\partial
    W_{\text{hom}}}{\partial h_i} (\tmmathbf{h})},
  \end{aligned} \label{eq:hi-tilde}
\end{equation}
where $C_i (\tmmathbf{h})$ is the $i$-th component of $\tmmathbf{C}
(\tmmathbf{h})$, {\tmem{i.e.}}, the coefficient in factor of $h_i^{\dag}$ in
$\tmmathbf{C} (\tmmathbf{h}) \cdot \tmmathbf{h}^{\dag}$, see
equation~(\ref{eq:Ci-in-practice}).

This completes the construction of the functional $\Phi_{(2)}^{\star}
[\tmmathbf{h}]$. The process is long but straightforward. It can be turned
into a fully automated procedure using symbolic calculations, something which
we will explore in future work. In the remainder of this paper, we illustrate
the procedure by carrying out the calculations for two problems that are
tractable analytically; the first problem is linear and the other one is
non-linear.

\section{Illustration in a linear setting: twisting of a thick
bar}\label{sec:twisting}

\begin{figure}
  \centerline{\includegraphics{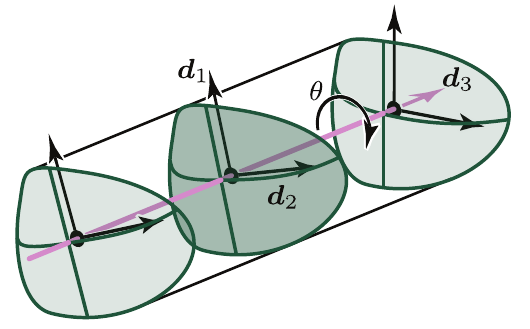}}
  \caption{Pure twisting of a thick bar: actual
  configuration.\label{eq:fig-twisting}}
\end{figure}

A number of authors have addressed higher-order effects in beam models for
prismatic solids in the limited context of linear elasticity
\citep{trabucho1989existence,nolde2018asymptotic,buannic2001higher,buannic2001higher2}.
In this section, we derive a higher-order model for the twisting of a
prismatic bar using our method; we show that its predictions are consistent
with prior work from the literature. This provides a first illustration of the
reduction procedure described in section~\ref{sec:asymptotic-1d-reduction}.

\subsection{Problem setting}

We consider the pure torsion of a linearly elastic bar including higher-order
effect. To simplify the calculations, we make some convenience assumptions.
First, the elastic material is assumed to be linear and isotropic with
homogeneous properties. This corresponds to a microscopic strain energy
density
\begin{equation}
  w (\tmmathbf{T}, \tmmathbf{E}) = \frac{1}{2} \mymultiply \left( \lambda
  \mymultiply \tmop{tr}^2 \tmmathbf{E}+ 2 \mymultiply \mu \mymultiply
  \tmmathbf{E} \doublecontract \tmmathbf{E} \right), \label{eq:linear-w}
\end{equation}
where $\lambda$ and $\mu$ are the Lam{\'e} elastic moduli. Second, we assume
that the cross-section $\Omega$ has two perpendicular axes of symmetry, and we
set up the cross-section coordinates $(T_1, T_2)$ along these axes; as a
result, the cross-section is invariant by the two mirror symmetries
$T_{\alpha} \longleftarrow (- T_{\alpha})$. This symmetry assumption decouples
the twisting mode from the bending and stretching
modes~\citep{trabucho1989existence}. It allows us to analyze twisting while
setting $\varepsilon = 0$ (no stretching) and $\kappa_{\alpha} = 0$ (no
bending). We will therefore have to deal with a single non-zero macroscopic
strain $\kappa_3$ (twisting), which we rename as $\tau = \kappa_3$.
Accordingly, we shrink the macroscopic strain $\tmmathbf{h}$ to a vector with
length 1,
\[ \tmmathbf{h}= (\tau) . \]

In the context of linear elasticity, the main unknown is the (true)
displacement $v_i (S, \tmmathbf{T})$ from the reference to the actual
configuration. It is connected to the microscopic positional unknown $y_i (S,
\tmmathbf{T})$ used so far by
\begin{equation}
  \begin{array}{rcccc}
    y_{\alpha} (S, \tmmathbf{T}) & = & T_{\alpha} & + & v_{\alpha} (S,
    \tmmathbf{T})\\
    y_3 (S, \tmmathbf{T}) & = &  &  & v_3 (S, \tmmathbf{T})
  \end{array} \label{eq:y-to-v}
\end{equation}
Inserting the equation above into equation~(\ref{eq:E-function}), one obtains
the strain $\tmmathbf{E}$ as
\begin{equation}
  \tmmathbf{E} (\tmmathbf{T}; \tau ; \tmmathbf{V}, \tmmathbf{V}^{\dag}) = -
  \tau \mymultiply \eta_{\alpha \nocomma \beta} \mymultiply T_{\beta}
  \mymultiply \tmmathbf{e}_{\alpha} \odot \tmmathbf{e}_3 + \partial_{\alpha}
  V_i (\tmmathbf{T}) \mymultiply \tmmathbf{e}_{\alpha} \odot \tmmathbf{e}_i +
  V_i^{\dag} (\tmmathbf{T}) \mymultiply \tmmathbf{e}_i \odot \tmmathbf{e}_3 .
  \label{eq:E-twist-linear}
\end{equation}
Here, $\tmmathbf{V}$ and $\tmmathbf{V}^{\dag}$ denote the restrictions of the
displacement and its longitudinal gradient to a particular cross-section,
\begin{equation}
  \begin{array}{ll}
    V_i = \nobracket v_i |_S & V_i^{\dag} = \nobracket v_i^{\dag} |_S
  \end{array} \label{eq:V-restrictions}
\end{equation}
In linear elasticity, the cross-sectional {\tmem{displacements}}
$\tmmathbf{V}$ and $\tmmathbf{V}^{\dag}$ are used as the unknowns
parameterizing the microscopic displacement, instead of the cross-sectional
positions $\tmmathbf{Y}$ and $\tmmathbf{Y}^{\dag}$ relevant to the non-linear
setting.

As we are working in the context of linear elasticity, we linearize all
quantities with respect to the twisting strain $\tau$, to the displacement
$\tmmathbf{V}$ and to its longitudinal gradient $\tmmathbf{V}^{\dag}$. Such a
linearization has been carried out silently in the right-hand side of
equation~(\ref{eq:E-twist-linear}), in particular.

In the forthcoming sections, we apply the method from
section~\ref{sec:asymptotic-1d-reduction} and derive a one-dimensional model
describing the twisting of the linearly elastic bar that includes the gradient
effect.

\subsection{Analysis of homogeneous solutions}

We first focus on homogeneous solutions, obtained by discarding the gradient
of the displacement in the local basis, $\tmmathbf{V}^{\dag} =\tmmathbf{0}$,
in equation~(\ref{eq:E-twist-linear}). This yields the strain of homogeneous
solutions as
\[ \tilde{\tmmathbf{E}} (\tmmathbf{T}, \tau, \tmmathbf{V}) = - \tau
   \mymultiply \eta_{\alpha \nocomma \beta} \mymultiply T_{\beta} \mymultiply
   \tmmathbf{e}_{\alpha} \odot \tmmathbf{e}_3 + \partial_{\alpha} V_i
   (\tmmathbf{T}) \mymultiply \tmmathbf{e}_{\alpha} \odot \tmmathbf{e}_i . \]
Homogeneous solutions $\tmmathbf{V}^{(\tau)}$ are the stationary points
$\tmmathbf{V}=\tmmathbf{V}^{(\tau)}$ of the strain energy $\frac{1}{2}
\mymultiply \iint_{\Omega} \left( \lambda \mymultiply \tmop{tr}^2
\tilde{\tmmathbf{E}} + 2 \mymultiply \mu \mymultiply \tilde{\tmmathbf{E}}
\doublecontract \tilde{\tmmathbf{E}} \right) \mymultiply \mathd A$ where
$\tilde{\tmmathbf{E}} = \tilde{\tmmathbf{E}} (\tmmathbf{T}, \tau,
\tmmathbf{V})$, subject to the kinematic constraint $\tmmathbf{q}
(\tmmathbf{V}) =\tmmathbf{0}$ in equation~(\ref{eq:q-vector}). With the help
of equation~(\ref{eq:app-red-variational-pb-homogeneous}), the variational
problem satisfied by $\tmmathbf{V}^{(\tau)}$ writes as
\begin{equation}
  \begin{gathered}
    \iint_{\Omega} V_i^{(\tau)} (\tmmathbf{T}) \mymultiply \mathd A
    =\tmmathbf{0}\\
    \iint_{\Omega} \left[ T_1 \mymultiply V_2^{(\tau)} (\tmmathbf{T}) -
    T_{2 \mymultiply} \mymultiply V_1^{(\tau)} (\tmmathbf{T}) \right]
    \mymultiply \mathd A =\tmmathbf{0}\\
    \forall \hat{\tmmathbf{V}} \quad \iint_{\Omega} \left[
    \tmmathbf{\Sigma} (\tmmathbf{T}, \tmmathbf{E}^{(\tau)} (\tmmathbf{T}))
    \doublecontract \widehat{\tilde{\tmmathbf{E}}}^{(\tau)} (\tmmathbf{T}) +
    F_i^{(\tau)} \mymultiply \hat{V}_i (\tmmathbf{T}) + Q^{(\tau)} \mymultiply
    \left( T_1 \mymultiply \hat{V}_2 (\tmmathbf{T}) - T_{2 \mymultiply}
    \mymultiply \hat{V}_1 (\tmmathbf{T}) \right) \right] \mymultiply \mathd A
    = 0.
  \end{gathered} \label{eq:twist-V-tau-pb}
\end{equation}
where $\tmmathbf{E}^{(\tau)} (\tmmathbf{T}) = \tilde{\tmmathbf{E}}
(\tmmathbf{T}, \tau, \tmmathbf{V}^{(\tau)})$ is the microscopic strain,
$\tmmathbf{\Sigma} (\tmmathbf{T}, \tmmathbf{E}^{(\tau)} (\tmmathbf{T})) =
\frac{\mathd w}{\mathd \tmmathbf{E}} (\tmmathbf{T}, \tmmathbf{E}^{(\tau)}
(\tmmathbf{T})) = \lambda \mymultiply \tmop{tr} \tmmathbf{E}^{(\tau)}
(\tmmathbf{T}) \times \tmmathbf{I}+ 2 \mymultiply \mu \mymultiply
\tmmathbf{E}^{(\tau)} (\tmmathbf{T}) = 2 \mymultiply \mu \mymultiply \left( -
\tau \mymultiply \eta_{\alpha \nocomma \beta} \mymultiply T_{\beta} +
\partial_{\alpha} V_3 (\tmmathbf{T}) \right) \mymultiply \tmmathbf{e}_{\alpha}
\odot \tmmathbf{e}_3 + \left( \lambda \mymultiply \partial_{\gamma} V_{\gamma}
(\tmmathbf{T}) \mymultiply \delta_{\alpha \nocomma \beta} + 2 \mymultiply \mu
\mymultiply \partial_{\alpha} V_{\beta} (\tmmathbf{T}) \right) \mymultiply
\tmmathbf{e}_{\alpha} \odot \tmmathbf{e}_{\beta} + \lambda \mymultiply
\partial_{\gamma} V_{\gamma} (\tmmathbf{T}) \mymultiply \tmmathbf{e}_3 \otimes
\tmmathbf{e}_3$ is the microscopic stress, $\tmmathbf{I}$ is the $3 \times 3$
identity matrix, $\widehat{\tilde{\tmmathbf{E}}}^{(\tau)} (\tmmathbf{T}) =
\partial_{\alpha} \hat{V}_i (\tmmathbf{T}) \mymultiply \tmmathbf{e}_{\alpha}
\odot \tmmathbf{e}_i$ is the virtual increment of strain, and
$\tmmathbf{F}^{(\tau)}$ and $Q^{(\tau)}$ are Lagrange multiplier enforcing the
constraints written in the first two lines of
equation~(\ref{eq:twist-V-tau-pb}).

By inserting these expressions into~(\ref{eq:twist-V-tau-pb}), one obtains two
decoupled problems, namely one for the cross-sectional displacement
$V_{\alpha} (\tmmathbf{T})$ which has no source term, and one for the
longitudinal displacement $V_3 (\tmmathbf{T})$ having a source term
proportional to the kinematic strain $\tau$. The solution writes
\begin{equation}
  V_{\alpha}^{(\tau)} (\tmmathbf{T}) = 0 \qquad V_3^{(\tau)} (\tmmathbf{T}) =
  \tau \mymultiply \omega (\tmmathbf{T}) \label{eq:twisting-homogeneous-sol}
\end{equation}
where $\omega (\tmmathbf{T})$ is the classical warping
function~\citep{trabucho1989existence}, defined as the solution of the
variational problem
\begin{equation}
  \iint_{\Omega} \omega \mymultiply \mathd A = 0 \text{\quad and\quad}
  \forall \hat{\omega} \quad \iint_{\Omega} \partial_{\alpha} \omega
  \mymultiply \partial_{\alpha} \hat{\omega} \mymultiply \mathd A = -
  \iint_{\Omega} \eta_{\alpha \nocomma \beta} \mymultiply T_{\alpha}
  \mymultiply \partial_{\beta} \hat{\omega} \mymultiply \mathd A.
  \label{eq:warping-function-variational}
\end{equation}
The function $\omega (\tmmathbf{T})$ depends on the geometry of the
cross-section only.

In terms of the solution~(\ref{eq:twisting-homogeneous-sol}), one can
reconstruct the microscopic strain and the microscopic stress as
\begin{equation}
  \begin{array}{rllll}
    \tmmathbf{E}^{(\tau)} (\tmmathbf{T}) & = & \tilde{\tmmathbf{E}}
    (\tmmathbf{T}, \tau, \tmmathbf{V}) & = & \tau \mymultiply \left( -
    \eta_{\alpha \nocomma \beta} \mymultiply T_{\beta} + \partial_{\alpha}
    \omega (\tmmathbf{T}) \right) \mymultiply \tmmathbf{e}_{\alpha} \odot
    \tmmathbf{e}_3\\
    \tmmathbf{\Sigma}^{(\tau)} (\tmmathbf{T}) & = & 2 \mymultiply \mu
    \mymultiply \tmmathbf{E}^{(\tau)} (\tmmathbf{T}) & = & \mu \mymultiply
    \tau \mymultiply \left( - \eta_{\alpha \nocomma \beta} \mymultiply
    T_{\beta} + \partial_{\alpha} \omega (\tmmathbf{T}) \right) \mymultiply
    (\tmmathbf{e}_{\alpha} \otimes \tmmathbf{e}_3 +\tmmathbf{e}_3 \otimes
    \tmmathbf{e}_{\alpha})
  \end{array} \label{eq:torsion-homogeneousStress}
\end{equation}
Next, the strain energy density $W_{\text{hom}} (\tau)$ defined
in~(\ref{eq:Wh-def}) is found by inserting the strain $\tmmathbf{E}^{(\tau)}
(\tmmathbf{T})$ into~(\ref{eq:linear-w}), which yields
\begin{equation}
  W_{\text{hom}} (\tau) = \frac{1}{2} \mymultiply \mu \mymultiply J
  \mymultiply \tau^2, \label{eq:torsion-Whom}
\end{equation}
where $J$ is the torsional constant, classically defined as
\begin{equation}
  \begin{split}
    J & =  \iint_{\Omega} \sum_{\alpha} \left( - \eta_{\alpha \nocomma
    \beta} \mymultiply T_{\beta} + \partial_{\alpha} \omega (\tmmathbf{T})
    \right)^2 \mymultiply \mathd A\\
    & =  \iint_{\Omega} (T_1^2 + T_2^2) \mymultiply \mathd A -
    \iint_{\Omega} ((\partial_1 \omega)^2 + (\partial_2 \omega)^2)
    \mymultiply \mathd A.
  \end{split} \label{eq:torsion-J}
\end{equation}
The last equality can be established by using of an identity obtained by
setting $\hat{\omega} = \omega$ in~(\ref{eq:warping-function-variational}).

In view of equation~(\ref{eq:phi-no-gradient}), the one-dimensional strain
energy is
\begin{equation}
  \Phi_{(0)}^{\star} [\tau] = \int_0^{\ell} \frac{1}{2} \mymultiply \mu
  \mymultiply J \mymultiply \tau^2(S) \mymultiply \mathd S.
  \label{eq:twisting-Phi0}
\end{equation}
We have recovered the classical linear model for the twisting of bars,
ignoring the gradient effect for the moment. Repeating a similar reduction but
for stretching and bending modes rather than for the twisting mode, one can
recover the strain energy potential governing the planar Euler-Bernoulli beam
model, $\Phi^{\star}_{(0)} [\varepsilon, \kappa] = \int \left[ \frac{Y
\mymultiply A}{2} \mymultiply \varepsilon^2 (S) + \frac{Y \mymultiply I}{2}
\mymultiply \kappa^2 (S) \right] \mymultiply \mathd S$, where $\varepsilon$
and $\kappa$ are the stretching and bending strain measures, respectively, and
$Y \mymultiply A$ and $Y \mymultiply I$ are the classical traction and bending
moduli, respectively. Here, we have considered rods made up of a uniform,
linearly elastic, isotropic material here; extensions accounting for
anisotropic or non-linear elastic materials \citep{cimetiere1988asymptotic},
for inhomogeneous elastic properties in the cross-section
\citep{hodges2006nonlinear}, or for a pre-strain distribution across the
cross-section
\citep{cicalese2017global,kohn2018bending,moulton2020morphoelastic} can be
obtained easily by following the same procedure.

\subsection{Gradient effect}

In \ref{app:twisting}, we derive the one-dimensional energy
functional capturing the gradient effect associated with a non-uniform
distribution of twist $\tau (S)$. We do so by applying the general recipe from
section~\ref{sec:gradient-effect} to the strain function $\tmmathbf{E}
(\tmmathbf{T}; \tau ; \tmmathbf{V}, \tmmathbf{V}^{\dag})$ in
equation~(\ref{eq:E-twist-linear}). The main results can be summarized as
follows.

A second torsional constant, classically called the warping constant, is
defined by
\begin{equation}
  J_{\omega} = \iint_{\Omega} \omega^2 (\tmmathbf{T}) \mymultiply \mathd A.
  \label{eq:Jw}
\end{equation}
The gradient of kinematic twist $\tau' (S)$ gives rise to a corrective
displacement along to the cross-section
\begin{equation}
  Z_{\alpha}^{\text{opt}} (\tau^{\dag}, \tmmathbf{T}) = \tau^{\dag}
  \mymultiply u_{\alpha} (\tmmathbf{T}) \qquad Z_3^{\text{opt}} (\tau^{\dag},
  \tmmathbf{T}) = 0 \label{eq:twisting-corrective-displacement}
\end{equation}
where $u_{\alpha} (\tmmathbf{T})$ for $\alpha = 1$, 2 are two functions
satisfying the variational problem
\begin{equation}
  \forall \hat{u}_{\alpha}  \iint_{\Omega} \left( \left\{ \lambda
  \mymultiply \partial_{\rho} u_{\rho} \mymultiply \delta_{\alpha \nocomma
  \beta} + 2 \mymultiply \mu \mymultiply \frac{\partial_{\alpha} u_{\beta} +
  \partial_{\beta} u_{\alpha}}{2} \right\} \mymultiply \partial_{\beta}
  \hat{u}_{\alpha} + \left( \lambda \mymultiply \omega \mymultiply
  \partial_{\alpha} \hat{u}_{\alpha} - \mu \mymultiply \partial_{\alpha}
  \omega \mymultiply \hat{u}_{\alpha} \right) - F_{\alpha} \mymultiply
  \hat{u}_{\alpha} - Q \mymultiply \eta_{\alpha \nocomma \beta} \mymultiply
  T_{\alpha} \mymultiply \hat{u}_{\beta} \right) \mymultiply \mathd A = 0
  \label{eq:twisting-psi-alpha-variational-pb}
\end{equation}
and the kinematic constraints
\begin{equation}
  \iint_{\Omega} u_{\alpha} \mymultiply \mathd A = 0 \qquad
  \iint_{\Omega} \eta_{\alpha \nocomma \beta} \mymultiply T_{\alpha}
  \mymultiply u_{\beta} \mymultiply \mathd A = 0.
  \label{eq:twist-corrective-displ-cstr}
\end{equation}
In equation~(\ref{eq:twisting-psi-alpha-variational-pb}), $\hat{u}_{\alpha}
(\tmmathbf{T})$ for $\alpha = 1$, 2 are test functions defined on the
cross-section, and $(F_1, F_2, Q)$ are three scalar multipliers enforcing the
kinematic constraints. The solutions $u_{\alpha} (\tmmathbf{T})$ depend on the
cross-section geometry and on Poisson's ratio $\nu = \frac{\lambda}{2
\mymultiply (\lambda + \mu)}$.

In terms of the corrective displacement $u_{\alpha}$ and of the warping
function $\omega (\tmmathbf{T})$ found earlier, see
equation~(\ref{eq:warping-function-variational}), one can define three
additional constants,
\begin{equation}
  \begin{split}
    D_{\lambda} & =  \lambda \mymultiply \iint_{\Omega} \omega
    (\tmmathbf{T}) \mymultiply \partial_{\alpha} u_{\alpha} (\tmmathbf{T})
    \mymultiply \mathd A\\
    D_{\mu} & = \mu \mymultiply \iint_{\Omega} \partial_{\alpha} \omega
    (\tmmathbf{T}) \mymultiply u_{\alpha} (\tmmathbf{T}) \mymultiply \mathd
    A\\
    D_{\omega} & = \left( \lambda + 2 \mymultiply \mu \right) \mymultiply
    J_{\omega} + D_{\lambda}
  \end{split} \label{eq:twist-D-sub-x}
\end{equation}

The final expression of the one-dimensional strain energy is
\begin{equation}
  \Phi_{(2)}^{\star} [\tau] = \int_0^{\ell} \left( \frac{1}{2} \mymultiply \mu
  \mymultiply J \mymultiply \left( \tau (S) + \frac{D_{\mu}}{\mu \mymultiply
  J} \mymultiply \frac{\mathd^2 \tau}{\mathd S^2} \right)^2 + \frac{1}{2}
  (D_{\omega} + D_{\mu}) \mymultiply \left( \frac{\mathd \tau}{\mathd S}
  \right)^2 \right) \mymultiply \mathd S. \label{eq:twisting-final-phi}
\end{equation}
To the best of our knowledge, this simple one-dimensional strain energy for
the twisting of a linearly elastic bar is not known from the literature. It
captures the gradient effect and is asymptotically correct. It underpins some
of the results of \citet{trabucho1989existence}, as discussed below, in an
accessible form.

By combining equations~(\ref{eq:x-centerline-based-crspondence}),
(\ref{eq:y-star-expansion}), (\ref{eq:y-to-v}),
(\ref{eq:twisting-homogeneous-sol})
and~(\ref{eq:twisting-corrective-displacement}), the solution in displacement
is found as $\tmmathbf{x} (S, \tmmathbf{T}) = S \mymultiply \tmmathbf{e}_3 +
\left( T_{\alpha} + \tau' (S) \mymultiply u_{\alpha} (\tmmathbf{T}) \right)
\mymultiply \tmmathbf{d}_{\alpha} (S) + \tau (S) \mymultiply \omega
(\tmmathbf{T}) \mymultiply \tmmathbf{e}_3 (S)$ where $\tmmathbf{d}_1 (S)
=\tmmathbf{e}_1 + \theta (S) \mymultiply \tmmathbf{e}_2$ and $\tmmathbf{d}_2
(S) =\tmmathbf{e}_2 - \theta (S) \mymultiply \tmmathbf{e}_1$ are the rotated
directors and $\theta (S)$ is the twisting angle, see
figure~\ref{eq:fig-twisting} (recall that we are working in the linear
setting). The usual relation defining the twisting strain $\tau (S) = \theta'
(S)$ as the gradient of the twisting angle $\theta (S)$ is recovered in the
process.

The particular case of an elliptical cross-section is worked out in
\ref{app-twist-elliptical}: with $a$ and $b$ as the
semi-major and semi-minor axes, in any order, the constants appearing
in the energy $\Phi_{(2)}^{\dag}$ are calculated as
\begin{equation}
  \begin{array}{llll}
    J = \mathpi \mymultiply \frac{a^3 \mymultiply b^3}{a^2 + b^2}\quad & J_{\omega}
    = \frac{1}{24} \mymultiply \frac{(b^2 - a^2)^2}{a^2 + b^2} \mymultiply J\quad &
    D_{\mu} = 8 \mymultiply \mu \mymultiply J_{\omega} \mymultiply \left(
    \frac{a \mymultiply b}{a^2 + b^2} \right)^2\quad & D_{\omega} = Y \mymultiply
    J_{\omega}
  \end{array} \text{{\hspace{1.2em}}(elliptical cross-section)}
  \label{eq:elliptical-X-section}
\end{equation}
where $Y$ is the Young modulus,
\begin{equation}
  Y = \mu \mymultiply \frac{3 \mymultiply \lambda + 2 \mymultiply \mu}{\lambda
  + \mu} . \label{eq:twisting-Young-modulus}
\end{equation}

\subsection{Equilibrium}

In the presence of a distributed external twisting moment $m_3 (S)$ per unit
length $\mathd S$, the total potential energy of the bar is
\[ \Psi^{\star} [\theta] = \Phi_{(2)}^{\star} [\theta'] - \int_0^{\ell} m_3
   (S) \mymultiply \theta (S) \mymultiply \mathd S, \]
see equation~(\ref{eq:ideal-1d-total-potential-energy}). The equations of
equilibrium of the bar can be found by making $\Psi^{\star} [\theta]$
stationary with respect to $\theta$. Upon integration by parts and after
several simplifications, one obtains the equilibrium equation in the interior
as
\begin{equation}
  \frac{\mathd}{\mathd S} \left( \mu \mymultiply J \mymultiply \tau -
  (D_{\omega} - D_{\mu}) \mymultiply \tau'' \right) + m_3 (S) = 0,
  \label{eq:twist-equil-eq-with-gradient}
\end{equation}
along with the applicable boundary conditions. Note the plus sign in front of
$D_{\mu}$ in equation~(\ref{eq:twisting-final-phi}) and the minus sign in
equation~(\ref{eq:twist-equil-eq-with-gradient}).

\subsection{Comments}

The equilibrium equation~(\ref{eq:twist-equil-eq-with-gradient}) underpins the
analysis of the gradient effect in twisted prismatic bar done
by~\citet{trabucho1989existence}, as shown in \ref{app:Trabucho};
however this simple and important equation did not appear explicitly in this
work.

In equation~(\ref{eq:twist-equil-eq-with-gradient}), the quantity inside the
derivative $M_3 = \mu \mymultiply J \mymultiply \tau - (D_{\omega} - D_{\mu})
\mymultiply \tau''$ can be interpreted as the internal twisting moment in the
bar; it is made up of the prediction $M_3 = \mu \mymultiply J \mymultiply
\tau$ \ of the classical model without gradient effect, and of a correction
coming from the gradient effect; it is a hallmark of higher-order gradient
models that the stress not only depends on the local strain but also on its
gradients. The quantity $M_3 = \mu \mymultiply J \mymultiply \tau -
(D_{\omega} - D_{\mu}) \mymultiply \tau''$ is identical to that obtained by
the general constitutive law in
equation~(\ref{eq:internal-stress-full-model}), as can be checked.

In the particular case of a circular cross-section, $a = b$, the gradient
effect is absent: in equation~(\ref{eq:elliptical-X-section}), $J_{\omega} =
0$ and therefore $D_{\mu} = D_{\omega} = 0$.

The gradient model for a twisted bar has been derived here in the context of
linear elasticity. A non-linear extension of this model can be obtained along
the same lines; in order to build the catalog of solutions having homogeneous
twist, a non-linear one-dimensional boundary-value problem must be solved,
which requires some numerical solution in general. In the non-linear model,
the constitutive law is of the form $M_3 = H (\tau) + Q (\tau) \mymultiply
\tau''$. Consistency with the linear model is warranted by the approximations
$H (\tau) \approx \mu \mymultiply J \mymultiply \tau$ and $Q (\tau) \approx -
(D_{\omega} - D_{\mu})$ which hold when linear elasticity is applicable,
{\tmem{i.e.}}, when the microscopic strain is small, $| \tau | \ll 1 / \max
(a, b)$. This shows that the linear model derived in this section is
applicable as long as the absolute value of the twisting strain $\tau$ remains
small.

\section{Illustration in a weakly non-linear setting: buckling of a thick
beam}\label{sec:Euler-buckling}

Euler buckling ({\tmem{i.e.}}, the buckling of an elastic cylinder subjected
to an axial compressive force) can be analyzed using the classical theory of
rods: this yields a prediction for the buckling load which is accurate for
infinitely slender beams. With the aim to characterize the buckling load of
thicker beams, several authors have derived corrections to the Euler buckling
load in powers of the aspect-ratio. This requires restarting from the
non-linear theory of elasticity in three dimensions, as both constitutive and
geometric nonlinearities affect these corrections.

In an early and remarkable work,
\citet{Fosdick-Shield-Small-bending-of-a-circular-1963} have carried out
what is essentially a linear bifurcation analysis of a hyper-elastic cylinder
having a finite length-to-radius ratio. They obtained a prediction of the
buckling load that connects with Euler's prediction in the limit of a slender
cylinder, thereby showing consistency of the buckling analyses based on
three-dimensional versus one-dimensional models. However, their solution
assumes that the internal moment in the cylinder is proportional to the local
value of the center-line curvature. This is questionable for thick beams: the
internal moment $M$ given by equation~(\ref{eq:internal-stress-full-model})
depends on higher derivatives of the curvature as well, as earlier in
equation~(\ref{eq:twist-equil-eq-with-gradient}). It is therefore unclear
whether their analysis is valid beyond the infinitely slender case.

In more recent work, \citet{scherzinger1998asymptotic} derived the first
buckling load of a stubby hyper-elastic cylinder in powers of its
aspect-ratio, starting from the full theory of three-dimensional elasticity
with finite strain---a similar analysis has been carried out independently
by~\citet{Goriely-Vandiver-EtAl-Nonlinear-Euler-buckling-2008} and \citet{de2011nonlinear}.
Here, we show that the results of \citet{scherzinger1998asymptotic} can be
recovered by (i)~deriving a non-linear {\tmem{one-dimensional}} model for the
stubby cylinder that captures the gradient effect, using our reduction method
and (ii)~by carrying out a linear bifurcation analysis of this one-dimensional
model.

By doing so, our goal is twofold: we provide another illustration of our
reduction method and we verify its predictions in a weakly non-linear setting.

\subsection{Problem setting}\label{sec:beam-problem-setting}

\begin{figure}
  \centerline{\includegraphics{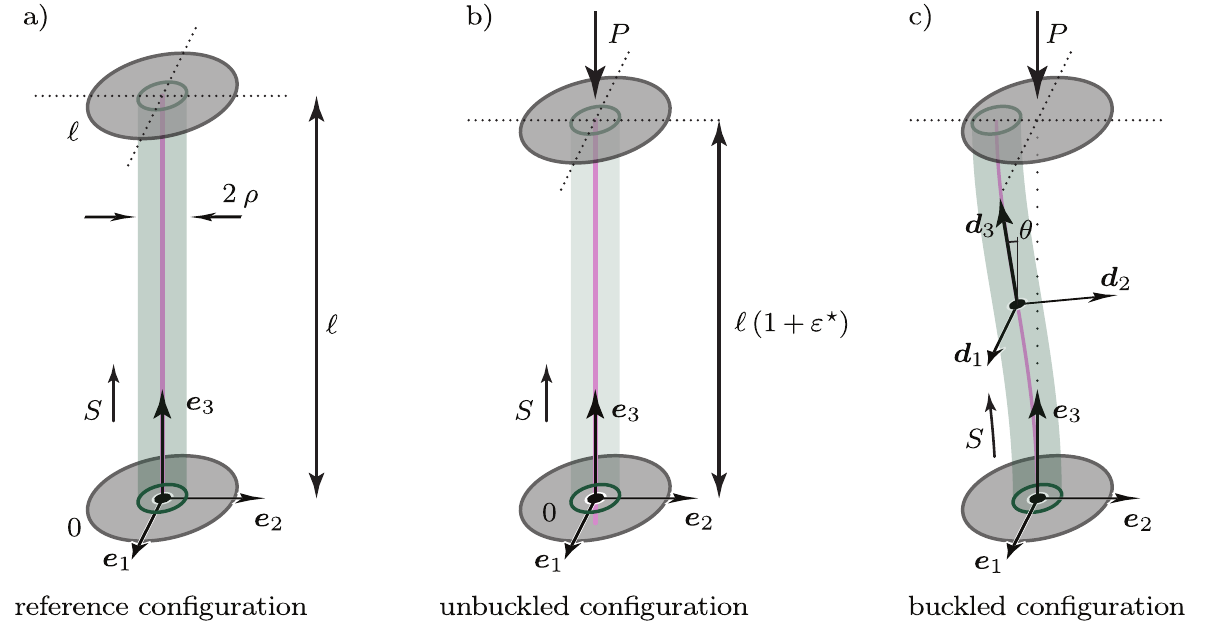}}
  \caption{Buckling of a thick circular cylinder with initial radius $\rho$,
  whose terminal faces slide onto two parallel planes, as analyzed by
  {\citet{scherzinger1998asymptotic}}
  and~{\citet{Goriely-Vandiver-EtAl-Nonlinear-Euler-buckling-2008}}. Our
  analysis addresses the slightly more general case of a prismatic body, whose
  cross-section $\Omega$ is mirror-symmetric with respect to be axes
  $\tmmathbf{e}_1$ and $\tmmathbf{e}_2$.\label{fig:thick-Euler-buckling}}
\end{figure}

We revisit the buckling problem of~\citet{scherzinger1998asymptotic} as
follows. We consider a prismatic elastic body having length is $\ell$ in the
undeformed configuration: in figure~\ref{fig:thick-Euler-buckling}a, the
particular case of a cylinder with initial radius $\rho$ is shown. We use
Cartesian coordinates such that the axis $\tmmathbf{e}_3$ is aligned with the
initial axis of the cylinder, and one of the terminal faces of the cylinder is
centered on the origin $O$ of the coordinate system. The two ends $S = 0$ and
$S = \ell$ are assumed to slide without friction on two planes perpendicular
to $\tmmathbf{e}_3$, {\tmem{i.e.}}, the displacement along $\tmmathbf{e}_3$ is
zero on the terminal faces of the cylinder; in particular the longitudinal
displacement is restrained on the terminal faces,
\begin{equation}
    \tmmathbf{x} (0, \tmmathbf{T}) \cdot \tmmathbf{e}_3 = 0 
	\qquad
	(\tmmathbf{x}
    (\ell, \tmmathbf{T}) -\tmmathbf{r} (\ell)) \cdot \tmmathbf{e}_3 = 0.
  \label{eq:Euler-buckling-sliding-condition}
\end{equation}
The distance between the planes is changed from $\ell$ in the natural
configuration, to $\ell \mymultiply (1 + \varepsilon)$ in the actual
configuration with $- 1 < \varepsilon < 0$. We seek the critical value of
$\varepsilon$ corresponding to the occurrence of the first buckling modes.

We assume that the prismatic body is made up of an isotropic material, having
uniform elastic properties. We also assume that the cross-section domain
$\Omega$ is mirror-symmetric about the axes $\tmmathbf{e}_1$ and
$\tmmathbf{e}_2$ in reference configuration, i.e., it is invariant by both
$(T_1, T_2) \leftarrow (T_1, - T_2)$ and $(T_1, T_2) \leftarrow (- T_1, T_2)$;
this warrants
\begin{equation}
  \left\langle (T_1)^i \mymultiply (T_2)^j \right\rangle = 0 \text{ whenever
  $i$ or $j$ is odd, or both are odd}
  \label{eq:Euler-buckling-cross-section-symmetry}
\end{equation}
With these assumptions, the two bending modes and the stretching modes are
uncoupled, and we limit attention to buckling modes such that the center-line
remains in the plane perpendicular to $\tmmathbf{e}_1$. We denote by $\theta
(S)$ the rotation of the material frame about the constant vector normal
$\tmmathbf{d}_1 (S) =\tmmathbf{e}_1$ to the plane of deformation.

The analysis of less symmetric cross-sections, non-isotropic materials, or
elastic properties that vary in the cross-section is more involved but it does
not raise any fundamental difficulty.

The prismatic body is homogeneous and made of an isotropic hyper-elastic
material. We can for instance use the same constitutive model $w
(\tmmathbf{T}, \tmmathbf{E}) = w_{\text{ST}} (\tmmathbf{E})$ as used
by~{\citet{scherzinger1998asymptotic}}, which reads, after restoring a missing
coefficient $1 / 4$ in their equation~[55], $w_{\text{ST}} (\tmmathbf{E}) =
A_{\text{ST}} \mymultiply \Big( \frac{I_1}{I_3^{1 / 3}} - 3 \Big) +
B_{\text{ST}} \mymultiply \Big( \frac{I_2}{I_3^{2 / 3}} - 3 \Big) +
\frac{1}{4} \mymultiply \frac{A_{\text{ST}} + B_{\text{ST}}}{24} \mymultiply
\frac{1 + \nu_{\text{ST}}}{1 - 2 \mymultiply \nu_{\text{ST}}} \mymultiply
\Big( I_3^2 - \frac{1}{I_3^2} \Big)^2$, where $I_1 = \tmop{tr}
\tmmathbf{C}$, $I_2 = \frac{1}{2} \mymultiply (I_1^2 - \tmop{tr}
(\tmmathbf{C}^2))$, $I_3 = \det \tmmathbf{C}$ and $\tmmathbf{C}=\tmmathbf{I}+
2 \mymultiply \tmmathbf{E}$. However, we do not specify the isotropic
constitutive law for the moment. The only constitutive assumptions which we
will use in the forthcoming analysis is that, in the unbuckled configuration,
the stress is uniaxial and the incremental constitutive law is transversely
isotropic: this holds for {\tmem{any}} isotropic constitutive law.
Specifically, our analysis makes use of the three constitutive functions
$w_{\text{tr}} (\varepsilon)$, $Y_{\text{t}} (\varepsilon)$, $p (\varepsilon)$
that characterize the non-linear material response in simple traction, where
$\varepsilon$ is the longitudinal engineering strain: $w_{\text{tr}}
(\varepsilon)$ is the strain energy density of the material in simple
traction, $Y_{\text{t}} (\varepsilon)$ is the tangent Young modulus and $p
(\varepsilon)$ is the transverse stretch resulting from Poisson's effect, see
section~\ref{app-sec:simple-stretching} in the appendix for details. In terms
of these material functions, we also define the initial Young modulus $Y_0$,
the initial Poisson's ratio $\nu_0$ and the initial curvature $Y_0'$ of the
load-displacement curve, see equation~(\ref{eq:beam-Y0-Y0p}) from the
appendix.

\subsection{One-dimensional reduction}\label{ssec:beam-1d-model}

In this section, we apply our reduction method to obtain the one-dimension
model for the planar beam capturing the gradient effect; it describes the
bending and stretching of a planar, higher-order Elastica.

For planar deformation, there are two relevant macroscopic strain measures,
namely the bending strain $\kappa = \kappa_1$ and the stretching strain
$\varepsilon$; they are grouped into a macroscopic strain vector
$\tmmathbf{h}= (\varepsilon, \kappa)$. In the planar twistless case, the
general form~(\ref{eq:phi-gr}) of the higher-order one-dimensional model
writes
\begin{multline}
    \Phi_{(2)}^{\star} [\varepsilon, \kappa] = \int_0^{\ell} \left[
    W_{\text{hom}} \left( \varepsilon + \xi_0 (\varepsilon, \kappa)
    \mymultiply \varepsilon'', \kappa + \xi_1 (\varepsilon, \kappa)
    \mymultiply \kappa'' \right) \right. \ldots\\
    \left. \nobracket \nobracket {}+ A_0 (\varepsilon, \kappa) \mymultiply
    \varepsilon' + A_1 (\varepsilon, \kappa) \mymultiply \kappa' + \frac{1}{2}
    \mymultiply \left(\begin{array}{c}
      \varepsilon'\\
      \kappa'
    \end{array}\right) \cdot \left(\begin{array}{cc}
      D_{0 \nocomma 0} (\varepsilon, \kappa) & D_{1 \nocomma 0} (\varepsilon,
      \kappa)\\
      D_{1 \nocomma 0} (\varepsilon, \kappa) & D_{1 \nocomma 1} (\varepsilon,
      \kappa)
    \end{array}\right) \cdot \left(\begin{array}{c}
      \varepsilon'\\
      \kappa'
    \end{array}\right) \right] \mymultiply \mathd S.
  \label{eq:bending-phi2-anticipation}
\end{multline}
We now proceed to calculate the quantities $W_{\text{hom}}$, $\xi_0$, $\xi_1$,
$A_0$, $A_1$ and $D_{i \nocomma j}$: we consider the case of a finite axial
strain $\varepsilon$ but limit attention to small values of the curvature
$\kappa$, anticipating that this is all that is needed for the forthcoming
buckling analysis.

\subsubsection{Symmetry properties}\label{sssec:beam-symmetries}

We first characterize the symmetry properties of the functions
$W_{\text{hom}}$, $\xi_0$, $\xi_1$, $A_0$, $A_1$ and $D_{i \nocomma j}$ as
they will save us from calculating some quantities that are zero by symmetry.
The cylinder is invariant both by a mirror symmetry with respect to the axis
$\tmmathbf{e}_3$, and by a reversal of the parameterization $S \leftarrow (-
S)$. These symmetries correspond to changing $(\varepsilon, \kappa,
\varepsilon', \kappa', \varepsilon'', \kappa'')$ into $(+ \varepsilon, -
\kappa, + \varepsilon', - \kappa', + \varepsilon'', - \kappa'')$ and $(+
\varepsilon, - \kappa, - \varepsilon', + \kappa', + \varepsilon'', -
\kappa'')$, respectively. For $\Phi_{(2)}^{\star} [\varepsilon, \kappa]$ to
remain invariant by both these transformations, both $A_0$ and $A_1$ must be
zero identically, $W_{\text{hom}}$, $\xi_0$, $\xi_1$, $D_{0 \nocomma 0}$ and
$D_{1 \nocomma 1}$ must be even with respect to $\kappa$, and $D_{1 \nocomma
0}$ must be odd with respect to $\kappa$, see
equation~(\ref{eq:bending-phi2-anticipation}). Therefore, for any
$\varepsilon$ and $\kappa$, and for any set of non-negative integers $i$ and
$j$, we have
\begin{equation}
  \begin{gathered}
    \begin{array}{ll}
      A_0 (\varepsilon, \kappa) = 0 & A_1 (\varepsilon, \kappa) = 0
    \end{array}\\
    \frac{\partial^{i + 2 \mymultiply j + 1} f}{\partial \varepsilon^i
    \mymultiply \partial \kappa^{2 \mymultiply j + 1}} (\varepsilon, 0) = 0
    \text{ for any choice of $f$ in $\left\{ W_{\text{hom}}, \xi_0, \xi_1,
    D_{0 \nocomma 0}, D_{1 \nocomma 1} \right\}$}\\
    \frac{\partial^{i + 2 \mymultiply j} D_{1 \nocomma 0}}{\partial
    \varepsilon^i \mymultiply \partial \kappa^{2 \mymultiply j}} (\varepsilon,
    0) = 0 \text{}
  \end{gathered} \label{eq:bending-phi2-symmetries}
\end{equation}
In particular, the matrix $D_{i \nocomma j} (\varepsilon, 0)$ is diagonal,
{\tmem{i.e.}}, $D_{1 \nocomma 0} (\varepsilon, 0) = 0$.

\subsubsection{Analysis of homogeneous solutions}

Homogeneous solutions are derived here by solving the equations from
\ref{app:compendium-homogeneous}, using the expression
of the strain from equation~(\ref{eq:homogeneous-strain}) relevant to the
non-linear and homogeneous setting, and by specializing to the planar
twistless case, $\kappa_2 = \kappa_3 = 0$. As we are ultimately interested in
weakly bent configurations of the cylinder, close to the bifurcation
threshold, we treat $\kappa$ as a small parameter. We refrain from setting
$\kappa = 0$ from the onset, however, as some derivatives of intermediate
quantities with respect to $\kappa$ will be needed in the course of the
derivation.

As shown in \ref{app-sec:beam-homogeneous-solutions},
the microscopic displacement corresponding to a homogeneous axial strain
$\varepsilon$ and curvature $\kappa$ is
\begin{equation}
  \begin{array}{ll}
    Y_{\alpha}^{(\varepsilon, \kappa)} = \nlPoisson (\varepsilon) \mymultiply
    \left( T_{\alpha} + \kappa \mymultiply \frac{\mathd \nlPoisson}{\mathd
    \varepsilon} (\varepsilon) \mymultiply \varphi_{\alpha} (\tmmathbf{T})
    \right) +\mathcal{O} (\kappa^2)\qquad & Y_3^{(\varepsilon, \kappa)}
    (\tmmathbf{T}) = 0
  \end{array} \label{eq:beam-homogeneous-Y}
\end{equation}
where $\varphi_1 (\tmmathbf{T})$ and $\varphi_2 (\tmmathbf{T})$ are two
functions depending on the cross-section geometry, which are the solutions of
the differential problem on the cross-section
\begin{equation}
  \begin{array}{lll}
    \forall \tmmathbf{T} \in \Omega \quad \frac{\partial_{\alpha}
    \varphi_{\beta} (\tmmathbf{T}) + \partial_{\beta} \varphi_{\alpha}
    (\tmmathbf{T})}{2} = T_2 \mymultiply \delta_{\alpha \nocomma \beta} &
    \langle \varphi_{\alpha} \rangle = 0 & \left\langle \eta_{\alpha \nocomma
    \beta} \mymultiply T_{\alpha} \mymultiply \varphi_{\beta} (\tmmathbf{T})
    \right\rangle = 0.
  \end{array} \label{eq:phi-alpha-fundamental-def}
\end{equation}
With the symmetry assumptions
in~(\ref{eq:Euler-buckling-cross-section-symmetry}), the solution is
\begin{equation}
  \varphi_1 (\tmmathbf{T}) = T_1 \mymultiply T_2 \qquad
  \varphi_2 (\tmmathbf{T}) = \frac{T_2^2 - T_1^2}{2} - \frac{\langle T_2^2
  \rangle - \langle T_1^2 \rangle}{2} .
  \label{eq:phi-alpha-general-cross-section}
\end{equation}
Up to a rigid-body displacement, the functions $\varphi_1$ and $\varphi_2$
match the functions $\phi_{2 \nocomma 1}$ and $\phi_{2 \nocomma 2}$
classically used in the {\tmem{linear}} analysis of bending,
respectively---see for instance equations~[2.5,3.8,3.9] in the work
of~{\citet{trabucho1989existence}}. Our analysis shows that they are relevant
to the analysis of finite-stretching and infinitesimal-bending as well.

As shown in the appendix, the displacement~(\ref{eq:beam-homogeneous-Y}) is
such that every point in the bar is in simple traction with a local
longitudinal strain $\varepsilon + \kappa \mymultiply p (\varepsilon)
\mymultiply T_2$ depending on the transverse coordinate $T_2$: the strain is
given by $\tmmathbf{E}^{(\varepsilon, \kappa)} (\tmmathbf{T})
=\tmmathbf{E}_{\text{tr}} \left( \varepsilon + \kappa \mymultiply \nlPoisson
(\varepsilon) \mymultiply T_2 \right) +\mathcal{O} (\kappa^2)$ and the stress
is uniaxial, $\tmmathbf{\Sigma}^{(\varepsilon, \kappa)} (\tmmathbf{T}) =
\Sigma_{\text{tr}} \left( \varepsilon + \kappa \mymultiply \nlPoisson
(\varepsilon) \mymultiply T_2 \right) \mymultiply \tmmathbf{e}_3 \otimes
\tmmathbf{e}_3 +\mathcal{O} (\kappa^2)$, where $\tmmathbf{E}_{\text{tr}}
(\varepsilon)$ and $\Sigma_{\text{tr}} (\varepsilon) \mymultiply
\tmmathbf{e}_3 \otimes \tmmathbf{e}_3$ are the strain and the stress in simple
traction for the particular material considered, see
equation~(\ref{eq:beam-simple-traction-E-Sigma}).

The strain energy per unit length associated with this homogeneous solution is
found in the appendix as
\begin{equation}
  W_{\text{hom}} (\varepsilon, \kappa) = A \mymultiply w_{\text{tr}}
  (\varepsilon) + \frac{1}{2} \mymultiply Y_{\text{t}} (\varepsilon)
  \mymultiply \nlPoisson^2 (\varepsilon) \mymultiply I_1^0 \mymultiply
  \kappa^2 +\mathcal{O} (\kappa^4) \label{eq:beam-Whom}
\end{equation}
where $A = \iint_{\Omega} \mathd A$ is the initial area and $I_1^0 =
\iint T_2^2 \mymultiply \mathd A$ is
the initial geometric moment of inertia. In the small-strain limit, the
potential $W_{\text{hom}}$ is consistent with the classical Euler beam model
$W_{\text{hom}} (\varepsilon, \kappa) \approx C + \mymultiply \frac{Y_0 A}{2}
\mymultiply \varepsilon^2 + \frac{Y_0 \mymultiply I_1^0}{2} \mymultiply
\kappa^2$ (where $C$ is a constant) as can be shown by inserting the
equivalents of $w_{\text{tr}}$, $Y_{\text{t}}$ and $p$ for small $\varepsilon$
derived in \ref{app-sec:simple-stretching}.

\subsubsection{Gradient effect}\label{sssec:beam-gradient-sol}

The corrective displacement associated with longitudinal gradients of axial
strain $\varepsilon^{\dag}$ and curvature $\kappa^{\dag}$ is found in the
appendix as
\begin{equation}
  \begin{array}{ll}
    Z_{\text{opt}, \alpha}^{(\varepsilon, 0)} ((\varepsilon^{\dag},
    \kappa^{\dag}), \tmmathbf{T}) = 0 \qquad& Z_{\text{opt}, 3}^{(\varepsilon, 0)}
    ((\varepsilon^{\dag}, \kappa^{\dag}), \tmmathbf{T}) = \varepsilon^{\dag}
    \mymultiply (\ldots) + \kappa^{\dag} \mymultiply \frac{\nlPoisson^2
    (\varepsilon) \mymultiply \frac{\mathd \nlPoisson}{\mathd \varepsilon}
    (\varepsilon)}{1 + \varepsilon} \mymultiply \left( \Theta (\tmmathbf{T}) +
    c_{\Gamma} (\varepsilon) \mymultiply \Gamma (\tmmathbf{T}) \right)
  \end{array} \label{eq:beam-optima-corrective-displacement}
\end{equation}
where the contribution associated with $\varepsilon^{\dag}$ is denoted by an
ellipsis and does not need to be calculated, $c_{\Gamma} (\varepsilon)$ is a
material parameter depending on the strain $\varepsilon$,
\begin{equation}
  c_{\Gamma} (\varepsilon) = \frac{Y_{\text{t}} (\varepsilon)}{2 \mymultiply
  G_{\text{t}} (\varepsilon) \mymultiply \nlPoisson (\varepsilon) \mymultiply
  \left( - \frac{\mathd \nlPoisson}{\mathd \varepsilon} (\varepsilon) \right)
  \mymultiply (1 + \varepsilon)}, \label{eq:beam-1d-reduction-coefficients}
\end{equation}
and $\Theta$ and $\Gamma$ are the cross-sectional functions satisfying the
variational problems
\begin{equation}
  \begin{array}{lll}
    \forall \hat{Z}_3 \qquad \iint_{\Omega} \left[
    \partial_{\alpha} \Theta (\tmmathbf{T}) \mymultiply \partial_{\alpha}
    \hat{Z}_3 (\tmmathbf{T}) + \varphi_{\alpha} (\tmmathbf{T}) \mymultiply
    \partial_{\alpha} \hat{Z}_3 (\tmmathbf{T}) \right] \mymultiply \mathd A =
    0 & \quad & \langle \Theta \rangle = 0\\[.5em]
    \forall \hat{Z}_3 \qquad \iint_{\Omega} \left[
    \partial_{\alpha} \Gamma (\tmmathbf{T}) \mymultiply \partial_{\alpha}
    \hat{Z}_3 (\tmmathbf{T}) + 2 \mymultiply T_2 \mymultiply \hat{Z}_3
    (\tmmathbf{T}) \right] \mymultiply \mathd A = 0 &  & \langle \Gamma
    \rangle = 0
  \end{array} \label{eq:beam-1d-reduc-pb-Theta-Gamma}
\end{equation}
The functions $\Theta$ and $\Gamma$ are denoted as $\theta_2$ and $\eta_2$ in
the work of~{\citet{trabucho1989existence}}, see their equations~[2.23]
and~[2.17].

Finally, we define four constants depending on the cross-section shape,
\begin{equation}
  M = \sum_{\alpha} \iint_{\Omega} \varphi_{\alpha}^2 (\tmmathbf{T})
  \mymultiply \mathd \Omega \qquad J_{\Theta \nocomma
  \Theta} = \iint_{\Omega} \partial_{\alpha} \Theta \mymultiply
  \partial_{\alpha} \Theta \mymultiply \mathd \Omega \qquad J_{\Theta \nocomma \Gamma} = \iint_{\Omega} \partial_{\alpha}
  \Theta \mymultiply \partial_{\alpha} \Gamma \mymultiply \mathd \Omega
  \qquad J_{\Gamma \nocomma \Gamma} = \iint_{\Omega}
  \partial_{\alpha} \Gamma \mymultiply \partial_{\alpha} \Gamma \mymultiply
  \mathd \Omega . \label{eq:beam-1d-reduc-geom-constants-general-cross-sect}
\end{equation}

As shown in appendix~\ref{ssec:beam-app-gradient-effect}, the reduction method
from section~\ref{sec:asymptotic-1d-reduction} yields the following
expressions for the quantities entering in the one-dimensional
model~(\ref{eq:bending-phi2-anticipation}) as
\begin{equation}
  \begin{array}{rll}
    \xi_0 (\varepsilon, 0) & = & 0\\
    \xi_1 (\varepsilon, 0) & = & A \mymultiply \frac{\nlPoisson (\varepsilon)
    \mymultiply \left( - \frac{\mathd \nlPoisson}{\mathd \varepsilon}
    (\varepsilon) \right)}{2 \mymultiply (1 + \varepsilon)} \mymultiply
    \frac{1}{I_1^0 / A^2} \left( \frac{J_{\Theta \nocomma \Gamma}}{A^3} +
    \frac{J_{\Gamma \nocomma \Gamma}}{A^3} \mymultiply c_{\Gamma}
    (\varepsilon) \right)\\
    D_{1 \nocomma 1} (\varepsilon, 0) & = & A^3 \mymultiply \left( \nlPoisson
    (\varepsilon) \mymultiply \frac{\mathd \nlPoisson}{\mathd \varepsilon}
    (\varepsilon) \right)^2 \mymultiply \left[ \frac{w_{\text{tr}}'
    (\varepsilon)}{1 + \varepsilon} \mymultiply \frac{M}{A^3} + \nlPoisson^2
    (\varepsilon) \mymultiply G_{\text{t}} (\varepsilon) \mymultiply \left(
    \frac{M - J_{\Theta \nocomma \Theta}}{A^3} + \frac{J_{\Gamma \nocomma
    \Gamma}}{A^3} \mymultiply c_{\Gamma}^2 (\varepsilon) \right) \right] .
  \end{array} \label{eq:beam-1d-reduction-result-generic}
\end{equation}
These are the only properties of the one-dimensional model which are needed in
the linear buckling analysis, as we will show.

For reference, the microscopic solution in displacement is found by combining
equations~(\ref{eq:x-centerline-based-crspondence}),
(\ref{eq:y-star-expansion}), (\ref{eq:beam-homogeneous-Y})
and~(\ref{eq:beam-optima-corrective-displacement}) as
\begin{equation}
  \begin{array}{l}
    \tmmathbf{x} (S, \tmmathbf{T}) =\\
    \nobracket \nobracket \quad \tmmathbf{r} (S) + \left( \nlPoisson
    (\varepsilon) \mymultiply \left( T_{\alpha} + \kappa \mymultiply
    \frac{\mathd \nlPoisson}{\mathd \varepsilon} (\varepsilon) \mymultiply
    \varphi_{\alpha} (\tmmathbf{T}) \right) \right) \mymultiply
    \tmmathbf{d}_{\alpha} (S) + \left( \varepsilon' \mymultiply (\ldots) +
    \kappa' \mymultiply \frac{\nlPoisson^2 (\varepsilon) \mymultiply
    \frac{\mathd \nlPoisson}{\mathd \varepsilon} (\varepsilon)}{1 +
    \varepsilon} \mymultiply \left( \Theta (\tmmathbf{T}) + c_{\Gamma}
    (\varepsilon) \mymultiply \Gamma (\tmmathbf{T}) \right) \right)
    \mymultiply \tmmathbf{d}_3 (S) + \cdots
  \end{array} \label{eq:Euler-buckling-microscopic-displacement}
\end{equation}

\subsubsection{Case of a circular cross-section}

When the cross-section is a disk with radius $\rho$, as shown in
figure~\ref{fig:thick-Euler-buckling}, the initial area is $A = \mathpi
\mymultiply \rho^2$, the initial moment of inertia is $I_1^0 = \iint T_2^2
\mymultiply \mathd A = \frac{\mathpi \mymultiply \rho^4}{4} = \frac{A^2}{4
\mymultiply \mathpi}$, the functions $\varphi_{\alpha}$ in
equation~(\ref{eq:phi-alpha-general-cross-section}) take the slightly simpler
form
\begin{equation}
  \begin{array}{ccc}
    \varphi_1 (\tmmathbf{T}) = T_1 \mymultiply T_2 & \nobracket \nobracket
    \qquad & \varphi_2 (\tmmathbf{T}) = \frac{T_2^2 - T_1^2}{2},
  \end{array} \label{eq:beam-phi-i}
\end{equation}
and the solutions $\Theta$ and $\Gamma$ to
equation~(\ref{eq:beam-1d-reduc-pb-Theta-Gamma}) are
\begin{equation}
  \begin{array}{rll}
    \Theta (\tmmathbf{T}) & = & - \frac{1}{4} \mymultiply (T_1^2 + T_2^2 -
    \rho^2) \mymultiply T_2\\
    \Gamma (\tmmathbf{T}) & = & + \frac{1}{4} \mymultiply \left( T_1^2 + T_2^2
    - 3 \mymultiply \rho^2 \right) \mymultiply T_2 .
  \end{array} \label{eq:beam-1d-reduction-Theta-Gamma-disk}
\end{equation}
After factoring out the cube area $A^3 = \left( \mathpi \mymultiply \rho^2
\right)^3$, the constants appearing in
equation~(\ref{eq:beam-1d-reduc-geom-constants-general-cross-sect}) can then
be expressed as
\[ M = \frac{A^3}{12 \mymultiply \mathpi^2} \qquad
   (J_{\Theta \nocomma \Theta}, J_{\Theta \nocomma \Gamma}, J_{\Gamma \nocomma
   \Gamma}) = \frac{A^3}{24 \mymultiply \mathpi^2} \mymultiply (+ 1, - 1, + 7)
   . \]
Finally, the quantities defining the one-dimensional model in
equation~(\ref{eq:beam-1d-reduction-result-generic}) are calculated as
\begin{equation}
  \begin{array}{rll}
    \xi_0 (\varepsilon, 0) & = & 0\\
    \xi_1 (\varepsilon, 0) & = & \rho^2 \mymultiply \nlPoisson (\varepsilon)
    \mymultiply \left( - \frac{\mathd \nlPoisson}{\mathd \varepsilon}
    (\varepsilon) \right) \mymultiply \frac{- 1 + 7 \mymultiply c_{\Gamma}
    (\varepsilon)}{12 \mymultiply (1 + \varepsilon)}\\
    D_{1 \nocomma 1} (\varepsilon, 0) & = & \left( \nlPoisson (\varepsilon)
    \mymultiply \frac{\mathd \nlPoisson}{\mathd \varepsilon} (\varepsilon)
    \right)^2 \mymultiply \frac{\left( \mathpi \mymultiply \rho^2
    \right)^3}{24 \mymultiply \mathpi^2} \left( \frac{2 \mymultiply
    w_{\text{tr}}' (\varepsilon)}{1 + \varepsilon} + \nlPoisson^2
    (\varepsilon) \mymultiply G_{\text{t}} (\varepsilon) \mymultiply \left( 1
    + 7 \mymultiply c_{\Gamma}^2 (\varepsilon) \right) \right),
  \end{array} \label{eq:beam-1d-reduction-result}
\end{equation}

\subsection{Buckling analysis of the one-dimensional model}

We turn to the analysis of the buckling problem based on the one-dimensional
model just derived. We denote by $\varepsilon^{\star}$ the axial strain in the
unbuckled configuration, and by $u (S)$ and $v (S)$ the longitudinal and
transverse displacement associated with the buckling mode. The center-line
position therefore writes, in the buckled configuration of
figure~\ref{fig:thick-Euler-buckling}c, $\tmmathbf{r} (S) = v (S) \mymultiply
\tmmathbf{e}_2 + \left( S \mymultiply (1 + \varepsilon^{\star}) + u (S)
\right) \mymultiply \tmmathbf{e}_1$.

The axial strain $\varepsilon (S)$ and the rotation $\theta (S)$ are defined
by $\tmmathbf{r}' (S) = (1 + \varepsilon (S)) \mymultiply \tmmathbf{d}_3
(\theta (S))$ where $(\tmmathbf{d}_2 (\theta), \tmmathbf{d}_3 (\theta)) =
\left( \cos \theta \mymultiply \tmmathbf{e}_2 + \sin \theta \mymultiply
\tmmathbf{e}_3, - \sin \theta \mymultiply \tmmathbf{e}_2 + \cos \theta
\mymultiply \tmmathbf{e}_3 \right)$ is the rotated basis, see
equation~(\ref{eq:rPrime-epsilon-d3}). The bending strain is $\kappa (S) =
\theta' (S)$ from equation~(\ref{eq:kappa-i}). This yields the strain in terms
of the displacement as
\begin{equation}
  \begin{array}{rll}
    \varepsilon (S) & = & - 1 + \sqrt{(1 + \varepsilon^{\star} + u' (S))^2 +
    v^{\prime 2} (S)}\\
    \kappa (S) & = & \frac{\mathd}{\mathd S} \left( \tan^{- 1}  \frac{v'
    (S)}{1 + \varepsilon^{\star} + u' (S)} \right) .
  \end{array} \label{eq:bending-strain}
\end{equation}

The buckling problem is governed by the total potential energy
\begin{equation}
  \Psi^{\star} [u, v] = \Phi_{(2)}^{\star} [\varepsilon, \kappa] - P
  \mymultiply u (\ell), \label{eq:beam-Psi}
\end{equation}
where $\Phi_{(2)}^{\star}$ is the one-dimensional strain energy obtained in
section~\ref{ssec:beam-1d-model} and $P$ is the buckling load applied on the
plane in sliding contact with the endpoint of the cylinder. For the sake of
definiteness, we analyze buckling under force control rather than displacement
control; this makes no difference for the calculation of the critical loads.

We proceed to identify the boundary conditions applicable to the
one-dimensional model. By inserting the microscopic
displacement~(\ref{eq:Euler-buckling-microscopic-displacement}) into the
sliding conditions~(\ref{eq:Euler-buckling-sliding-condition}), we find that
the following boundary conditions must hold on both ends:
$\tmmathbf{d}_{\alpha} \cdot \tmmathbf{e}_3 = 0$ (which is equivalent to
$\theta = 0$), $\varepsilon' = 0$ and $\kappa' = 0$. In addition, the bottom
support is fixed, which yields $u (0) = 0$. The following kinematic boundary
conditions are therefore applicable,
\begin{equation}
  \begin{array}{lllllll}
    u (0) = 0 & \theta (0) = 0 & \theta (\ell) = 0 & \theta'' (0) = 0 &
    \theta'' (\ell) = 0 & \varepsilon' (0) = 0 & \varepsilon' (\ell) = 0.
  \end{array} \label{eq:stubby-cyl-boundary-conditions}
\end{equation}
The high-order boundary conditions on $\kappa' = \theta''$ are legal in the
variational problem of equilibrium as the
energy~(\ref{eq:bending-phi2-anticipation}) depends on $\kappa'' = \theta'''$
when $\xi_1 \neq 0$. The high-order boundary conditions on $\varepsilon'$ are
normally not legal since $\xi_0 = 0$ and the energy depends on $\varepsilon'$
but not on $\varepsilon''$; this points to the fact that boundary layers occur
generically near the boundaries as is known since the work of St-Venant. Such
layers are nevertheless absent for the particular choice of sliding boundaries
made here; this will enable us to ultimately satisfy all boundary conditions,
even if the problem looks ill-posed from a variational standpoint.

A principle of virtual work is obtained by inserting the
strain~(\ref{eq:bending-strain}) into the total potential energy
$\Psi^{\star}$, and by calculating the first variation with respect to the
unknowns $u$ and $v$.

To characterize the fundamental solution, we require that $\Psi^{\star} [u,
v]$ is stationary at $u \equiv 0$ and $v \equiv 0$: this yields the condition
$\int_0^{\ell} \left( \frac{\partial W_{\text{hom}}}{\partial \varepsilon}
(\varepsilon^{\star}, 0) - P \right) \mymultiply \hat{u}' (S) \mymultiply
\mathd S = 0$ for any $\hat{u} (S)$ such that $\hat{u} (0) = 0$, after taking
into account the identity $\xi_0 (\varepsilon^{\star}, 0) = 0$ in
equation~(\ref{eq:beam-1d-reduction-result}) and the symmetry
properties~(\ref{eq:bending-phi2-symmetries}). Therefore, the fundamental
solution selects the axial strain $\varepsilon^{\star}$ such that
\begin{equation}
  P = \frac{\partial W_{\text{hom}}}{\partial \varepsilon}
  (\varepsilon^{\star}, 0) . \label{eq:beam-fundamental-equilibrium}
\end{equation}
We have just recovered the force-displacement relation of our particular
material in simple traction.

The bifurcation condition is found by setting to zero the second variation of
$\Psi^{\star} [u, v]$ about the fundamental solution $u \equiv 0$ and $v
\equiv 0$. With the help of a symbolic calculation language, we obtain the
following variational problem for the critical strain $\varepsilon^{\star} =
\varepsilon_{\text{cr}}$ and the buckling mode $\overline{v} (S)$: for any
$\forall \hat{v} \text{ such that } \hat{v}' (0) = \hat{v}''' (0) = \hat{v}'
(\ell) = \hat{v}''' (\ell) = 0$,
\begin{equation}
  \int_0^{\ell} \left[ \left( 1 + \varepsilon_{\text{cr}} \right)  \left(
  \frac{\partial W_{\text{hom}}}{\partial \varepsilon} \right)_{\text{cr}}
  \mymultiply \overline{v}' \mymultiply \hat{v}' + \left( \frac{\partial^2
  W_{\text{hom}}}{\partial \kappa^2} \right)_{\text{cr}} \mymultiply \left(
  \overline{v}'' + (\xi_1)_{\text{cr}} \mymultiply \overline{v}'''' \right)
  \mymultiply \left( \hat{v}'' + (\xi_1)_{\text{cr}} \mymultiply \hat{v}''''
  \right) + (D_{1 \nocomma 1})_{\text{cr}} \mymultiply \overline{v}'''
  \mymultiply \hat{v}''' \right] \mymultiply \mathd S = 0.
  \label{eq:beam-bifurcation-eq}
\end{equation}
A decoupled eigenvalue problem is obtained for the longitudinal displacement
$u (S)$ as well but it is not reported here as it characterizes necking
instabilities, which we ignore. In equation above, all quantities bearing the
subscript `$\text{cr}$' are evaluated in the fundamental solution, i.e.,
$(f)_{\text{cr}} = f \left( \varepsilon_{\text{cr}}, 0 \right)$.

It is interesting to contrast equation~(\ref{eq:beam-bifurcation-eq}) with the
bifurcation equation predicted by a classical beam model, which ignores the
gradient effect. The latter can be be recovered by setting $\Phi^{\star}_{(2)}
[\varepsilon, \kappa] = \int_0^{\ell} W_{\text{hom}} (\varepsilon, \kappa)
\mymultiply \mathd S$ in equation~(\ref{eq:bending-phi2-anticipation}) and
hence corresponds to $(\xi_1)_{\text{cr}} = 0$ and $(D_{1 \nocomma
1})_{\text{cr}} = 0$; this yields a different bifurcation equation, namely
\[ \int_0^{\ell} \left[ \left( 1 + \varepsilon_{\text{cr}} \right)  \left(
   \frac{\partial W_{\text{hom}}}{\partial \varepsilon} \right)_{\text{cr}}
   \mymultiply \overline{v}' \mymultiply \hat{v}' + \left( \frac{\partial^2
   W_{\text{hom}}}{\partial \kappa^2} \right)_{\text{cr}} \mymultiply
   \overline{v}'' \mymultiply \hat{v}'' \right] \mymultiply \mathd S = 0
   \text{\quad (classical beam model)} . \]
Here, $\left( \frac{\partial W_{\text{hom}}}{\partial \varepsilon}
\right)_{\text{cr}} = P$ is the applied load, see
equation~(\ref{eq:beam-fundamental-equilibrium}), and $\left( \frac{\partial^2
W_{\text{hom}}}{\partial \kappa^2} \right)_{\text{cr}}$ is the incremental
bending modulus. Comparison with equation~(\ref{eq:beam-bifurcation-eq}) shows
that our asymptotic one-dimensional model corrects the classical buckling
analysis of beams in two ways, which are important for thick cylinders: it
makes use of the modified bending strain, $\tilde{\kappa} = \overline{v}'' +
(\xi_1)_{\text{cr}} \mymultiply \overline{v}''''$ instead of the standard
bending strain $\overline{v}''$, and it takes into account the energy cost
associated with the {\tmem{gradient}} of curvature $\overline{v}'''$, through
the term proportional to $D_{11}$.

We return to the asymptotically correct model, and proceed to solve the
bifurcation equation~(\ref{eq:beam-bifurcation-eq}). An ordinary differential
equation with constant coefficients can be obtained by integration by parts
and elimination of the virtual quantity $\hat{v}$. In view of the kinematic
boundary conditions $v' (0) = v''' (0) = v' (\ell) = v''' (\ell) = 0$, a
simple calculation shows that the first buckling mode is $\overline{v} (S) =
\frac{1 - \cos \left( k \mymultiply S \right)}{2}$ where $k =
\frac{\mathpi}{\ell}$, and that the critical strain $\varepsilon_{\text{cr}}$
is selected by the dispersion equation
\begin{equation}
  \left( 1 + \varepsilon_{\text{cr}} \right) \mymultiply \frac{\partial
  W_{\text{hom}}}{\partial \varepsilon} \left( \varepsilon_{\text{cr}}, 0
  \right) + \frac{\partial^2 W_{\text{hom}}}{\partial \kappa^2} \left(
  \varepsilon_{\text{cr}}, 0 \right) \mymultiply \left( 1 - \xi_1 \left(
  \varepsilon_{\text{cr}}, 0 \right) \mymultiply k^2 \right)^2 \mymultiply k^2
  + D_{1 \nocomma 1} \left( \varepsilon_{\text{cr}}, 0 \right) \mymultiply k^4
  = 0. \label{eq:beam-dispersion}
\end{equation}
This implicit equation for the first buckling load $\varepsilon_{\text{cr}}$
is valid for finite $\varepsilon_{\text{cr}}$. For a long beam, {\tmem{i.e.}},
when $\ell / \rho$ is large and $\left(k \mymultiply \rho\right)$ is small, we
can seek an expansion of the critical strain $\varepsilon_{\text{cr}}$ in
powers of the aspect-ratio parameter $e = \frac{k \mymultiply
\rho}{\sqrt{\mathpi}} = \frac{\sqrt{\mathpi} \mymultiply \rho}{\ell}$,
{\tmem{i.e.}},
\begin{equation}
  e^2 = \frac{\mathpi \mymultiply \rho^2}{\ell^2} . \label{eq:beam-e2}
\end{equation}

With the help of a symbolic calculation language, the series
$\varepsilon_{\text{cr}}$ satisfying the dispersion
equation~(\ref{eq:beam-dispersion}) is found as
\begin{equation}
  \varepsilon_{\text{cr}} = - \frac{\mathpi \mymultiply \chi_0}{4} \mymultiply
  e^2 - \mathpi^2 \mymultiply \left( \chi_1 - (\chi_2 + \chi_4) \mymultiply
  \chi_0 + \frac{2 + \chi_3}{32} \mymultiply \chi_0^2 \right) \mymultiply e^4
  +\mathcal{O} (e^6) \label{eq:beam-1d-model-epsilon-cr}
\end{equation}
where the $\chi_i$'s are dimensionless parameters from the one-dimensional
model,
\begin{equation}
  \begin{array}{lllll}
    \chi_0 = \frac{4}{\rho^2 \mymultiply \tilde{\chi}} \mymultiply \left(
    \frac{\partial^2 W_{\text{hom}}}{\partial \kappa^2} \right)_0 & \chi_1 =
    \frac{1}{\rho^4 \mymultiply \tilde{\chi}} \mymultiply (D_{1 \nocomma 1})_0
    & \chi_2 = \frac{\xi_1 (0, 0)}{2 \mymultiply \rho^2} & \chi_3 =
    \frac{1}{\tilde{\chi}} \mymultiply \left( \frac{\partial^2
    W_{\text{hom}}}{\partial \varepsilon^2} \right)_0 & \chi_4 = \frac{1}{4
    \mymultiply \rho^2 \mymultiply \tilde{\chi}} \mymultiply \left(
    \frac{\partial^3 W_{\text{hom}}}{\partial \varepsilon \mymultiply \partial
    \kappa^2} \right)_0
  \end{array}, \label{eq:chi-i}
\end{equation}
where $\tilde{\chi} = \left( \frac{\partial^2 W_{\text{hom}}}{\partial
\varepsilon^2} \right)_0$. Here the `$0$' is subscript means that the quantity
inside the corresponding parentheses must be evaluated in the undeformed
configuration, $(f)_0 = f (0, 0)$. To derive the
expansion~(\ref{eq:beam-1d-model-epsilon-cr}), we have used the fact that
there is no pre-stress in the reference configuration, $\frac{\partial
W_{\text{hom}}}{\partial \varepsilon} (0, 0) = 0$, as shown by combining
equations~(\ref{eq:beam-Whom}) and~(\ref{eq:beam-wtr-prime-0}): this warrants
that $\varepsilon_{\text{cr}} \rightarrow 0$ for $e \rightarrow 0$.

\subsection{Expansion of the critical load}

With the help of equations~(\ref{eq:beam-wtr-prime-0}--\ref{eq:beam-Y0-Y0p})
from the appendix, equations~(\ref{eq:beam-1d-reduction-coefficients})
and~(\ref{eq:beam-1d-reduction-result}) for a circular cross-section 
yield, in the limit $\varepsilon \rightarrow 0$,
\[ \begin{array}{lll}
     c_{\Gamma} (0) = \frac{1 + \nu_0}{\nu_0} \qquad & \xi_1 (0, 0) = \frac{7 + 6
     \mymultiply \nu_0}{12} \mymultiply \rho^2 \qquad & D_{1 \nocomma 1} (0, 0) = Y_0
     \mymultiply \frac{\left( \mathpi \mymultiply \rho^2
     \right)^3}{48 \mymultiply \mathpi^2} \mymultiply \frac{7 + 14 \mymultiply
     \nu_0 + 8 \mymultiply \nu_0^2}{1 + \nu_0} .
   \end{array} \]
Using this, the
relations~(\ref{eq:beam-simple-traction-E-Sigma}--\ref{eq:beam-Y0-Y0p}) and
the expression of $W_{\text{hom}}$ found in~(\ref{eq:beam-Whom}), we can
calculate the coefficients appearing in~(\ref{eq:chi-i}) as
\begin{equation}
  \begin{array}{lllll}
    \chi_0 = 1 \quad & \chi_1 = \mymultiply \mymultiply \mymultiply \frac{7 + 14
    \mymultiply \nu_0 + 8 \mymultiply \nu_0^2}{48 \mymultiply (1 + 
	\nu_0)} \quad  &
    \chi_2 = \frac{7 + 6 \mymultiply \nu_0}{24} \quad  & \chi_3 = \frac{Y_0'}{Y_0} &
    \quad  \chi_4 = \frac{1}{16} \mymultiply \left( \frac{Y_0'}{Y_0} - 2 \mymultiply
    \nu_0 \right) .
  \end{array}
\end{equation}
Inserting into equation~(\ref{eq:beam-1d-model-epsilon-cr}), we obtain our
final expression for the first critical load of the cylinder as a function of
the aspect-ratio parameter,
\begin{equation}
  \varepsilon_{\text{cr}} = - \frac{\mathpi \mymultiply e^2}{4} +
  \frac{\mathpi^2 \mymultiply e^4}{48} \mymultiply \left( \frac{3 \mymultiply
  Y_0'}{2 \mymultiply Y_0} + 4 - \frac{\nu_0 \mymultiply \left( 1 + 2
  \mymultiply \nu_0 \right)}{1 + \nu_0} \right) +\mathcal{O} (e^6) .
  \label{eq:beam-epsilon-expansion}
\end{equation}
This is identical to the result of~\citet{scherzinger1998asymptotic}; we
refer the reader to their paper for a comparison of this expansion with
finite-element simulations for a finite aspect-ratio $e$. Note that the
correction to the classical Euler prediction $\varepsilon_{\text{cr}} = -
\frac{\mathpi \mymultiply e^2}{4}$, {\tmem{i.e.}}, the term proportional to
$e^4$ in equation above, depends on both material nonlinearity (through the
non-linear elastic modulus $Y_0'$) and on geometric nonlinearity (through the
other terms in the parentheses).

\citet{scherzinger1998asymptotic} observed that classical models such as the
Euler beam model fail to capture the correction of order $e^4$ in
equation~(\ref{eq:beam-epsilon-expansion}) and are therefore inappropriate for
the analysis of stubby or thick structures; we concur with this statement. It
has apparently gone unnoticed that this difficulty can be overcome by using a
refined one-dimensional model capturing the gradient effect, as we just did:
when this is done in an asymptotically correct way, the expansion of the
critical buckling strain is correctly predicted in terms of the aspect-ratio.

Unlike in earlier work, we have split the analysis of this buckling problem
into two distinct tasks: deriving a one-dimensional model on the one hand, and
carrying out a bifurcation analysis on the other hand. Keeping these two tasks
separate is not only arguably more elegant, it also avoids the need to
reinvent the wheel for every buckling problem: if one were to study the
buckling of a stubby circular {\tmem{ring}} or the {\tmem{post-buckling}} of a
stubby Elastica in compression, for instance, one could reuse the
one-dimensional structural model from section~\ref{ssec:beam-1d-model} and
simply update the buckling analysis to reflect the geometry of interest.

\section{Discussion}

We have proposed an asymptotic method for constructing one-dimensional models,
see section~\ref{sec:asymptotic-1d-reduction}. The method achieves dimension
reduction by relaxing the microscopic displacement. Concretely, it is
implemented as a straightforward (albeit lengthy) series of steps, as
described in \ref{app:compendium}. It builds up on the general recipe
for dimension reduction published in our previous
work~\citep{LESTRINGANT2020103730}. The method yields an asymptotically
exact, variational model that accounts for the gradient effect. It also
accounts for geometric and material nonlinearities, and thereby may help
broaden the range of applicability of rod theories significantly. With a view
to illustrating the method, we have treated the linear twisting of an elastic
cylinder, including higher-order effects; in
equation~(\ref{eq:twisting-final-phi}) we have derived a simple
one-dimensional strain energy potential that governs the equilibrium, which is
new to the best of our knowledge. We have also applied our reduction method to
the Euler buckling of a beam having a moderate aspect-ratio: the expansion of
the critical load in powers of the aspect-ratio from earlier work has been
recovered based on a high-order rod model. The capabilities of the method go
much beyond these two simple examples; it can be readily applied to structures
involving finite strain, arbitrary pre-stress distributions or low material
symmetries, which will be the subject of future work.

For more complex geometries, the analytical approach adopted here may no
longer be tractable, and the quantities $W_{\text{hom}} (\tmmathbf{h})$,
$\tmmathbf{B} (\tmmathbf{h})$, etc.~defining the one-dimensional model may
have to be found by solving variational problems on the cross-section
numerically; the finite-element method is perfectly suited to this task. In
this approach, the 3D elasticity problem for the slender elastic body is split
into 2D+1D problems, where the 2D problems are microscopic and are formulated
in the cross-section while the 1D one is a macroscopic structural problem.
This splitting approach makes the solution considerably easier, which is
precisely the point of dimension reduction.

The rod models that are derived by our method include a kinematic constraint
which ensures that the tangent to the center-line $\tmmathbf{r}' (S)$ stays
aligned with the director $\tmmathbf{d}_3 (S)$, see
equation~(\ref{eq:rPrime-epsilon-d3}). One-dimensional model of this kind are
referred to as {\tmem{unshearable}} but this qualifier is misleading: shear
can take place at the microscopic scale in our approach, even if it is not
exposed in the one-dimensional model. As discussed at the very end
of~{\textsection}\ref{ssec:centerline-based-parameterization}, the directors
capture the deformed cross-sections in an average sense only: the
cross-sections are by no ways constrained to remain aligned with the directors
$\tmmathbf{d}_1 (S)$ and $\tmmathbf{d}_2 (S)$, {\tmem{i.e.}}, to remain
perpendicular to the center-line, not even in an average sense. For example,
in a rod made up of an anisotropic material that is very stiff in a direction
making a 45$^{\circ}$ angle with the axis of the rod, the cross-sections
rotate about $\tmmathbf{d}_2$ (and therefore tilt along the axis) by an angle
$a \mymultiply \varepsilon$ proportional to axial stretch $\varepsilon$. This
microscopic shear is accounted for in our approach but it is not reflected in
the directors $\tmmathbf{d}_i (S)$: their assigned role is to keep track of
the twisting of the cross-sections about the tangent, {\tmem{not}} to provide
a faithful representation of the microscopic solution. To some extent, our
approach therefore has the same capabilities as Timoshenko models, except that
the microscopic shear is dealt with {\tmem{internally}}. The benefit is that a
minimal set of degrees of freedom are presented to the user. The only minor
complication is that, in order to block the rotation at the endpoints, one has
to look up the average orientation of the terminal cross-sections from the
microscopic solution, as the vectors $\tmmathbf{d}_1 (S)$ and $\tmmathbf{d}_2
(S)$ cannot be used directly.

In this paper, we have carried out dimension reduction without making scaling
assumptions on the intensity of the loading. This is not a standard way of
proceeding. It is the special form of the external force potential in
equation~(\ref{eq:full-problem-total-potential-energy}) that plays the role of
the standard scaling assumptions, as discussed at the end of
{\textsection}\ref{sec:nonlinear-energy-formulation}. Let us briefly expose
how our method can be extended to handle non-standard scaling assumptions for
the loading. Consider for instance the case where the distributed applied
torque is so large that it can induce shear, by tilting the cross-sections
towards the center-line. Such a load cannot be represented by
equation~(\ref{eq:full-problem-total-potential-energy}) since by design the
directors $\tmmathbf{d}_1$ and $\tmmathbf{d}_2$ remain perpendicular to the
center-line. The solution is to introduce an additional kinematic variable,
similar to Timoshenko's shear angle, and to modify the procedure as follows.
The new internal degree of freedom is appended to the set of macroscopic
strain $\tmmathbf{h}$, meaning that it is fixed during the relaxation
procedure and is a variable of the one-dimensional model; it is coupled to the
large applied torque in the potential energy $\Psi$. The result of this
modified relaxation procedure is an {\tmem{asymptotic}} Timoshenko-like model,
in which the average microscopic shear is explicitly represented.

It is also possible to extend the method to the case where the geometric or
elastic properties of the body vary slowly in the longitudinal
direction---such as the case of rods having non-uniform cross-sectional
dimensions---, with little additional work. This extension is discussed at the
very end of our previous paper~\citep{LESTRINGANT2020103730}: in this case
the operator $\tmmathbf{C}_{\tmmathbf{h}}^{(1)}$ gets an explicit dependence
on the axial variable $S$, and an additional term proportional to $\left.
\frac{\partial \tmmathbf{C}_{\tmmathbf{h}}^{(1)} (S)}{\partial S}
\right|_{\tmmathbf{h}=\tmmathbf{h} (S)}$ appears in the one-dimensional
potential $\Phi_{(2)}^{\star}$. This, along with other extensions, will be
described in follow-up work.

\paragraph{Acknowledgments}This paper was prepared using {\TeXmacs}, an
outstanding and freely available scientific text
editor~\citep{Hoeven-The-jolly-writer.-Your-2020}.

\appendix\section{Equilibrium of the original (3D) and ideal (1D)
models}\label{app:equilibrium-full-ideal}

\subsection{Equilibrium of the ideal
model}\label{app-sec:equilibrium-ideal-model}

In this section, we derive the equations of equilibrium for the ideal
one-dimensional model from section~\ref{sec:ideal-model} using a variational
method. Doing so, we prove the results announced in
section~\ref{ssec:ideal-equilibrium}: the ideal model is governed by the
classical Kirchhoff equations for the equilibrium thin rods and by the
constitutive laws~(\ref{eq:internal-stress-full-model}).

Combining equations~(\ref{eq:ideal-1d-total-potential-energy}) for the total
potential $\Psi^{\star}$, and~(\ref{eq:phi-star-by-relaxation}) for the
condensed strain energy, we obtain
\[ \Psi^{\star} [\tmmathbf{r}, \tmmathbf{d}_i] = \Phi [\tmmathbf{h},
   \tmmathbf{y}^{\star} [\tmmathbf{h}]] + \int_0^{\ell} V (\tmmathbf{r} (S),
   \tmmathbf{d}_i (S)) \mymultiply \mathd S. \]
We introduce perturbations $\hat{\tmmathbf{r}}$ and $\hat{\tmmathbf{\theta}}$
to the macroscopic fields $\tmmathbf{r}$ and $\tmmathbf{d}_i$, as in
section~\ref{ssec:ideal-equilibrium}, that satisfy the incremental form of the
kinematic constraint~(\ref{eq:rPrime-epsilon-d3}),
\begin{equation}
  \hat{\tmmathbf{r}}' (S) = \hat{\varepsilon} (S) \mymultiply \tmmathbf{d}_3
  (S) + \hat{\tmmathbf{\theta}} (S) \times \tmmathbf{r}' (S) .
  \label{eq:incremental-adaptation}
\end{equation}
Using the incremental form of the equations defining the macroscopic strain
from section~\ref{ssec:apparent-strain}, one can define the strain
perturbations $\hat{\tmmathbf{h}} = (\hat{\varepsilon}, \hat{\kappa}_1,
\hat{\kappa}_2, \hat{\kappa}_3)$ in terms of $\hat{\tmmathbf{r}}$ and
$\hat{\tmmathbf{\theta}}$. This leads to a classical result from rod theory,
\begin{equation}
  \hat{\kappa}_i (S) = \hat{\tmmathbf{\theta}} (S) \cdot \tmmathbf{d}_i (S),
  \label{eq:kappa-i-hat}
\end{equation}
see for instance \citet{Audoly-Pomeau-Elasticity-and-geometry:-from-2010}.

Using the first variation of the potential of external load
in~(\ref{eq:V-peturb}), one can write the perturbation to the total potential
as
\[ \hat{\Psi}^{\star} = \frac{\partial \Phi}{\partial \tmmathbf{y}}
   [\tmmathbf{h}, \tmmathbf{y}^{\star} [\tmmathbf{h}]] \cdot \left(
   \frac{\mathd \tmmathbf{y}^{\star}}{\mathd \tmmathbf{h}} [\tmmathbf{h}]
   \cdot \hat{\tmmathbf{h}} \right) + \frac{\partial \Phi}{\partial
   \tmmathbf{h}} [\tmmathbf{h}, \tmmathbf{y}^{\star} [\tmmathbf{h}]] \cdot
   \hat{\tmmathbf{h}} - \int_0^{\ell} (\tmmathbf{p} (S) \cdot
   \hat{\tmmathbf{r}} (S) +\tmmathbf{m} (S) \cdot \hat{\tmmathbf{\theta}} (S))
   \mymultiply \mathd S. \]
In the first term, the perturbation $\left( \frac{\mathd
\tmmathbf{y}^{\star}}{\mathd \tmmathbf{h}} [\tmmathbf{h}] \cdot
\hat{\tmmathbf{h}} \right)$ to the microscopic solution satisfies the
kinematic constraint $\tmmathbf{q} (\tmmathbf{Y}) =\tmmathbf{0}$, as can be
shown by differentiating $\tmmathbf{q} (\tmmathbf{y}^{\star} [\tmmathbf{h}])
=\tmmathbf{0}$ with respect to $\tmmathbf{h}$, recalling that $\tmmathbf{q}$
is linear. In view of equation~(\ref{eq:variational-eq-for-y-star}), we
conclude that this first term is zero, so that
\begin{equation}
  \hat{\Psi}^{\star} = \frac{\partial \Phi}{\partial \tmmathbf{h}}
  [\tmmathbf{h}, \tmmathbf{y}^{\star} [\tmmathbf{h}]] \cdot \hat{\tmmathbf{h}}
  - \int_0^{\ell} (\tmmathbf{p} (S) \cdot \hat{\tmmathbf{r}} (S) +\tmmathbf{m}
  (S) \cdot \hat{\tmmathbf{\theta}} (S)) \mymultiply \mathd S.
  \label{eq:psi-star-hat}
\end{equation}

The first term in the right-hand side of equation~(\ref{eq:psi-star-hat}) can
be rewritten as follows, using equations~(\ref{eq:canonicalForm}),
(\ref{eq:microscopic-stress}), (\ref{eq:internal-stress-full-model})
and~(\ref{eq:kappa-i-hat}) successively,
\[ \begin{array}{lll}
     \frac{\partial \Phi}{\partial \tmmathbf{h}} [\tmmathbf{h},
     \tmmathbf{y}^{\star} [\tmmathbf{h}]] & = & \frac{\partial}{\partial
     \tmmathbf{h}} \left( \int_0^{\ell} \iint_{\Omega} w (\tmmathbf{T},
     \tmmathbf{E} (\tmmathbf{T}; \tmmathbf{h} (S) ; \nobracket
     \tmmathbf{y}^{\star} [\tmmathbf{h}] |_S, \nobracket \tmmathbf{y}^{\star}
     [\tmmathbf{h}]' |_S)) \mymultiply \mathd A \mymultiply \mathd S \right)
     \cdot \hat{\tmmathbf{h}}\\
     & = & \int_0^{\ell} \iint_{\Omega} \Sigma_{i \nocomma j}
     (\tmmathbf{T}, \tmmathbf{E} (\tmmathbf{T}; \tmmathbf{h} (S) ; \nobracket
     \tmmathbf{y}^{\star} [\tmmathbf{h}] |_S, \nobracket \tmmathbf{y}^{\star}
     [\tmmathbf{h}]' |_S)) \mymultiply \mymultiply \frac{\partial E_{i
     \nocomma j}}{\partial \tmmathbf{h}} (\tmmathbf{T}; \tmmathbf{h} (S) ;
     \nobracket \tmmathbf{y}^{\star} [\tmmathbf{h}] |_S, \nobracket
     \tmmathbf{y}^{\star} [\tmmathbf{h}]' |_S) \cdot \hat{\tmmathbf{h}} (S)
     \mymultiply \mathd A \mymultiply \mathd S\\
     & = & \int_0^{\ell} \left[ N (S) \mymultiply \hat{\varepsilon} + M_i (S)
     \mymultiply \hat{\kappa}_i \right] \mymultiply \mathd S\\
     & = & \int_0^{\ell} \left[ N (S) \mymultiply \hat{\varepsilon}
     +\tmmathbf{M} (S) \cdot \hat{\tmmathbf{\theta}}' (S) \right] \mymultiply
     \mathd S
   \end{array} \]
where $\tmmathbf{M} (S) = M_i (S) \mymultiply \tmmathbf{d}_i (S)$ is the
internal moment.

Treating the kinematic constraint~(\ref{eq:incremental-adaptation}) using a
Lagrange multiplier~$\tmmathbf{R} (S)$, we obtain by setting
$\hat{\Psi}^{\star} = 0$ in equation~(\ref{eq:psi-star-hat}) the principle of
virtual work of the ideal one-dimensional model as
\begin{multline}
    \forall (\hat{\tmmathbf{r}}, \hat{\tmmathbf{\theta}}) \quad \int_0^{\ell}
    \left( N (S) \mymultiply \hat{\varepsilon} (S) +\tmmathbf{M} (S) \cdot
    \hat{\tmmathbf{\theta}}' (S) \right) \mathd S - \int_0^{\ell}
    (\tmmathbf{p} (S) \cdot \hat{\tmmathbf{r}} (S) +\tmmathbf{m} (S) \cdot
    \hat{\tmmathbf{\theta}} (S)) \mymultiply \mathd S \ldots\\
    + \int_0^{\ell} \tmmathbf{R} (S) \cdot \left( \hat{\tmmathbf{r}}' (S) -
    \hat{\varepsilon} (S) \mymultiply \tmmathbf{d}_3 (S) -
    \hat{\tmmathbf{\theta}} (S) \times \tmmathbf{r}' (S) \right) \mymultiply
    \mathd S = 0
\end{multline}

We have recovered the classical form of the principle of virtual for thin
elastic rods. Upon integration by parts, one recovers the Kirchhoff
equations~(\ref{eq:rod-strong-equilibrium}) governing the equilibrium of rods,
together with the relevant equilibrium conditions.

\subsection{Equivalence with the equilibrium of the three-dimensional
model}\label{app-sec:original-ideal-same-equilibrium}

In this section, we derive the equilibrium equations of the three-dimensional
model in center-line representation, as formulated in
section~\ref{sec:full-model}. We show that these equilibrium equations are
mathematically equivalent to those of the ideal one-dimensional model.

In the three-dimensional model, the microscopic displacement $\tmmathbf{y} (S,
\tmmathbf{T})$ and the macroscopic variables $\tmmathbf{r} (S)$ and
$\tmmathbf{d}_i (S)$ are treated as independent unknowns.

The principle of virtual work for the prismatic solid is obtained by setting
to zero the first variation of the total potential energy $\Psi [\tmmathbf{r},
\tmmathbf{d}_i, \tmmathbf{y}]$ in
equation~(\ref{eq:full-problem-total-potential-energy}), with respect to both
the microscopic and the macroscopic variables. The variation with respect to
the microscopic displacement $\tmmathbf{y}$, which is subject to the
constraint~(\ref{eq:constraint-q}) yields
\[ \left( \forall \hat{\tmmathbf{y}} \text{ such that $\forall S, \tmmathbf{q}
   (\nobracket \hat{\tmmathbf{y}} |_S) =\tmmathbf{0}$} \right) \qquad
   \frac{\partial \Phi}{\partial \tmmathbf{y}} [\tmmathbf{h}, \tmmathbf{y}]
   \cdot \tmmathbf{y}= 0. \]
We have recovered the relaxation problem from
equation~(\ref{eq:variational-eq-for-y-star}), whose solution is
\[ \tmmathbf{y}=\tmmathbf{y}^{\star} [\tmmathbf{h}] . \]
The variation of the total potential
energy~(\ref{eq:full-problem-total-potential-energy}) with respect to the
macroscopic variables writes
\[ \hat{\Psi} [\tmmathbf{r}, \tmmathbf{d}_i, \tmmathbf{y}] = \frac{\partial
   \Phi}{\partial \tmmathbf{h}} [\tmmathbf{h}, \tmmathbf{y}] \cdot
   \hat{\tmmathbf{h}} - \int_0^{\ell} (\tmmathbf{p} (S) \cdot
   \hat{\tmmathbf{r}} (S) +\tmmathbf{m} (S) \cdot \hat{\tmmathbf{\theta}} (S))
   \mymultiply \mathd S \]
after identifying the external load from equation~(\ref{eq:V-peturb}).
Inserting $\tmmathbf{y}=\tmmathbf{y}^{\star} [\tmmathbf{h}]$ in this equation,
and taking into account the constraints in
equation~(\ref{eq:di-orthonormal-frame}) and~(\ref{eq:rPrime-epsilon-d3}), we
obtain the same variational problem as that of the ideal one-dimensional
model, see equation~(\ref{eq:psi-star-hat}). By the same reasoning as earlier,
one can show that the equilibrium of the three-dimensional model is governed
by the Kirchhoff equations for thin rods, and by the same constitutive laws as
those identified earlier in equation~(\ref{eq:internal-stress-full-model}).

We conclude that the original three-dimensional model and its ideal reduction
have the same equations of equilibrium.

\subsection{Microscopic interpretation of the one-dimensional
stress}\label{app:microscopic-interpretation-1d-internal-stress}

In this section, we show that the one-dimensional stress quantities $N (S)$
and $\tmmathbf{M} (S) = M_i (S) \mymultiply \tmmathbf{d}_i (S)$ that appear in
equation~(\ref{eq:internal-stress-full-model}) can be interpreted
microscopically as the normal force ({\tmem{i.e.}}, the tangent component of
the internal force $\tmmathbf{R} (S)$) and the internal moment, respectively.

The integrand in the right hand side of (\ref{eq:internal-stress-full-model})
can be rewritten as $\Sigma_{i \nocomma j} \mymultiply \hat{E}_{i \nocomma
j}$, where $\hat{E}_{i \nocomma j} = \frac{\partial E_{i \nocomma j}}{\partial
\tmmathbf{h}} (\tmmathbf{T}; \tmmathbf{h} (S) ; \nobracket \tmmathbf{y} |_S,
\nobracket \tmmathbf{y}' |_S) \cdot \hat{\tmmathbf{h}} (S)$ is the
perturbation to the microscopic strain due to a change $\hat{\tmmathbf{h}}$ in
the macroscopic strain $\tmmathbf{h}$. By taking the first variation of the
general strain in equation~(\ref{eq:E-function}) with respect to
$\tmmathbf{h}$, we obtain
\[ \Sigma_{i \nocomma j} \mymultiply \hat{E}_{i \nocomma j} = \left( t_k
   \mymultiply \Sigma_{3 \nocomma 3} + \partial_{\alpha} Y_k (\tmmathbf{T})
   \mymultiply \Sigma_{\alpha \nocomma 3} \right) \mymultiply \hat{t}_k \]
where $\hat{t}_k = \hat{\varepsilon} \mymultiply \delta_{i \nocomma 3} +
\eta_{i \nocomma j \nocomma k} \mymultiply \hat{\kappa}_j \mymultiply y_k
(\tmmathbf{T})$. In the equation above, we can identify the components of the
transformation gradient from equation~(\ref{eq:transformation-gradient}) as
$t_k = F_{k \nocomma 3}$ and $\partial_{\alpha} Y_k (\tmmathbf{T}) = F_{k
\nocomma \alpha}$, so that
\[ \begin{array}{lll}
     \Sigma_{i \nocomma j} \mymultiply \hat{E}_{i \nocomma j} & = & \left(
     F_{k \nocomma 3} \mymultiply \Sigma_{3 \nocomma 3} + F_{k \nocomma
     \alpha} \mymultiply \Sigma_{\alpha \nocomma 3} \right) \mymultiply
     \hat{t}_k\\
     & = & (\tmmathbf{F} \cdot \tmmathbf{\Sigma} \cdot \tmmathbf{e}_3)_k
     \mymultiply \left( \hat{\varepsilon} \mymultiply \delta_{i \nocomma 3} +
     \eta_{k \nocomma i \nocomma j} \mymultiply \hat{\kappa}_i \mymultiply
     \mymultiply y_j (\tmmathbf{T}) \right)
   \end{array} \]
According to the standard interpretation of the Piola-Kirchhoff stress
$\tmmathbf{\Sigma}$, the quantity $\tmmathbf{F} \cdot \tmmathbf{\Sigma} \cdot
\tmmathbf{e}_3$ is the microscopic internal force $\mathd \tmmathbf{\sigma}$
per unit area $\mathd A$ transmitted across a cross-section (the unit normal
to the cross-section is $\tmmathbf{e}_3$ in reference configuration). Denoting
as $\mathd \sigma_k$ the component $\mathd \sigma_k = \mathd \tmmathbf{\sigma}
\cdot \tmmathbf{d}_k$ of the microscopic internal force on the orthonormal
basis of directors, we introduce the density of internal force across a
cross-section as
\[ \frac{\mathd \sigma_k}{\mathd A} =\tmmathbf{d}_k \cdot \tmmathbf{F} \cdot
   \tmmathbf{\Sigma} \cdot \tmmathbf{e}_3 = t_k \mymultiply \Sigma_{3 \nocomma
   3} + \partial_{\alpha} Y_k (\tmmathbf{T}) \mymultiply \Sigma_{\alpha
   \nocomma 3} . \]
Inserting this into the expression of the integrand $\Sigma_{i \nocomma j}
\mymultiply \hat{E}_{i \nocomma j}$, we can rewrite
equation~(\ref{eq:internal-stress-full-model}) as
\[ N (S) \mymultiply \hat{\varepsilon} + M_i (S) \mymultiply \hat{\kappa}_i
   \equiv \iint_{\Omega} \left( \hat{\varepsilon} \mymultiply \delta_{i
   \nocomma 3} + \eta_{k \nocomma i \nocomma j} \mymultiply \hat{\kappa}_i
   \mymultiply \mymultiply y_j (\tmmathbf{T}) \right) \mymultiply \frac{\mathd
   \sigma_k}{\mathd A} \mymultiply \mathd A. \]
Identifying the coefficients of $\hat{\varepsilon}$ and $\hat{\kappa}_i$ on
both sides, we obtain
\begin{equation}
  \begin{aligned}
    N & =  \tmmathbf{d}_3 \cdot \iint_{\Omega} \mathd \tmmathbf{\sigma}\\
    \tmmathbf{M} & =  \iint_{\Omega} \left( y_j (\tmmathbf{T}) \mymultiply
    \tmmathbf{d}_j (S) \right) \times \mathd \tmmathbf{\sigma}=
    \iint_{\Omega} (\tmmathbf{x}-\tmmathbf{r} (S)) \times \mathd
    \tmmathbf{\sigma},
  \end{aligned} \label{eq:app-NM-interpretation}
\end{equation}
which is the classical microscopic interpretation of the internal stress $N$
and $\tmmathbf{M}$ from the one-dimensional model: in these equations
$\iint_{\Omega} \mathd \tmmathbf{\sigma}=\tmmathbf{R}$ is the internal
force resultant, $\tmmathbf{r} (S)$ is the point of the center-line that is
formally associated with the cross-section with coordinate $S$, and
$\tmmathbf{x}$ is a current point in the deformed cross-section.

\section{A compendium of the reduction method}\label{app:compendium}

This section is a self-contained summary of the reduction method from our
previous paper, with ({\tmem{i}})~small changes of notation, (ii)~a new method
for removing boundary terms, see the end
of~{\textsection}\ref{sec-app:elimination-of-boundary-terms},
({\tmem{iii}})~some additional simplifications arising when the equations are
specialized to the case relevant to three-dimensional elasticity where the
strain $\tmmathbf{E}$ depends on the displacement $\tmmathbf{Y}$, on the
longitudinal gradient of displacement $\tmmathbf{Y}^{\dag}$ and on the
macroscopic strain $\tmmathbf{h}$ only but not on higher-order derivatives.
The reader is referred to the original paper~\citep{LESTRINGANT2020103730}
for a detailed justification of the method.

The reduction procedure uses the definition of the full model in equations
(\ref{eq:canonicalForm}--\ref{eq:constraint-q}) as a starting point.

\subsection{Analysis of homogeneous
solutions}\label{app:compendium-homogeneous}

We first focus on homogeneous solutions, such that $\tmmathbf{y}' \equiv
\tmmathbf{0}$. The strain is then $\tilde{\tmmathbf{E}} (\tmmathbf{T},
\tmmathbf{h}, \tmmathbf{Y}) =\tmmathbf{E} (\tmmathbf{T}; \tmmathbf{h};
\tmmathbf{Y}, \tmmathbf{0})$, where $\tmmathbf{Y}= \nobracket \tmmathbf{y}
|_S$ denotes the microscopic displacement restricted to a particular
cross-section, which is independent of $S$ in the homogeneous case.

As explained earlier in equation~(\ref{eq:variational-eq-for-y-star}), we
identify the relaxed displacement of the homogeneous solutions by making
stationary the strain energy per unit length $\iint_{\Omega} w
(\tmmathbf{T}, \tilde{\tmmathbf{E}} (\tmmathbf{T}, \tmmathbf{h},
\tmmathbf{Y})) \mymultiply \mathd A$ with respect to $\tmmathbf{Y}$ for
prescribed $\tmmathbf{h}$, taking the kinematic
constraint~(\ref{eq:constraint-q}) into account. We denote as
$\tmmathbf{Y}^{\tmmathbf{h}} (\tmmathbf{T})$ this solution. This variational
problem can be formulated using Lagrange multipliers as follows. Given
$\tmmathbf{h}$, we seek a cross-sectional function
$\tmmathbf{Y}^{\tmmathbf{h}}$ and Lagrange multipliers
$\tmmathbf{M}^{\tmmathbf{h}} = (F_1^{\tmmathbf{h}}, F_2^{\tmmathbf{h}},
F_3^{\tmmathbf{h}}, Q^{\tmmathbf{h}})$ dual to the constraints $\tmmathbf{q}$,
such that
\begin{equation}
  \begin{gathered}
    \tmmathbf{q} (\tmmathbf{Y}^{\tmmathbf{h}}) =\tmmathbf{0}\\
    \forall \hat{\tmmathbf{Y}} \quad \iint_{\Omega} \tmmathbf{\Sigma}
    (\tmmathbf{T}, \tmmathbf{E}^{\tmmathbf{h}} (\tmmathbf{T})) \doublecontract
    \widehat{\tilde{\tmmathbf{E}}}^{\tmmathbf{h}} (\tmmathbf{T}) \mymultiply
    \mathd A +\tmmathbf{M}^{\tmmathbf{h}} \cdot \tmmathbf{q}
    (\hat{\tmmathbf{Y}}^{\tmmathbf{h}}) = 0.
  \end{gathered} \label{eq:Yh-variational-abstract}
\end{equation}
Here, $\tmmathbf{\Sigma} (\tmmathbf{T}, \tmmathbf{E}) = \frac{\partial
w}{\partial \tmmathbf{E}} (\tmmathbf{T}, \tmmathbf{E})$ denotes the
microscopic stress, $\tmmathbf{E}^{\tmmathbf{h}} (\tmmathbf{T}) =
\tilde{\tmmathbf{E}} (\tmmathbf{T}; \tmmathbf{h},
\tmmathbf{Y}^{\tmmathbf{h}})$ is the distribution of microscopic strain in the
homogeneous solution having macroscopic strain $\tmmathbf{h}$, and
$\widehat{\tilde{\tmmathbf{E}}}^{\tmmathbf{h}} (\tmmathbf{T}) = \frac{\partial
\tilde{\tmmathbf{E}}}{\partial \tmmathbf{Y}} (\tmmathbf{T}, \tmmathbf{h},
\tmmathbf{Y}^{\tmmathbf{h}}) \cdot \hat{\tmmathbf{Y}}$ is the virtual
increment of microscopic strain.

Having solved for the microscopic displacement, one can define the strain
energy per unit length $W_{\text{hom}} (\tmmathbf{h})$ from
equation~(\ref{eq:Wh-def}), as well as the homogeneous strain
$\tmmathbf{E}^{\tmmathbf{h}}$, stress $\tmmathbf{\Sigma}^{\tmmathbf{h}}$ and
tangent elastic moduli $\tmmathbf{K}^{\tmmathbf{h}}$ from
equation~(\ref{eq:gr-effect-homogeneous-qties}).

\subsection{Gradient effect}\label{sec-app:general-method-gradient}

The microscopic displacement solving the relaxation problem~(\ref{eq:relax-y})
is sought in the form
\begin{equation}
  \tmmathbf{y}^{\star} [\tmmathbf{h}] (S, \tmmathbf{T})
  =\tmmathbf{Y}^{\tmmathbf{h} (S)} (\tmmathbf{T}) +\tmmathbf{z} (S,
  \tmmathbf{T}), \label{eq:y-expansion}
\end{equation}
{\tmem{i.e.}}, as the solution $\tmmathbf{Y}^{\tmmathbf{h} (S)}$ predicted by
the catalog of homogeneous solutions based on the local value $\tmmathbf{h}
(S)$ of the macroscopic strain, plus a small correction $\tmmathbf{z} (S,
\tmmathbf{T})$ to be calculated by solving the equilibrium equations.

\subsubsection{Structure operators}\label{app-sec:structure-operators}

The restrictions of the displacement and its gradient in
equation~(\ref{eq:y-expansion}) can be calculated as $\nobracket
\tmmathbf{y}^{\star} [\tmmathbf{h}] |_S =\tmmathbf{Y}^{\tmmathbf{h}}
+\tmmathbf{Z}$ and $\nobracket \tmmathbf{y}^{\star} [\tmmathbf{h}]' |_S
=\tmmathbf{h}^{\dag} \cdot \nabla \tmmathbf{Y}^{\tmmathbf{h}}
+\tmmathbf{Z}^{\dag}$, respectively, where
\begin{equation}
  \begin{array}{lll}
    \tmmathbf{h}^{\dag} =\tmmathbf{h}' (S) & \tmmathbf{Z}= \nobracket
    \tmmathbf{z} |_S & \tmmathbf{Z}^{\dag} = \nobracket \tmmathbf{z}' |_S,
  \end{array} \label{eq:dag-quantities}
\end{equation}
and we use the $\nabla$ notation for gradients with respect to $\tmmathbf{h}$,
{\tmem{i.e.}}, $\tmmathbf{h}^{\dag} \cdot \nabla \tmmathbf{Y}^{\tmmathbf{h}} =
\frac{\mathd (\tmmathbf{Y}^{\tmmathbf{h}})}{\mathd \tmmathbf{h}} \cdot
\tmmathbf{h}^{\dag}$ as earlier in equation~(\ref{eq:nabla-notation}).

We define the structure operators $\tmmathbf{e}_{j \nocomma k}^i
(\tmmathbf{T}, \tmmathbf{h})$ as the coefficients entering in the expansion of
the strain $\tmmathbf{E} (\tmmathbf{T}; \tmmathbf{h};
\tmmathbf{Y}^{\tmmathbf{h}} +\tmmathbf{Z}, \tmmathbf{h}^{\dag} \cdot \nabla
\tmmathbf{Y}^{\tmmathbf{h}} +\tmmathbf{Z}^{\dag})$ associated with the
solution~(\ref{eq:y-expansion}), in powers of $\tmmathbf{h}^{\dag}$,
$\tmmathbf{Z}$ and $\tmmathbf{Z}^{\dag}$
\begin{multline}
    \tmmathbf{E} (\tmmathbf{T}; \tmmathbf{h}; \tmmathbf{Y}^{\tmmathbf{h}}
    +\tmmathbf{Z}, \tmmathbf{h}^{\dag} \cdot \nabla
    \tmmathbf{Y}^{\tmmathbf{h}} +\tmmathbf{Z}^{\dag}) =
    \tmmathbf{E}^{\tmmathbf{h}} (\tmmathbf{T}) +\tmmathbf{e}_{0 \nocomma 0}^1
    (\tmmathbf{T}, \tmmathbf{h}) \cdot \tmmathbf{h}^{\dag} +\tmmathbf{e}_{1
    \nocomma 0}^0 (\tmmathbf{T}, \tmmathbf{h}) \cdot
    \tmmathbf{Z}+\tmmathbf{e}_{0 \nocomma 1}^0 (\tmmathbf{T}, \tmmathbf{h})
    \cdot \tmmathbf{Z}^{\dag} \ldots\\
    \nobracket \nobracket \nobracket \nobracket {} + \frac{1}{2}
    \mymultiply \left( \tmmathbf{h}^{\dag} \cdot \tmmathbf{e}_{0 \nocomma 0}^2
    (\tmmathbf{T}, \tmmathbf{h}) \cdot \tmmathbf{h}^{\dag} + 2 \mymultiply
    \tmmathbf{h}^{\dag} \cdot \tmmathbf{e}_{1 \nocomma 0}^1 (\tmmathbf{T},
    \tmmathbf{h}) \cdot \tmmathbf{Z}+\tmmathbf{Z} \cdot \tmmathbf{e}_{2
    \nocomma 0}^0 (\tmmathbf{T}, \tmmathbf{h}) \cdot \tmmathbf{Z}+ \cdots
    \right) .
    \label{eq:expansion-for-structure-operators}
\end{multline}
The structure operators $\tmmathbf{e}_{j \nocomma k}^i (\tmmathbf{T},
\tmmathbf{h})$ are calculated explicitly in
\ref{app:structure-operators}.

Note that the arguments $\tmmathbf{Z}$ and $\tmmathbf{Z}^{\dag}$ are unknown
so far: both of them is a triple of functions $Z_i$ and $Z_i^{\dag}$ defined
on the cross-section.

\subsubsection{Optimal displacement}\label{ssec:compendium-ACB}

In terms of the structure operators $e_{i \nocomma j}^k (\tmmathbf{T},
\tmmathbf{h})$ just identified, we define the operator
\begin{equation}
  \mathcal{E}^{\tmmathbf{h}} (\tmmathbf{T}, \tmmathbf{h}^{\dag}, \tmmathbf{Z})
  =\tmmathbf{e}_{0 \nocomma 0}^1 (\tmmathbf{T}, \tmmathbf{h}) \cdot
  \tmmathbf{h}^{\dag} +\tmmathbf{e}_{1 \nocomma 0}^0 (\tmmathbf{T},
  \tmmathbf{h}) \cdot \tmmathbf{Z}, \label{eq:perturbed-strain-Ecal}
\end{equation}
which corresponds to the difference between the actual microscopic strain
associated with the displacement~(\ref{eq:y-expansion}) and the crude
prediction $\tmmathbf{E}^{\tmmathbf{h} (S)} (\tmmathbf{T})$ obtained by
looking up the catalog of homogeneous solutions with the local value of the
macroscopic strain $\tmmathbf{h} (S)$. We also define the operators
\begin{equation}
  \begin{array}{rcl}
    \tmmathbf{C}_{\tmmathbf{h}}^{(1)} \cdot \tmmathbf{Z}^{\dag} & = &
    \iint_{\Omega} \tmmathbf{\Sigma}^{\tmmathbf{h}} (\tmmathbf{T})
    \doublecontract (\tmmathbf{e}^0_{0 \nocomma 1} (\tmmathbf{T},
    \tmmathbf{h}) \cdot \tmmathbf{Z}^{\dag}) \mymultiply \mathd A\\
    \mathcal{B}^{\tmmathbf{h}} (\tmmathbf{h}^{\dag}, \tmmathbf{Z}) & = &
    \iint_{\Omega} \frac{1}{2} \mymultiply \mathcal{E}^{\tmmathbf{h}}
    (\tmmathbf{T}, \tmmathbf{h}^{\dag}, \tmmathbf{Z}) \doublecontract
    \tmmathbf{K}^{\tmmathbf{h}} (\tmmathbf{T}) \doublecontract
    \mathcal{E}^{\tmmathbf{h}} (\tmmathbf{T}, \tmmathbf{h}^{\dag},
    \tmmathbf{Z}) \mymultiply \mathd A \ldots\\
    &  & \text{{\hspace{4em}}} + \iint_{\Omega} \frac{1}{2} \mymultiply
    \tmmathbf{\Sigma}^{\tmmathbf{h}} (\tmmathbf{T}) \doublecontract \left(
    \tmmathbf{h}^{\dag} \cdot \tmmathbf{e}^2_{0 \nocomma 0} (\tmmathbf{T},
    \tmmathbf{h}) \cdot \tmmathbf{h}^{\dag} + 2 \mymultiply
    \tmmathbf{h}^{\dag} \cdot \tmmathbf{e}^1_{1 \nocomma 0} (\tmmathbf{T},
    \tmmathbf{h}) \cdot \tmmathbf{Z}+\tmmathbf{Z} \cdot \tmmathbf{e}^0_{2
    \nocomma 0} (\tmmathbf{T}, \tmmathbf{h}) \cdot \tmmathbf{Z} \right)
    \mymultiply \mathd A \ldots\\
    &  & \text{{\hspace{4em}}} -\tmmathbf{h}^{\dag} \cdot \nabla
    \tmmathbf{C}_{\tmmathbf{h}}^{(1)} \cdot \tmmathbf{Z}
  \end{array} \label{eq:ACB-operators}
\end{equation}
Note that the operator $\tmmathbf{C}_{\tmmathbf{h}}^{(0)}$ defined
in~\citep{LESTRINGANT2020103730} is zero here, as the strain $\tmmathbf{E}$
does not depend on the second gradient of the macroscopic strain
$\tmmathbf{h}''$ in the theory of elasticity of bulk materials.

In equation~(\ref{eq:ACB-operators}), the notation $-\tmmathbf{h}^{\dag} \cdot
\nabla \tmmathbf{C}_{\tmmathbf{h}}^{(1)} \cdot \tmmathbf{Z}$ in the right-hand
side of $\mathcal{B}^{\tmmathbf{h}} (\tmmathbf{h}^{\dag}, \tmmathbf{Z})$
refers to the quantity obtained by integrating by parts
$\tmmathbf{C}_{\tmmathbf{h} (S)}^{(1)} \cdot \tmmathbf{z}' (S)$ with respect
to the variable $S$,
\begin{equation}
  -\tmmathbf{h}^{\dag} \cdot \nabla \tmmathbf{C}^{(1)}_{\tmmathbf{h}} \cdot
  \tmmathbf{Z}= - \left( \frac{\mathd
  \tmmathbf{C}_{\tmmathbf{h}}^{(1)}}{\mathd \tmmathbf{h}} \cdot
  \tmmathbf{h}^{\dag} \right) \cdot \tmmathbf{Z}.
  \label{eq:minus-grad-C1h-abstract}
\end{equation}

As shown in our previous work, the optimal displacement
$\tmmathbf{Z}_{\text{opt}}^{\tmmathbf{h}} (\tmmathbf{h}^{\dag})$ is a
stationary point of $\mathcal{B}^{\tmmathbf{h}} (\tmmathbf{h}^{\dag},
\tmmathbf{Z})$ over all $\tmmathbf{Z}$'s satisfying the kinematic constraint,
$\tmmathbf{q} (\tmmathbf{Z}) =\tmmathbf{0}$, for fixed $\tmmathbf{h}^{\dag}$.
It is therefore the solution of the variational problem
\begin{equation}
  \text{$(\tmmathbf{Z}, \tmmathbf{f}) = \left(
  \tmmathbf{Z}_{\text{opt}}^{\tmmathbf{h}} (\tmmathbf{h}^{\dag}),
  \tmmathbf{f}_{\text{opt}} (\tmmathbf{h}^{\dag}) \right) $ is the solution
  of } \left\{ \begin{array}{l}
    \tmmathbf{q} (\tmmathbf{Z}) =\tmmathbf{0}\\
    \forall \hat{\tmmathbf{Z}} \quad \frac{\partial
    \mathcal{B}^{\tmmathbf{h}}}{\partial \tmmathbf{Z}} (\tmmathbf{h}^{\dag},
    \tmmathbf{Z}) \cdot \hat{\tmmathbf{Z}} -\tmmathbf{f} \cdot \tmmathbf{q}
    (\hat{\tmmathbf{Z}}) = 0
  \end{array} \right. \label{eq:Z-variational-pb-abstract}
\end{equation}
As the operator $\mathcal{B}^{\tmmathbf{h}} (\tmmathbf{h}^{\dag},
\tmmathbf{Z})$ is quadratic with respect to both $\tmmathbf{h}^{\dag}$ and
$\tmmathbf{Z}$, this is a linear problem of elasticity in the cross-section
with residual strain proportional to $\tmmathbf{h}^{\dag}$.

\subsubsection{Equivalent one-dimensional
model}\label{sec-app:elimination-of-boundary-terms}

The microscopic solution in equation~(\ref{eq:y-expansion}) is now available
as
\[ \tmmathbf{y}^{\star} [\tmmathbf{h}] (S, \tmmathbf{T})
   =\tmmathbf{Y}^{\tmmathbf{h} (S)} (\tmmathbf{T})
   +\tmmathbf{Z}_{\text{opt}}^{\tmmathbf{h} (S)} (\tmmathbf{h}' (S),
   \tmmathbf{T}) + \ldots \]
where $\tmmathbf{Y}^{\tmmathbf{h}}$ is the catalog of homogeneous solutions,
$\tmmathbf{Z}_{\text{opt}}^{\tmmathbf{h}} (\tmmathbf{h}^{\dag})$ is the
corrective displacement just found, and the ellipsis stands for higher order
terms that do not enter in the approximation $\Phi_{(2)}^{\star}
[\tmmathbf{h}]$.

The one-dimensional strain energy $\Phi_{(2)}^{\star} [\tmmathbf{h}]$ can then
be obtained by inserting the expansion for $\tmmathbf{y}^{\star}
[\tmmathbf{h}] (S, \tmmathbf{T})$ above into $\Phi^{\star} [\tmmathbf{h}] =
\Phi [\tmmathbf{h}, \tmmathbf{y}^{\star} [\tmmathbf{h}]]$. As shown in
previous work, the result of this calculation is
\[ \Phi_{(2)}^{\star} [\tmmathbf{h}] = \int_0^{\ell} W_{\text{hom}}
   (\tmmathbf{h} (S)) \mymultiply \mathd S + \int_0^{\ell} \tmmathbf{A}
   (\tmmathbf{h} (S)) \cdot \tmmathbf{h}' (S) \mymultiply \mathd S +
   [\tmmathbf{C} (\tmmathbf{h} (S)) \cdot \tmmathbf{h}' (S)]_0^{\ell} +
   \frac{1}{2} \mymultiply \int_0^{\ell} \tmmathbf{h}' (S) \cdot \tmmathbf{B}
   (\tmmathbf{h} (S)) \cdot \tmmathbf{h}' (S) \mymultiply \mathd S, \]
where $\tmmathbf{A} (\tmmathbf{h})$, $\tmmathbf{B} (\tmmathbf{h})$ and
$\tmmathbf{C} (\tmmathbf{h})$ are the one-dimensional operators
\begin{equation}
  \begin{array}{rcl}
    \tmmathbf{A} (\tmmathbf{h}) \cdot \tmmathbf{h}^{\dag} & = &
    \iint_{\Omega} \tmmathbf{\Sigma}^{\tmmathbf{h}} (\tmmathbf{T})
    \doublecontract \mathcal{E}^{\tmmathbf{h}} (\tmmathbf{T},
    \tmmathbf{h}^{\dag}, \tmmathbf{0}) \mymultiply \mathd A,\\
    \frac{1}{2} \mymultiply \tmmathbf{h}^{\dag} \cdot \tmmathbf{B}
    (\tmmathbf{h}) \cdot \tmmathbf{h}^{\dag} & = & \mathcal{B}^{\tmmathbf{h}}
    \left( \tmmathbf{h}^{\dag}, \tmmathbf{Z}_{\text{opt}}^{\tmmathbf{h}}
    (\tmmathbf{T}, \tmmathbf{h}^{\dag}) \right),\\
    \tmmathbf{C} (\tmmathbf{h}) \cdot \tmmathbf{h}^{\dag} & = &
    \tmmathbf{C}_{\tmmathbf{h}}^{(1)} \cdot
    \tmmathbf{Z}_{\text{opt}}^{\tmmathbf{h}} (\tmmathbf{T},
    \tmmathbf{h}^{\dag}) .
  \end{array} \label{eq:B-C-app}
\end{equation}

We complement these known results with an original method that eliminates the
boundary terms in the expression of $\Phi_{(2)}^{\star} [\tmmathbf{h}]$ above,
as follows. We rewrite
\[ [\tmmathbf{C} (\tmmathbf{h} (S)) \cdot \tmmathbf{h}' (S)]_0^{\ell} =
   \int_0^{\ell} \left[ \tmmathbf{h}' (S) \cdot \frac{\mathd
   \tmmathbf{C}}{\mathd \tmmathbf{h}} (\tmmathbf{h} (S)) \cdot \tmmathbf{h}'
   (S) +\tmmathbf{C} (\tmmathbf{h} (S)) \cdot \tmmathbf{h}'' (S) \right]
   \mymultiply \mathd S \]
so that
\begin{equation}
  \Phi_{(2)}^{\star} [\tmmathbf{h}] = \int_0^{\ell} \left[ \left\{
  W_{\text{hom}} (\tmmathbf{h} (S)) +\tmmathbf{C} (\tmmathbf{h} (S)) \cdot
  \tmmathbf{h}'' (S) \right\} +\tmmathbf{A} \left( \tmmathbf{h} \left( S
  \right) \right) \cdot \tmmathbf{h}' (S) + \frac{1}{2} \mymultiply
  \tmmathbf{h}' (S) \cdot \tmmathbf{D} (\tmmathbf{h} (S)) \cdot \tmmathbf{h}'
  (S) \right] \mymultiply \mathd S \label{eq:phi-gr-with-boundary-terms}
\end{equation}
where $\tmmathbf{D} (\tmmathbf{h}) =\tmmathbf{B} (\tmmathbf{h}) + 2
\mymultiply \frac{\mathd \tmmathbf{C}}{\mathd \tmmathbf{h}} (\tmmathbf{h})$ as
announced in equation~(\ref{eq:D-of-h}). The terms in curly braces can be
rewritten as
\[ \begin{array}{lll}
    W_{\text{hom}} (\tmmathbf{h}) +\tmmathbf{C} (\tmmathbf{h}) \cdot
    \tmmathbf{h}'' & = & W_{\text{hom}} (h_0, h_1, \ldots, h_{n - 1}) +
    \sum_{i = 0}^{n - 1} C_i (\tmmathbf{h}) \mymultiply h_i'' (S)\\
    & = & W_{\text{hom}} \left( h_0 + \frac{C_0 (\tmmathbf{h})}{\frac{\mathd
    W_{\text{hom}}}{\mathd h_0} (\tmmathbf{h})} \mymultiply h_0'', \ldots,
    h_{n - 1} + \frac{C_{n - 1} (\tmmathbf{h})}{\frac{\mathd
    W_{\text{hom}}}{\mathd h_{n - 1}} (\tmmathbf{h})} \mymultiply h_{n - 1}''
    \right) +\mathcal{O} (\zeta^4),
  \end{array} \label{eq:Ci-in-practice} \]
where $n = 4$ is the number of macroscopic strain measures in the vector
$\tmmathbf{h}$. The term $\mathcal{O} (\zeta^4)$ is beyond the order $\zeta^2$
of approximation which $\Phi_{(2)}^{\star} [\tmmathbf{h}]$ can resolve, and it
will be ignored. Inserting the above result into the
expression~(\ref{eq:phi-gr-with-boundary-terms}) of $\Phi_{(2)}^{\star}
[\tmmathbf{h}]$, we obtain the expression of $\Phi_{(2)}^{\star}
[\tmmathbf{h}]$ announced in equation~(\ref{eq:phi-gr}). In addition, the
modified strain is identified as $\tilde{h}_i (S) = h_i (S) + \xi_i
(\tmmathbf{h} (S)) \mymultiply h_i'' (S)$ (no implicit sum on $i$) where
\begin{equation}
  \xi_i (\tmmathbf{h}) = \frac{C_i (\tmmathbf{h})}{\frac{\partial
  W_{\text{hom}}}{\partial h_i} (\tmmathbf{h})},
\end{equation}
as announced in equation~(\ref{eq:hi-tilde}).

\section{Calculation of the structure
operators}\label{app:structure-operators}

We derive the structure operators $\tmmathbf{e}_{j \nocomma k}^i
(\tmmathbf{T}, \tmmathbf{h})$ by identifying the two sides of
equation~(\ref{eq:expansion-for-structure-operators}). To do so, we expand
$\tmmathbf{E} (\tmmathbf{T}; \tmmathbf{h}; \tmmathbf{Y}^{\tmmathbf{h}}
+\tmmathbf{Z}, \tmmathbf{h}^{\dag} \cdot \nabla \tmmathbf{Y}^{\tmmathbf{h}}
+\tmmathbf{Z}^{\dag})$ in powers of $\tmmathbf{h}^{\dag}$, $\tmmathbf{Z}$ and
$\tmmathbf{Z}^{\dag}$. We start by setting
$\tmmathbf{Y}=\tmmathbf{Y}^{\tmmathbf{h}} +\tmmathbf{Z}$ and
$\tmmathbf{Y}^{\dag} =\tmmathbf{h}^{\dag} \cdot \nabla
\tmmathbf{Y}^{\tmmathbf{h}} +\tmmathbf{Z}^{\dag}$ in the auxiliary quantity
$t_i$ in~(\ref{eq:E-function}),
\[ \begin{array}{lcl}
     t_i & = & (1 + \varepsilon) \mymultiply \delta_{i \nocomma 3} + \eta_{i
     \nocomma j \nocomma k} \mymultiply \kappa_j \mymultiply
     Y_k^{\tmmathbf{h}} (\tmmathbf{T}) +\tmmathbf{h}^{\dag} \cdot \nabla
     Y^{\tmmathbf{h}}_i (\tmmathbf{T}) + \eta_{i \nocomma j \nocomma k}
     \mymultiply \kappa_j \mymultiply Z_k (\tmmathbf{T}) + Z_i^{\dag}
     (\tmmathbf{T})\\
     & = & F^{\tmmathbf{h}}_{i \nocomma 3} (\tmmathbf{T})
     +\tmmathbf{h}^{\dag} \cdot \nabla Y^{\tmmathbf{h}}_i (\tmmathbf{T}) +
     \eta_{i \nocomma j \nocomma k} \mymultiply \kappa_j \mymultiply Z_k
     (\tmmathbf{T}) + Z_i^{\dag} (\tmmathbf{T})
   \end{array} \]
where $F^{\tmmathbf{h}}_{i \nocomma 3} (\tmmathbf{T})$ has been defined
earlier in~(\ref{eq:gr-effect-homogeneous-qties}).

Inserting this into the expression of $\tmmathbf{E} (\tmmathbf{T};
\tmmathbf{h}; \tmmathbf{Y}, \tmmathbf{Y}^{\dag})$ in~(\ref{eq:E-function}), we
obtain
\[ \begin{array}{l}
     \tmmathbf{E} (\tmmathbf{T}; \tmmathbf{h}; \tmmathbf{Y}^{\tmmathbf{h}}
     +\tmmathbf{Z}, \tmmathbf{h}^{\dag} \cdot \nabla
     \tmmathbf{Y}^{\tmmathbf{h}} +\tmmathbf{Z}^{\dag}) =\\
     \begin{array}{l}
       \frac{1}{2} \mymultiply \left( - 1 + \left[ F^{\tmmathbf{h}}_{i
       \nocomma 3} (\tmmathbf{T}) +\tmmathbf{h}^{\dag} \cdot \nabla
       Y^{\tmmathbf{h}}_i (\tmmathbf{T}) + \eta_{i \nocomma j \nocomma k}
       \mymultiply \kappa_j \mymultiply Z_k (\tmmathbf{T}) + Z_i^{\dag}
       (\tmmathbf{T}) \right]^2 \right) \mymultiply \tmmathbf{e}_3 \otimes
       \tmmathbf{e}_3 \ldots\\
       \nobracket \nobracket \qquad + \left[ F^{\tmmathbf{h}}_{i \nocomma 3}
       (\tmmathbf{T}) +\tmmathbf{h}^{\dag} \cdot \nabla Y^{\tmmathbf{h}}_i
       (\tmmathbf{T}) + \eta_{i \nocomma j \nocomma k} \mymultiply \kappa_j
       \mymultiply Z_k (\tmmathbf{T}) + Z_i^{\dag} (\tmmathbf{T}) \right]
       \mymultiply [F^{\tmmathbf{h}}_{i \nocomma \alpha} (\tmmathbf{T}) +
       \partial_{\alpha} Z_i (\tmmathbf{T})] \mymultiply \tmmathbf{e}_{\alpha}
       \odot \tmmathbf{e}_3 \ldots\\
       \nobracket \nobracket \qquad + \frac{[F^{\tmmathbf{h}}_{i \nocomma
       \alpha} (\tmmathbf{T}) + \partial_{\alpha} Z_i (\tmmathbf{T})]
       \mymultiply [F^{\tmmathbf{h}}_{i \nocomma \beta} (\tmmathbf{T}) +
       \partial_{\beta} Z_i (\tmmathbf{T})] - \delta_{\alpha \nocomma
       \beta}}{2} \mymultiply \tmmathbf{e}_{\alpha} \otimes
       \tmmathbf{e}_{\beta}
     \end{array}
   \end{array} \]
Expanding and identifying with~(\ref{eq:expansion-for-structure-operators}),
we obtain the explicit expression of the operators as
\begin{equation}
  \begin{array}{ccl}
    \tmmathbf{e}_{0 \nocomma 0}^1 (\tmmathbf{T}, \tmmathbf{h}) \cdot
    \tmmathbf{h}^{\dag} & = & (\tmmathbf{h}^{\dag} \cdot \nabla
    Y^{\tmmathbf{h}}_i (\tmmathbf{T})) \mymultiply F^{\tmmathbf{h}}_{i
    \nocomma j} (\tmmathbf{T}) \mymultiply \tmmathbf{e}_j \odot
    \tmmathbf{e}_3\\
    \tmmathbf{e}_{1 \nocomma 0}^0 (\tmmathbf{T}, \tmmathbf{h}) \cdot
    \tmmathbf{Z} & = & \eta_{i \nocomma j \nocomma k} \mymultiply \kappa_k
    \mymultiply F^{\tmmathbf{h}}_{j \nocomma l} (\tmmathbf{T}) \mymultiply Z_i
    (\tmmathbf{T}) \mymultiply \tmmathbf{e}_l \odot \tmmathbf{e}_3 +
    F^{\tmmathbf{h}}_{i \nocomma j} (\tmmathbf{T}) \mymultiply
    \partial_{\alpha} Z_i (\tmmathbf{T}) \tmmathbf{e}_j \odot
    \tmmathbf{e}_{\alpha}\\
    \tmmathbf{e}_{0 \nocomma 1}^0 (\tmmathbf{T}, \tmmathbf{h}) \cdot
    \tmmathbf{Z}^{\dag} & = & F^{\tmmathbf{h}}_{i \nocomma j} (\tmmathbf{T})
    \mymultiply Z_i^{\dag} (\tmmathbf{T}) \mymultiply \tmmathbf{e}_j \odot
    \tmmathbf{e}_3\\
    \frac{1}{2} \mymultiply \tmmathbf{h}^{\dag} \cdot \tmmathbf{e}_{0 \nocomma
    0}^2 (\tmmathbf{T}, \tmmathbf{h}) \cdot \tmmathbf{h}^{\dag} & = &
    \frac{1}{2} \mymultiply (\tmmathbf{h}^{\dag} \cdot \nabla
    Y^{\tmmathbf{h}}_i (\tmmathbf{T}))^2 \mymultiply \tmmathbf{e}_3 \otimes
    \tmmathbf{e}_3\\
    \tmmathbf{h}^{\dag} \cdot \tmmathbf{e}_{1 \nocomma 0}^1 (\tmmathbf{T},
    \tmmathbf{h}) \cdot \tmmathbf{Z} & = & \tmmathbf{h}^{\dag} \cdot \nabla
    Y^{\tmmathbf{h}}_i (\tmmathbf{T}) \mymultiply \left( \eta_{i \nocomma j
    \nocomma k} \mymultiply \kappa_j \mymultiply Z_k (\tmmathbf{T})
    \mymultiply \tmmathbf{e}_3 + \partial_{\beta} Z_i (\tmmathbf{T})
    \mymultiply \tmmathbf{e}_{\beta} \right) \odot \tmmathbf{e}_3\\
    \frac{1}{2} \mymultiply \tmmathbf{Z} \cdot \tmmathbf{e}_{2 \nocomma 0}^0
    (\tmmathbf{T}, \tmmathbf{h}) \cdot \tmmathbf{Z} & = & \frac{1}{2}
    \mymultiply \left( \left( \delta_{i \nocomma j} \mymultiply \kappa_l
    \mymultiply \kappa_l - \kappa_i \mymultiply \kappa_j \right) \mymultiply
    Z_i (\tmmathbf{T}) \mymultiply Z_j (\tmmathbf{T}) \mymultiply
    \tmmathbf{e}_3 \otimes \tmmathbf{e}_3 \ldots \right.\\
    &  & \nobracket \nobracket \hspace{3em} \left. + 2 \mymultiply \eta_{i
    \nocomma j \nocomma k} \mymultiply \kappa_j \mymultiply Z_k (\tmmathbf{T})
    \mymultiply \partial_{\alpha} Z_i (\tmmathbf{T}) \mymultiply
    \tmmathbf{e}_{\alpha} \odot \tmmathbf{e}_3 + \partial_{\alpha} Z_i
    (\tmmathbf{T}) \mymultiply \partial_{\beta} Z_i (\tmmathbf{T}) \mymultiply
    \tmmathbf{e}_{\alpha} \odot \tmmathbf{e}_{\beta} \right)
  \end{array} \label{eq:structure-operators}
\end{equation}
\section{Gradient effect for a twisted bar}\label{app:twisting}

\subsection{Derivation of the higher-model for a twisted bar}

In this section, we apply the general method from
section~\ref{sec:gradient-effect} to derive the higher-order model
$\Phi_{(2)}^{\star} [\tau]$ for a twisted bar.

In view of the strain definition in equation~(\ref{eq:E-twist-linear}) and of
homogeneous solution in equation~(\ref{eq:twisting-homogeneous-sol}), the
strain reads
\begin{equation}
  \tmmathbf{E} \left( \tmmathbf{T}; \tau ; \tmmathbf{V}^{(\tau)}
  +\tmmathbf{Z}, \tau^{\dag} \mymultiply \nabla \tmmathbf{V}^{(\tau)}
  +\tmmathbf{Z}^{\dag} \right) =\tmmathbf{E}^{(\tau)} (\tmmathbf{T}) +
  \partial_{\alpha} Z_i (\tmmathbf{T}) \mymultiply \tmmathbf{e}_{\alpha} \odot
  \tmmathbf{e}_i + \left( \tau^{\dag} \mymultiply \nabla V_i^{(\tau)}
  (\tmmathbf{T}) + Z_i^{\dag} (\tmmathbf{T}) \right) \mymultiply
  \tmmathbf{e}_i \odot \tmmathbf{e}_3 \label{eq:twist-E-expansion}
\end{equation}
where $\nabla V_i^{(\tau)} (\tmmathbf{T}) = \frac{\mathd V_i^{(\tau)}
(\tmmathbf{T})}{\mathd \tau}$ is the gradient of the homogeneous solution with
respect to the homogeneous strain $\tau$,
\[ \begin{array}{ll}
     \nabla V_{\alpha}^{(\tau)} (\tmmathbf{T}) = 0,\qquad & \nabla V_3^{(\tau)}
     (\tmmathbf{T}) = \omega (\tmmathbf{T}) .
   \end{array} \]
By identifying equation~(\ref{eq:twist-E-expansion}) with the generic
expansion~(\ref{eq:expansion-for-structure-operators}), we obtain the
structure operators as
\begin{equation}
  \begin{array}{rcl}
    \tmmathbf{e}_{0 \nocomma 0}^1 (\tmmathbf{T}, \tau) \cdot (\tau^{\dag}) & =
    & \tau^{\dag} \mymultiply \omega (\tmmathbf{T}) \mymultiply \tmmathbf{e}_3
    \otimes \tmmathbf{e}_3\\
    \tmmathbf{e}_{1 \nocomma 0}^0 (\tmmathbf{T}, \tau) \cdot \tmmathbf{Z} & =
    & \partial_{\alpha} Z_i (\tmmathbf{T}) \mymultiply \tmmathbf{e}_{\alpha}
    \odot \tmmathbf{e}_i\\
    \tmmathbf{e}_{0 \nocomma 1}^0 (\tmmathbf{T}, \tau) \cdot
    \tmmathbf{Z}^{\dag} & = & Z_i^{\dag} (\tmmathbf{T}) \mymultiply
    \tmmathbf{e}_i \odot \tmmathbf{e}_3\\
    \frac{1}{2} \mymultiply (\tau^{\dag}) \cdot \tmmathbf{e}_{0 \nocomma 0}^2
    (\tmmathbf{T}, \tau) \cdot (\tau^{\dag}) & = & \tmmathbf{0}\\
    (\tau^{\dag}) \cdot \tmmathbf{e}_{1 \nocomma 0}^1 (\tmmathbf{T}, \tau)
    \cdot \tmmathbf{Z} & = & \tmmathbf{0}\\
    \frac{1}{2} \mymultiply \tmmathbf{Z} \cdot \tmmathbf{e}_{2 \nocomma 0}^0
    (\tmmathbf{T}, \tau) \cdot \tmmathbf{Z} & = & \tmmathbf{0}
  \end{array} \label{eq:twist-operators}
\end{equation}
This is a special form of the non-linear expressions derived in
appendix~\ref{app:structure-operators}, relevant to linear elasticity.

The operators defined in equations~(\ref{eq:Lh}) and~(\ref{eq:A-C1h}) read,
respectively,

\[ \begin{array}{lll}
     \mathcal{E}^{(\tau)} (\tmmathbf{T}, \tau^{\dag}, \tmmathbf{Z}) & = &
     \tmmathbf{e}_{0 \nocomma 0}^1 (\tmmathbf{T}, \tau) \cdot (\tau^{\dag})
     +\tmmathbf{e}_{1 \nocomma 0}^0 (\tmmathbf{T}, \tau) \cdot \tmmathbf{Z}\\
     & = & \tau^{\dag} \mymultiply \omega (\tmmathbf{T}) \mymultiply
     \tmmathbf{e}_3 \otimes \tmmathbf{e}_3 + \partial_{\alpha} Z_i
     (\tmmathbf{T}) \mymultiply \tmmathbf{e}_{\alpha} \odot \tmmathbf{e}_i
   \end{array} \]
and
\[ \begin{array}{lll}
     \tmmathbf{C}_{(\tau)}^{(1)} \cdot \tmmathbf{Z}^{\dag} & = &
     \iint_{\Omega} \tmmathbf{\Sigma}^{(\tau)} (\tmmathbf{T})
     \doublecontract (\tmmathbf{e}^0_{0 \nocomma 1} (\tmmathbf{T}, \tau) \cdot
     \tmmathbf{Z}^{\dag}) \mymultiply \mathd A\\
     & = & \iint_{\Omega} 2 \mymultiply \mu \mymultiply \tau \mymultiply
     \left( - \eta_{\alpha \nocomma \beta} \mymultiply T_{\beta} +
     \partial_{\alpha} \omega (\tmmathbf{T}) \right) \mymultiply
     (\tmmathbf{e}_{\alpha} \otimes \tmmathbf{e}_3) \doublecontract \left(
     Z_{\beta}^{\dag} (\tmmathbf{T}) \mymultiply \tmmathbf{e}_{\beta} \odot
     \tmmathbf{e}_3 \right) \mymultiply \mathd A\\
     & = & \mu \mymultiply \tau \mymultiply \iint_{\Omega} \left( -
     \eta_{\alpha \nocomma \beta} \mymultiply T_{\beta} + \partial_{\alpha}
     \omega (\tmmathbf{T}) \right) \mymultiply Z_{\alpha}^{\dag}
     (\tmmathbf{T}) \mymultiply \mathd A.
   \end{array} \]
The operator $- \tau^{\dag} \mymultiply \nabla
\tmmathbf{C}_{\tmmathbf{h}}^{(1)} \cdot \tmmathbf{Z}$ is obtained by an
integration by parts, see equation~(\ref{eq:minus-grad-C1h}),
\[ - \tau^{\dag} \mymultiply \nabla \tmmathbf{C}_{\tmmathbf{h}}^{(1)} \cdot
   \tmmathbf{Z}= - \mu \mymultiply \tau^{\dag} \mymultiply \iint_{\Omega}
   \left( - \eta_{\alpha \nocomma \beta} \mymultiply T_{\beta} +
   \partial_{\alpha} \omega \right) \mymultiply Z_{\alpha} \mymultiply \mathd
   A. \]

We proceed to calculate the operator $\mathcal{B}^{(\tau)} (\tau^{\dag},
\tmmathbf{Z})$ from equation~(\ref{eq:ACB-operators}). Dropping the operators
$\tmmathbf{e}^i_{j \nocomma k}$ with $i + j + k \geqslant 2$ which are zero in
the linear setting, see equation~(\ref{eq:twist-operators}), we have
\[ \mathcal{B}^{(\tau)} (\tau^{\dag}, \tmmathbf{Z}) = \iint_{\Omega}
   \frac{1}{2} \mymultiply \mathcal{E}^{(\tau)} (\tmmathbf{T}, \tau^{\dag},
   \tmmathbf{Z}) \doublecontract \tmmathbf{K}^{(\tau)} (\tmmathbf{T})
   \doublecontract \mathcal{E}^{(\tau)} (\tmmathbf{T}, \tau^{\dag},
   \tmmathbf{Z}) \mymultiply \mathd A - \tau^{\dag} \mymultiply \nabla
   \tmmathbf{C}_{\tmmathbf{h}}^{(1)} \cdot \tmmathbf{Z}. \]
In linear elasticity, $\frac{1}{2} \mymultiply \mathcal{E} \doublecontract
\tmmathbf{K}^{(\tau)} (\tmmathbf{T}) \doublecontract \mathcal{E}= w
(\tmmathbf{T}, \mathcal{E})$ is the (quadratic) strain energy density, as
defined in equation~(\ref{eq:linear-w}). This yields
\[ \begin{array}{lll}
     \mathcal{B}^{(\tau)} (\tau^{\dag}, \tmmathbf{Z}) & = & \frac{1}{2}
     \mymultiply \iint_{\Omega} \left( \lambda \mymultiply \tmop{tr}^2
     \mathcal{E}^{(\tau)} (\tmmathbf{T}, \tau^{\dag}, \tmmathbf{Z}) + 2
     \mymultiply \mu \mymultiply \mathcal{E}^{(\tau)} (\tmmathbf{T},
     \tau^{\dag}, \tmmathbf{Z}) \doublecontract \mathcal{E}^{(\tau)}
     (\tmmathbf{T}, \tau^{\dag}, \tmmathbf{Z}) \right) \mymultiply \mathd A
     \ldots\\
     &  & \nobracket \nobracket \nobracket \nobracket \hspace{8.8em} - \mu
     \mymultiply \tau^{\dag} \mymultiply \iint_{\Omega} \left( -
     \eta_{\alpha \nocomma \beta} \mymultiply T_{\beta} + \partial_{\alpha}
     \omega (\tmmathbf{T}) \right) \mymultiply Z_{\alpha} (\tmmathbf{T})
     \mymultiply \mathd A\\
     & = & \frac{1}{2} \mymultiply \iint_{\Omega} \left( \lambda
     \mymultiply \left( \tau^{\dag} \mymultiply \omega + \partial_{\alpha}
     Z_{\alpha} \right)^2 + 2 \mymultiply \mu \mymultiply \left( \tau^{\dag 2}
     \mymultiply \omega^2 + \sum_{\alpha} \frac{(\partial_{\alpha} Z_3)^2}{2}
     + \sum_{\alpha \nocomma \beta} \left( \frac{\partial_{\alpha} Z_{\beta} +
     \partial_{\beta} Z_{\alpha}}{2} \right)^2 \right) \right) \mymultiply
     \mathd A \ldots\\
     &  & \nobracket \nobracket \nobracket \nobracket \hspace{9em} - \mu
     \mymultiply \tau^{\dag} \mymultiply \iint_{\Omega} \left( -
     \eta_{\alpha \nocomma \beta} \mymultiply T_{\beta} + \partial_{\alpha}
     \omega \right) \mymultiply Z_{\alpha} \mymultiply \mathd A\\
     & = & \frac{1}{2} \mymultiply \iint_{\Omega} \left( \left( 2
     \mymultiply \mu + \lambda \right) \mymultiply \tau^{\dag 2} \mymultiply
     \omega^2 + 2 \mymultiply \tau^{\dag} \mymultiply \left( \lambda
     \mymultiply \omega \mymultiply \partial_{\alpha} Z_{\alpha}
     (\tmmathbf{T}) - \mu \mymultiply \partial_{\alpha} \omega \mymultiply
     Z_{\alpha} + \mu \mymultiply \eta_{\alpha \nocomma \beta} \mymultiply
     T_{\beta} \mymultiply Z_{\alpha} \right) \ldots \right.\\
     &  & \nobracket \nobracket \nobracket \nobracket \hspace{9em} \left. +
     \left\{ \lambda \mymultiply \left( \sum_{\alpha} \partial_{\alpha}
     Z_{\alpha} \right)^2 + \mu \mymultiply \sum_{\alpha} (\partial_{\alpha}
     Z_3)^2 + 2 \mymultiply \mu \mymultiply \sum_{\alpha \nocomma \beta}
     \left( \frac{\partial_{\alpha} Z_{\beta} + \partial_{\beta}
     Z_{\alpha}}{2} \right)^2 \right\} \right) \mymultiply \mathd A
   \end{array} \]

The next step is to find the stationary point $\tmmathbf{Z}$ of
$\mathcal{B}^{(\tau)} (\tau^{\dag}, \tmmathbf{Z})$ subject to the constraint
$\tmmathbf{q} (\tmmathbf{Z}) =\tmmathbf{0}$, see
equation~(\ref{eq:Z-variational-pb}). We can drop the term $\mu \mymultiply
\eta_{\alpha \nocomma \beta} \mymultiply T_{\beta} \mymultiply Z_{\alpha}
(\tmmathbf{T})$ which integrates to zero thanks to the second constraint
in~(\ref{eq:Z-variational-pb}). In addition, we observe that the longitudinal
correction $Z_3$ appears only in the term $\mu \mymultiply \sum_{\alpha}
(\partial_{\alpha} Z_3)^2$: the stationarity condition with respect to $Z_3$
is that $Z_3 (\tmmathbf{T})$ is constant, this constant being zero by the
first constraint in~(\ref{eq:Z-variational-pb}),
\begin{equation}
  Z_3^{\text{opt}} (\tmmathbf{T}) = 0. \label{eq:twist-Zopt-3}
\end{equation}
We are left with
\[ \begin{array}{l}
     \mathcal{B}^{(\tau)} (\tau^{\dag}, Z_{\alpha}) = \frac{2 \mymultiply \mu
     + \lambda}{2} \mymultiply J_{\omega} \mymultiply \tau^{\dag 2} \ldots\\
     \nobracket \nobracket + \frac{1}{2} \mymultiply \iint_{\Omega} \left(
     \left\{ \lambda \mymultiply \left( \sum_{\alpha} \partial_{\alpha}
     Z_{\alpha} (\tmmathbf{T}) \right)^2 + 2 \mymultiply \mu \mymultiply
     \sum_{\alpha \nocomma \beta} \left( \frac{\partial_{\alpha} Z_{\beta} +
     \partial_{\beta} Z_{\alpha}}{2} \right)^2 \right\} + 2 \mymultiply
     \tau^{\dag} \mymultiply \left( \lambda \mymultiply \omega (\tmmathbf{T})
     \mymultiply \partial_{\alpha} Z_{\alpha} (\tmmathbf{T}) - \mu \mymultiply
     \partial_{\alpha} \omega (\tmmathbf{T}) \mymultiply Z_{\alpha}
     (\tmmathbf{T}) \right) \right) \mymultiply \mathd A
   \end{array} \]
where the curly bracket is the quadratic strain energy operator in
two-dimensional elasticity, and $J_{\omega}$ is the warping constant defined
in equation~(\ref{eq:Jw}).

In view of the linearity of the problem with respect to the strain gradient
$\tau^{\dag}$, the solution $Z_{\alpha} (\tmmathbf{T}) =
Z_{\alpha}^{\text{opt}} (\tau', \tmmathbf{T})$ that renders
$\mathcal{B}^{(\tau)} (\tau^{\dag}, Z_{\alpha})$ stationary subject to the
conditions $\tmmathbf{q} (Z_{\alpha}) =\tmmathbf{0}$ is of the form
\begin{equation}
  Z_{\alpha}^{\text{opt}} (\tau^{\dag}, \tmmathbf{T}) = \tau^{\dag}
  \mymultiply u_{\alpha} (\tmmathbf{T}) . \label{eq:twist:Zopt-alpha}
\end{equation}
This leads to the variational
problem~(\ref{eq:twisting-psi-alpha-variational-pb}--\ref{eq:twist-corrective-displ-cstr})
stated in the main text.

Inserting the optimal displacement given by~(\ref{eq:twist-Zopt-3})
and~(\ref{eq:twist:Zopt-alpha}) into the expression of $\mathcal{B}^{(\tau)}$,
we find from equation~(\ref{eq:B-C})
\[ \begin{array}{lll}
     \frac{1}{2} \mymultiply B (\tau) \mymultiply \tau^{\dag 2} & = &
     \mathcal{B}^{(\tau)} \left( \tau^{\dag},
     \tmmathbf{Z}_{\text{opt}}^{(\tau)} (\tmmathbf{T}, \tau^{\dag}) \right)\\
     & = & \frac{2 \mymultiply \mu + \lambda}{2} \mymultiply J_{\omega}
     \mymultiply \tau^{\dag 2} + \frac{\tau^{\dag 2}}{2} \mymultiply \left(
     \iint_{\Omega} \left\{ \lambda \mymultiply \left( \sum_{\alpha}
     \partial_{\alpha} u_{\alpha} \right)^2 + 2 \mymultiply \mu \mymultiply
     \sum_{\alpha \nocomma \beta} \left( \frac{\partial_{\alpha} u_{\beta} +
     \partial_{\beta} u_{\alpha}}{2} \right)^2 \right\} \mymultiply \mathd A +
     2 \mymultiply (D_{\lambda} - D_{\mu}) \right)
   \end{array} \]
where $D_{\lambda}$ and $D_{\mu}$ are the quantities defined in
equation~(\ref{eq:twist-D-sub-x}).

This expression can be simplified using an identity for $u_{\alpha}$ found by
inserting $\hat{u}_{\alpha} = u_{\alpha}$
into~(\ref{eq:twisting-psi-alpha-variational-pb}), namely
\[ \iint_{\Omega} \left\{ \lambda \mymultiply \left( \sum_{\alpha}
   \partial_{\alpha} u_{\alpha} \right)^2 + 2 \mymultiply \mu \mymultiply
   \sum_{\alpha \nocomma \beta} (\partial_{\alpha} u_{\beta} +
   \partial_{\beta} u_{\alpha})^2 \right\} \mymultiply \mathd A = -
   (D_{\lambda} - D_{\mu}), \]
which yields
\[ B (\tau) = D_{\omega} - D_{\mu}, \]
where $D_{\omega}$ is the quantity defined in~(\ref{eq:twist-D-sub-x}).

The second-gradient modulus $B (\tau)$ appears to be independent of $\tau$,
which is always the case in the framework of linearized elasticity.

In addition, we have from equation~(\ref{eq:B-C})
\[ \begin{array}{lll}
     C (\tau) \mymultiply \tau^{\dag} & = & \tmmathbf{C}_{(\tau)}^{(1)} \cdot
     \tmmathbf{Z}_{\text{opt}}^{(\tau)} (\tmmathbf{T}, \tau^{\dag})\\
     & = & \mu \mymultiply \tau \mymultiply \iint_{\Omega} \left( -
     \eta_{\alpha \nocomma \beta} \mymultiply T_{\beta} + \partial_{\alpha}
     \omega \right) \mymultiply Z^{\tmop{opt}}_{\alpha} \mymultiply \mathd A\\
     & = & \mu \mymultiply \tau \mymultiply \mymultiply \tau^{\dag}
     \mymultiply \iint_{\Omega} \left( - \eta_{\alpha \nocomma \beta}
     \mymultiply T_{\beta} + \partial_{\alpha} \omega \right) \mymultiply
     u_{\alpha} \mymultiply \mathd A\\
     & = & \mu \mymultiply \tau \mymultiply \mymultiply \tau^{\dag}
     \mymultiply \iint_{\Omega} \partial_{\alpha} \omega \mymultiply
     u_{\alpha} \mymultiply \mathd A\\
     & = & D_{\mu} \mymultiply \tau \mymultiply \tau^{\dag}
   \end{array} \]
and from equation~(\ref{eq:D-of-h}), and using the expression for the
homogeneous stress $\tmmathbf{\Sigma}^{(\tau)}$ in
(\ref{eq:torsion-homogeneousStress}),
\[ \begin{array}{lll}
     \tmmathbf{A} (\tau) \cdot (\tau^{\dag}) & = & \iint_{\Omega}
     \tmmathbf{\Sigma}^{(\tau)} (\tmmathbf{T}) \doublecontract
     \mathcal{E}^{(\tau)} (\tmmathbf{T}, \tau^{\dag}, \tmmathbf{0})
     \mymultiply \mathd A\\
     & = & \iint_{\Omega} \tau^{\dag} \mymultiply \omega (\tmmathbf{T})
     \mymultiply \tmmathbf{\Sigma}^{(\tau)} (\tmmathbf{T}) \doublecontract
     \tmmathbf{e}_3 \otimes \tmmathbf{e}_3 \mymultiply \mathd A\\
     & = & 0
   \end{array} \]

Using equation~(\ref{eq:hi-tilde}), we obtain
\[ D (\tau) = B (\tau) + 2 \mymultiply \frac{\mathd C (\tau)}{\mathd \tau} =
   (D_{\omega} - D_{\mu}) + 2 \mymultiply D_{\mu} = D_{\omega} + D_{\mu} . \]
In view of equation~(\ref{eq:hi-tilde}), the modified strain measure reads
$\hat{\tau} = \tau + \xi_0 (\tau) \mymultiply \frac{\mathd^2 \tau}{\mathd
S^2}$ where
\[ \xi_0 (\tau) = \frac{C (\tau)}{\frac{\mathd W_{\text{hom}}}{\mathd \tau}} =
   \frac{D_{\mu} \mymultiply \tau}{\mu \mymultiply J \mymultiply \tau} =
   \frac{D_{\mu}}{\mu \mymultiply J} . \]
Inserting into~(\ref{eq:phi-gr}), we arrive at the expression of
$\Phi_{(2)}^{\star} [\tmmathbf{h}]$ announced in
equation~(\ref{eq:twisting-final-phi}).

\subsection{Case of an elliptical cross-section}\label{app-twist-elliptical}

We consider the case of an elliptical cross-section with semi-minor and major
axes $a$ and $b$, in any order. The elliptical cross-section satisfies the
symmetry assumptions listed at the beginning of section~\ref{sec:twisting}.

Analytical solutions for the various quantities defined in
section~\ref{sec:twisting} can be obtained as follows. First, the warping
function is found as $\omega (\tmmathbf{T}) = \frac{b^2 - a^2}{a^2 + b^2}
\mymultiply T_1 \mymultiply T_2$, which yields the expressions of $J$ and
$J_{\omega}$ announced in equation~(\ref{eq:elliptical-X-section}). The
corrective displacement is found by solving the variational problem in
equation~(\ref{eq:twisting-psi-alpha-variational-pb}--\ref{eq:twist-corrective-displ-cstr})
as
\[ \begin{array}{rll}
     u_1 (\tmmathbf{T}) & = & \frac{1}{24} \mymultiply \frac{b^2 - a^2}{a^2 +
     b^2} \mymultiply T_2 \mymultiply \left( 4 \mymultiply a^2 \mymultiply
     \frac{\left( 5 \mymultiply a^2 \mymultiply b^2 + b^4 - 2 \mymultiply (a^2
     + b^2) \mymultiply T_2^2 \right)}{(a^2 + b^2)^2} + \frac{\lambda}{\mu +
     \lambda} \mymultiply \left( a^2 - b^2 - 6 \mymultiply T_1^2 + 2
     \mymultiply T_2^2 \right) \right)\\
     u_2 (\tmmathbf{T}) & = & \frac{1}{24} \mymultiply \frac{b^2 - a^2}{a^2 +
     b^2} \mymultiply T_1 \mymultiply \left( 4 \mymultiply b^2 \mymultiply
     \frac{\left( a^4 + 5 \mymultiply a^2 \mymultiply b^2 - 2 \mymultiply (a^2
     + b^2) \mymultiply T_1^2 \right)}{(a^2 + b^2)^2} + \frac{\lambda}{\mu +
     \lambda} \mymultiply \left( b^2 - a^2 + 2 \mymultiply T_1^2 - 6
     \mymultiply T_2^2 \right) \right)\\
     D_{\mu} & = & 8 \mymultiply \mu \mymultiply \left( \frac{a \mymultiply
     b}{a^2 + b^2} \right)^2 \mymultiply J_{\omega}\\
     D_{\omega} & = & Y \mymultiply J_{\omega},
   \end{array} \]
where $Y$ is the Young modulus, see
equation~(\ref{eq:twisting-Young-modulus}). The Lagrange multipliers are found
as $F_{\alpha} = 0$ and $Q = \left( \frac{b^2 - a^2}{a^2 + b^2} \right)^2
\mymultiply \mu$.

The quantities defined in equation~(\ref{eq:twist-D-sub-x}) are obtained as
$D_{\lambda} = - \frac{\lambda^2}{\lambda + \mu} \mymultiply J_{\omega}$, and
as stated in equation~(\ref{eq:elliptical-X-section}).

\section{higher-order model for a twisting bar: consistency with prior
work}\label{app:Trabucho}

In this appendix, we verify that our
equation~(\ref{eq:twist-equil-eq-with-gradient}) for the twisting of a bar
accounting for the gradient effect is equivalent to the equations derived by
Trabucho and Via{\~n}o~\cite{trabucho1989existence}. Mathematical quantities
defined in Trabucho and Via{\~n}o's paper such as $[\underline{u}_{\alpha}^4]$
are enclosed in square brackets; equation number in square brackets refer to
equations in their paper.

We focus on the case where the external load is of order $\varepsilon^2$,
where $\varepsilon$ is the small aspect-ratio, which they use as an expansion
parameter: in their notation, the external load is represented by the force
$\varepsilon^2 \mymultiply [f_i^2]$ per unit volume in the bulk, and the force
$\varepsilon^2 \mymultiply [g_i^2]$ per unit area applied on the lateral
surface.

Trabucho and Via{\~n}o seek the average rotation of the cross-section with
coordinate $S = [x_3]$ as a power series of the aspect-ratio parameter
$\varepsilon$, $[- v (S)] = - \varepsilon^2 \mymultiply [v^2 (S)] -
\varepsilon^4 \mymultiply [v^4 (S)] - \cdots$ where the minus signs arise
because of their non-standard sign conventions. This corresponds to a
kinematic twist $\tau (S) = - \varepsilon^2 \mymultiply \frac{\mathd
[v^2]}{\mathd S} - \varepsilon^4 \mymultiply \frac{\mathd [v^4]}{\mathd S} -
\cdots$

They obtain the principle of virtual work for the twisting of the bar order by
order: summing up their equation~[3.11] at order $\varepsilon^2$ and their
equation~[5.11] at order $\varepsilon^4$, their principle of virtual work
reads
\begin{equation}
  \forall \hat{\theta} \qquad - \int \mu \mymultiply J \mymultiply \tau (S)
  \mymultiply \hat{\theta}' (S) \mymultiply \mathd S = \int \left( - Y
  \mymultiply J_{\omega} \mymultiply \tau'' (S) + [M_4^3] \right) \mymultiply
  \hat{\theta}' (S) \mymultiply \mathd S - \int m_3 (S) \mymultiply
  \hat{\theta} (S) \mymultiply \mathd S, \label{eq:appTV-Trabucho-equil}
\end{equation}
where $m_3 = \iint_{\Omega} \left( [x_1] \mymultiply [f_2^2] - [x_2]
\mymultiply [f_1^2] \right) \mymultiply \mathd A + \int_{\partial \Omega}
\left( [x_1] \mymultiply [g_2^2] - [x_2] \mymultiply [g_1^2] \right)
\mymultiply \mathd \ell$ is the external twisting moment applied per unit
length, $[M_4^3]$ is the auxiliary quantity defined in~[5.12] as
\begin{equation}
  [M_4^3] = - \nu \mymultiply \frac{\mathd}{\mathd S} \left( \iint_{\Omega}
  2 \mymultiply (\lambda + \mu) \mymultiply \partial_{\alpha}
  [\underline{u}_{\alpha}^4]_{} \mymultiply \omega \mymultiply \mathd A
  \right) - \mu \mymultiply \tau'' (S) \iint_{\Omega} (\partial_2 u_1 -
  \partial_1 u_2) \mymultiply [\Psi] \mymultiply \mathd A,
  \label{eq:appTV-equil-tau-Trabucho}
\end{equation}
and $[\Psi (\tmmathbf{T})]$ is the auxiliary function satisfying,
see~[2.11--12],
\[ \begin{array}{ll}
     \partial_1 [\Psi] = - \partial_2 \omega - T_1 & \partial_2 [\Psi] =
     \partial_1 \omega - T_2
   \end{array} \]
as well as
\[ [\Psi] = 0 \text{ on $\partial \Omega$} . \]
The second integrand in equation~(\ref{eq:appTV-equil-tau-Trabucho}) is a
rewriting of that given by Trabucho and Via{\~n}o that uses the identity
\begin{equation}
  [\underline{u}^4_{\alpha}] + [\underline{\tilde{u}}^4_{\alpha}] = \tau'
  \mymultiply u_{\alpha} (\tmmathbf{T}) . \label{eq:appTV-our-u-to-Trabuchos}
\end{equation}
This identity can be established by noting that our variational
problem~(\ref{eq:twisting-psi-alpha-variational-pb}) for $u_{\alpha}$ is a
linear combination of the variational problems~[3.23] for
$[\underline{u}_{\alpha}^4]$ and~[5.8] for
$[\underline{\tilde{u}}^4_{\alpha}]$.

To obtain equation~(\ref{eq:appTV-equil-tau-Trabucho}), we have also used the
fact that the quantities related to the stretching and bending modes
$[u_{\alpha}^0]$, $[\underline{u}_3^0]$, $[z_{\alpha}^2]$ as well as the
quantity $[w^0]$ are all zero as a consequence of our symmetry
assumptions~\cite{trabucho1989existence}.

Using the properties of $[\Psi]$ just listed, the second term in the
right-hand side of~(\ref{eq:appTV-equil-tau-Trabucho}) can be integrated by
parts as
\[ \begin{array}{lll}
     - \mu \mymultiply \tau'' \iint_{\Omega} (\partial_2 u_1 - \partial_1
     u_2) \mymultiply [\Psi] \mymultiply \mathd A & = & \mu \mymultiply \tau''
     \mymultiply \iint_{\Omega} \left( u_1 \mymultiply \partial_2 [\Psi] -
     u_2 \mymultiply \partial_1 [\Psi] \right) \mymultiply \mathd A\\
     & = & \mu \mymultiply \tau'' \mymultiply \iint_{\Omega} u_{\alpha}
     \mymultiply \partial_{\alpha} \omega \mymultiply \mathd A + \mu
     \mymultiply \tau'' \mymultiply \iint_{\Omega} \eta_{\alpha \nocomma
     \beta} \mymultiply T_{\alpha} \mymultiply u_{\beta} \mymultiply \mathd
     A\\
     & = & \tau'' \mymultiply D_{\mu}
   \end{array} \]
To obtain the last equality, we have identified the quantity $D_{\mu}$ defined
in~(\ref{eq:twist-D-sub-x}), and have observed that the integrand of the
second term is zero by equation~(\ref{eq:twist-corrective-displ-cstr}).
Inserting this into~(\ref{eq:appTV-equil-tau-Trabucho}), and using the
identity $\nu = \frac{\lambda}{2 \mymultiply (\lambda + \mu)}$, we have
\[ [M_4^3] = - \lambda \mymultiply \frac{\mathd}{\mathd S} \left(
   \iint_{\Omega} \partial_{\alpha} [\underline{u}_{\alpha}^4] \mymultiply
   \omega \mymultiply \mathd A \right) - \tau'' \mymultiply D_{\mu} \]
Using equation~(\ref{eq:appTV-our-u-to-Trabuchos}) one more time, one can
eliminate $[\underline{u}_{\alpha}^4]$ in favor of
$[\underline{\tilde{u}}_{\alpha}^4]$,
\[ \begin{array}{lll}
     {}[M_4^3] & = & \lambda \mymultiply \frac{\mathd}{\mathd S} \left(
     \iint_{\Omega} \partial_{\alpha} [\underline{\tilde{u}}_{\alpha}^4]
     \mymultiply \omega \mymultiply \mathd A \right) - \lambda \mymultiply
     \frac{\mathd}{\mathd S} \left( \tau' (S) \mymultiply \iint_{\Omega}
     \partial_{\alpha} u_{\alpha} \mymultiply \omega \mymultiply \mathd A
     \right) + \tau'' \mymultiply D_{\mu}\\
     & = & \lambda \mymultiply \frac{\mathd}{\mathd S} \left(
     \iint_{\Omega} \partial_{\alpha} [\underline{\tilde{u}}_{\alpha}^4]
     \mymultiply \omega \mymultiply \mathd A \right) - \tau'' \mymultiply
     (D_{\lambda} - D_{\mu})
   \end{array} \]

At this point, we introduce two auxiliary functions $\tilde{u}_{\alpha}
(\tmmathbf{T})$ such that
\begin{equation}
  \begin{array}{lll}
    \partial_1 \tilde{u}_1 (\tmmathbf{T}) = \omega (\tmmathbf{T}) & \partial_2
    \tilde{u}_2 (\tmmathbf{T}) = \omega (\tmmathbf{T}) & \partial_2
    \tilde{u}_1 (\tmmathbf{T}) + \partial_1 \tilde{u}_2 (\tmmathbf{T}) = 0
  \end{array} \label{eq:u-tilde-identities}
\end{equation}
The existence of the functions $\tilde{u}_{\alpha} (\tmmathbf{T})$ is
warranted by the kinematic compatibility condition
\[ \partial_{12} (\partial_2 \tilde{u}_1 + \partial_1 \tilde{u}_2) -
   \partial_{22} (\partial_1 \tilde{u}_1) - \partial_{11} (\partial_2
   \tilde{u}_2) = \partial_{12} (0) - \partial_{22} \omega - \partial_{11}
   \omega = - \partial_{\alpha \nocomma \alpha} \omega = 0 \]
as can be checked by applying the Euler-Lagrange method to the weak form of
the variational problem for $\omega (\tmmathbf{T})$
in~(\ref{eq:warping-function-variational}).

The functions $\tilde{u}_{\alpha} (\tmmathbf{T})$ are defined up to a
rigid-body motion that can be set by requiring
\[ \begin{array}{ll}
     \iint_{\Omega} \tilde{u}_{\alpha} \mymultiply \mathd A = 0 &
     \iint_{\Omega} \eta_{\alpha \nocomma \beta} \mymultiply T_{\alpha}
     \mymultiply \tilde{u}_{\beta} \mymultiply \mathd A = 0
   \end{array} \]

The variational problem~[5.8] for $[\underline{\tilde{u}}_{\alpha}^4]$ writes
\[ \forall \hat{v}_{\alpha} \qquad \iint_{\Omega} \left\{ \left( \lambda
   \mymultiply \partial_{\rho} [\tilde{u}^4_{\rho}] \mymultiply \delta_{\alpha
   \nocomma \beta} + \mu \mymultiply \left( \partial_{\alpha}
   [\tilde{u}^4_{\beta}] + \partial_{\beta} [\tilde{u}^4_{\alpha}] \right)
   \right) \mymultiply \partial_{\beta} \hat{v}_{\alpha} + \lambda \mymultiply
   \tau' (S) \mymultiply \omega \mymultiply \partial_{\alpha} \hat{v}_{\alpha}
   \right\} \mymultiply \mathd A = 0 \]
Setting $\hat{v}_{\alpha} = \tilde{u}_{\alpha}$ in this variational problem,
we obtain, after making use of the identities in
equation~(\ref{eq:u-tilde-identities}),
\[ 2 \mymultiply (\lambda + \mu) \mymultiply \iint_{\Omega}
   \partial_{\alpha} [\tilde{u}^4_{\alpha}] \mymultiply \omega \mymultiply
   \mathd A + 2 \mymultiply \lambda \mymultiply \tau' \mymultiply J_{\omega} =
   0. \]
This identity can be used to rewrite $[M_3^4]$ as
\[ [M_4^3] = - \left( \frac{\lambda^2}{\lambda + \mu} \mymultiply J_{\omega}
   \mymultiply + D_{\lambda} - D_{\mu} \right) \mymultiply \tau'' = - \left(
   \left( \left( \lambda + 2 \mymultiply \mu \right) - Y \right) \mymultiply
   J_{\omega} \mymultiply + D_{\lambda} - D_{\mu} \right) \mymultiply \tau'' =
   - \left( - Y \mymultiply J_{\omega} + D_{\omega} - D_{\mu} \right)
   \mymultiply \tau'' \]
after making use successively of the identity $\frac{\lambda^2}{\lambda + \mu}
= \left( \lambda + 2 \mymultiply \mu \right) - Y$, see the definition of the
Young modulus in~(\ref{eq:twisting-Young-modulus}), and of the definition of
$D_{\omega}$ in~(\ref{eq:twist-D-sub-x}).

The principle of virtual work~(\ref{eq:appTV-Trabucho-equil}) derived by
Trabucho and Via{\~n}o can therefore be rewritten as
\[ \forall \hat{\theta} \qquad \int \left\{ \mu \mymultiply J \mymultiply \tau
   (S) - (D_{\omega} - D_{\mu}) \mymultiply \tau'' (S) \right\} \mymultiply
   \hat{\theta}' (S) \mymultiply \mathd S = \int m_3 (S) \mymultiply
   \hat{\theta} (S) \mymultiply \mathd S. \]
Integrating by parts and eliminating the virtual rotation $\hat{\theta}$, one
recovers the equation of equilibrium derived by our method, see
equation~(\ref{eq:twist-equil-eq-with-gradient}).

\section{A non-linear cylinder that bends and stretches}

\subsection{Simple traction of an isotropic hyper-elastic
material}\label{app-sec:simple-stretching}

Here, we characterize the finite-strain material model introduced in
section~\ref{sec:beam-problem-setting}. We are particularly interested in
simple traction along the material axis $\tmmathbf{e}_3$, since this is the
state of stress corresponding to the unbuckled solution. We denote by
$\varepsilon$ the axial strain, such that the axial stretch ratio is $1 +
\varepsilon$. The simple traction is such that the strain
$\tmmathbf{E}_{\text{tr}} (\varepsilon)$ is equi-biaxial and the stress
$\tmmathbf{\Sigma} \left( \tmmathbf{E}_{\text{tr}} (\varepsilon) \right)$ is
uniaxial:,
\begin{equation}
  \begin{array}{rcl}
    \tmmathbf{E}_{\text{tr}} (\varepsilon) & = & \frac{(1 + \varepsilon)^2 -
    1}{2} \mymultiply \tmmathbf{e}_3 \otimes \tmmathbf{e}_3 +
    \frac{\nlPoisson^2 (\varepsilon) - 1}{2} \mymultiply \tmmathbf{e}_{\alpha}
    \otimes \tmmathbf{e}_{\alpha}\\
    \tmmathbf{\Sigma} \left( \tmmathbf{E}_{\text{tr}} (\varepsilon) \right) &
    = & \Sigma_{\text{tr}} (\varepsilon) \mymultiply \tmmathbf{e}_3 \otimes
    \tmmathbf{e}_3,
  \end{array} \label{eq:beam-simple-traction-E-Sigma}
\end{equation}
where $\nlPoisson (\varepsilon)$ is the transverse stretch ratio due to
Poisson's effect.

The first derivative of the strain energy density in simple traction
\begin{equation}
  w_{\text{tr}} (\varepsilon) = w \left( \tmmathbf{E}_{\text{tr}}
  (\varepsilon) \right)
\end{equation}
is related to the stress in simple traction by
\begin{equation}
  (1 + \varepsilon) \mymultiply \Sigma_{\text{tr}} (\varepsilon) =
  \frac{\mathd w_{\text{tr}}}{\mathd \varepsilon} (\varepsilon),
  \label{eq:beam-Sigma-tr-wtr-prime}
\end{equation}
while the second derivative of $w_{\text{tr}} (\varepsilon)$ defines the
tangent Young modulus $Y_{\text{t}} (\varepsilon)$,
\begin{equation}
  Y_{\text{t}} (\varepsilon) = \frac{\mathd^2 w_{\text{tr}}}{\mathd
  \varepsilon^2} (\varepsilon) . \label{eq:beam-Yt}
\end{equation}

When in simple traction, an isotropic material has an incremental elastic
behavior that is transversely isotropic. As a result, the tangent elastic
moduli
\begin{equation}
  \tmmathbf{K}^{\text{tr}} (\varepsilon) =\tmmathbf{K} \left(
  \tmmathbf{E}_{\text{tr}} (\varepsilon) \right) = \frac{\mathd^2 w}{\mathd
  \tmmathbf{E}^2} \left( \tmmathbf{E}_{\text{tr}} (\varepsilon) \right)
  \label{eq:beam-K-tr}
\end{equation}
are of the form
\begin{equation}
  \begin{array}{ll}
    K^{\text{tr}}_{\alpha \nocomma 3, \beta \nocomma \gamma} (\varepsilon) = 0
    & K^{\text{tr}}_{\alpha \nocomma 3, \beta \nocomma 3} = G_{\text{t}}
    (\varepsilon) \mymultiply \delta_{\alpha \nocomma \beta},
  \end{array} \label{eq:beam-transverse-isotropic-K}
\end{equation}
where $G_{\text{t}} (\varepsilon)$ is the tangent shear modulus. In
equation~(\ref{eq:beam-1d-reduction-coefficients}), the ratio of the shear to
the Young modulus is defined in terms of a dimensionless constitutive
parameter $c_{\Gamma} (\varepsilon) = Y_{\text{t}} (\varepsilon) / \left[ 2
\mymultiply G_{\text{t}} (\varepsilon) \mymultiply \nlPoisson (\varepsilon)
\mymultiply \left( - \frac{\mathd \nlPoisson}{\mathd \varepsilon}
(\varepsilon) \right) \mymultiply (1 + \varepsilon) \right]$.

Let us now focus on the natural configuration, $\varepsilon = 0$. This
configuration being free of stress, we have
\begin{equation}
  \frac{\mathd w_{\text{tr}}}{\mathd \varepsilon} (0) = 0.
  \label{eq:beam-wtr-prime-0}
\end{equation}
The following relations warrant that the configuration $\varepsilon = 0$ is
undeformed, and define the initial Poisson's ratio $\nu_0$,
\begin{equation}
  \begin{array}{ll}
    p (0) = 1 & \frac{\mathd p}{\mathd \varepsilon} (0) = - \nu_0 .
  \end{array} \label{eq:beam-p-identities}
\end{equation}
In this initial configuration, Hookean elasticity applies and the
{\tmem{initial}} shear modulus \ is given by
\begin{equation}
  G_{\text{t}} (0) = \frac{Y_0}{2 \mymultiply (1 + \nu_0)} .
\end{equation}
We denote by $Y_0 = Y_{\text{t}} (0)$ the initial Young modulus and by $Y_0' =
\frac{\mathd Y_{\text{t}}}{\mathd \varepsilon} (0)$ the initial curvature of
the traction curve that yields the Kirchhoff stress $w_{\text{tr}}'
(\varepsilon)$ as a function of the strain $\varepsilon$,
\begin{equation}
  \begin{array}{lll}
    Y_0 = \frac{\mathd^2 w_{\text{tr}}}{\mathd \varepsilon^2} (0) & Y_0' =
    \frac{\mathd^3 w_{\text{tr}}}{\mathd \varepsilon^3} (0) & \nu_0 = -
    \frac{\mathd p}{\mathd \varepsilon} (0) .
  \end{array} \label{eq:beam-Y0-Y0p}
\end{equation}

For the particular constitutive law $w (\tmmathbf{T}, \tmmathbf{E}) =
w_{\text{ST}} (\tmmathbf{E})$ used by \citet{scherzinger1998asymptotic} and
given in section~\ref{sec:beam-problem-setting}, these material constants read
\[ \begin{array}{rll}
     Y_0 & = & 4 \mymultiply \left( A_{\text{ST}} + B_{\text{ST}} \right)
     \mymultiply \left( 1 + \nu_{\text{ST}} \right)\\
     \nu_0 & = & \nu_{\text{ST}}\\
     Y_0' & = & \frac{4 \mymultiply \left( 1 + \nu_{\text{ST}} \right)}{9}
     \mymultiply \left( A_{\text{ST}} \mymultiply \left( - 23 + 8 \mymultiply
     \nu_{\tmop{ST}} + 4 \mymultiply \nu_{\tmop{ST}}^2 \right) - B_{\text{ST}}
     \mymultiply \left( 31 + 8 \mymultiply \nu_{\tmop{ST}} + 4 \mymultiply
     \nu_{\tmop{ST}}^2 \right) \right) .
   \end{array} \]
\subsection{Analysis of homogeneous
solutions}\label{app-sec:beam-homogeneous-solutions}

In this section, we check that the solution proposed in
equation~(\ref{eq:beam-homogeneous-Y}) satisfies all the equations applicable
to the homogeneous solution, which have been listed in
section~\ref{ssec:non-regularized-model-outline}. As the
displacement~(\ref{eq:beam-homogeneous-Y}) satisfies $Y_3^{(\varepsilon,
\kappa)} (\tmmathbf{T}) = 0$, the non-linear
expression~(\ref{eq:homogeneous-strain}) of the strain simplifies to
$\tmmathbf{E}^{\tmmathbf{h}} (\tmmathbf{T}) =\tmmathbf{E}^{(\varepsilon,
\kappa)} (\tmmathbf{T}) = \tilde{\tmmathbf{E}} (\tmmathbf{T}, \tmmathbf{h},
\tmmathbf{Y}^{(\varepsilon, \kappa)})$ where
\begin{equation}
  \tmmathbf{E}^{(\varepsilon, \kappa)} (\tmmathbf{T}) = \frac{\left( 1 +
  \varepsilon + \kappa \mymultiply Y_2^{(\varepsilon, \kappa)} (\tmmathbf{T})
  \right)^2 - 1}{2} \mymultiply \tmmathbf{e}_3 \otimes \tmmathbf{e}_3 +
  \frac{\partial_{\alpha} Y_{\gamma}^{(\varepsilon, \kappa)} (\tmmathbf{T})
  \mymultiply \partial_{\beta} Y_{\gamma}^{(\varepsilon, \kappa)}
  (\tmmathbf{T}) - \delta_{\alpha \nocomma \beta}}{2} \mymultiply
  \tmmathbf{e}_{\alpha} \odot \tmmathbf{e}_{\beta} .
  \label{eq:beam-Ehom-exact}
\end{equation}
To derive this equality, we have used the fact that the shear term
$\tilde{t}_i \mymultiply \partial_{\alpha} Y_i^{(\varepsilon, \kappa)}
(\tmmathbf{T}) \mymultiply \tmmathbf{e}_{\alpha} \odot \tmmathbf{e}_3$ in
equation~(\ref{eq:homogeneous-strain}) is zero, as $\tilde{t}_i = (1 +
\varepsilon) \mymultiply \delta_{i \nocomma 3} + \eta_{i \nocomma j \nocomma
k} \mymultiply \kappa_j \mymultiply Y^{(\varepsilon, \kappa)}_k = (1 +
\varepsilon) \mymultiply \delta_{i \nocomma 3} + \eta_{i \nocomma 1 \nocomma
\alpha} \mymultiply \kappa \mymultiply Y^{(\varepsilon, \kappa)}_{\alpha} = (1
+ \varepsilon + Y^{(\varepsilon, \kappa)}_2) \mymultiply \delta_{i \nocomma
3}$ and so $\tilde{t}_i \mymultiply \partial_{\alpha} Y_i^{(\varepsilon,
\kappa)} (\tmmathbf{T}) \mymultiply \tmmathbf{e}_{\alpha} \odot \tmmathbf{e}_3
= \tilde{t}_3 \mymultiply \partial_{\alpha} \not{Y_3^{(\varepsilon, \kappa)}
(\tmmathbf{T})} \mymultiply \tmmathbf{e}_{\alpha} \odot \tmmathbf{e}_3 = 0$.

Inserting the expansion of $Y_{\alpha}^{(\varepsilon, \kappa)}$ in powers of
$\kappa$ from equation~(\ref{eq:beam-homogeneous-Y}), we have
\begin{equation}
  \begin{array}{lcl}
    \tmmathbf{E}^{(\varepsilon, \kappa)} (\tmmathbf{T}) & = & \frac{\left( 1 +
    \varepsilon + \kappa \mymultiply \nlPoisson (\varepsilon) \mymultiply T_2
    \right)^2 - 1}{2} \mymultiply \tmmathbf{e}_3 \otimes \tmmathbf{e}_3\\
    &  & \ldots + \frac{1}{2} \mymultiply \left( \nlPoisson^2 (\varepsilon)
    \mymultiply \left( \delta_{\alpha \nocomma \gamma} + \kappa \mymultiply
    \frac{\mathd \nlPoisson}{\mathd \varepsilon} (\varepsilon) \mymultiply
    \partial_{\alpha} \varphi_{\gamma} \right) \mymultiply \left(
    \delta_{\beta \nocomma \gamma} + \kappa \mymultiply \frac{\mathd
    \nlPoisson}{\mathd \varepsilon} (\varepsilon) \mymultiply \partial_{\beta}
    \varphi_{\gamma} \right) - \delta_{\alpha \nocomma \beta} \right)
    \mymultiply \tmmathbf{e}_{\alpha} \odot \tmmathbf{e}_{\beta} +\mathcal{O}
    (\kappa^2)\\
    & = & \frac{\left( 1 + \varepsilon + \kappa \mymultiply \nlPoisson
    (\varepsilon) \mymultiply T_2 \right)^2 - 1}{2} \mymultiply \tmmathbf{e}_3
    \otimes \tmmathbf{e}_3 + \frac{1}{2} \mymultiply \left( \nlPoisson^2
    (\varepsilon) \mymultiply \left( 1 + 2 \mymultiply \kappa \mymultiply
    \frac{\mathd \nlPoisson}{\mathd \varepsilon} (\varepsilon) \mymultiply T_2
    \right) - 1 \right) \mymultiply \tmmathbf{e}_{\alpha} \otimes
    \tmmathbf{e}_{\alpha} +\mathcal{O} (\kappa^2)\\
    & = & \frac{\left( 1 + \varepsilon + \kappa \mymultiply \nlPoisson
    (\varepsilon) \mymultiply T_2 \right)^2 - 1}{2} \mymultiply \tmmathbf{e}_3
    \otimes \tmmathbf{e}_3 + \frac{1}{2} \mymultiply \left( \left( \nlPoisson
    (\varepsilon) + \frac{\mathd \nlPoisson}{\mathd \varepsilon} (\varepsilon)
    \times \left[ \kappa \mymultiply \nlPoisson (\varepsilon) \mymultiply T_2
    \right] \right)^2 - 1 \right) \mymultiply \tmmathbf{e}_{\alpha} \otimes
    \tmmathbf{e}_{\alpha} +\mathcal{O} (\kappa^2)\\
    & = & \frac{\left( 1 + \varepsilon + \kappa \mymultiply \nlPoisson
    (\varepsilon) \mymultiply T_2 \right)^2 - 1}{2} \mymultiply \tmmathbf{e}_3
    \otimes \tmmathbf{e}_3 + \frac{\left( \nlPoisson \left( \varepsilon +
    \left[ \kappa \mymultiply \nlPoisson (\varepsilon) \mymultiply T_2 \right]
    \right) \right)^2 - 1}{2} \mymultiply \tmmathbf{e}_{\alpha} \otimes
    \tmmathbf{e}_{\alpha} +\mathcal{O} (\kappa^2)\\
    & = & \tmmathbf{E}_{\text{tr}} \left( \varepsilon + \kappa \mymultiply
    \nlPoisson (\varepsilon) \mymultiply T_2 \right) +\mathcal{O} (\kappa^2)
  \end{array} \label{eq:beam-E-expansion}
\end{equation}
Here, we have used the identity $\partial_{\alpha} \varphi_{\beta} +
\partial_{\beta} \varphi_{\alpha} = 2 \mymultiply T_2 \mymultiply
\delta_{\alpha \nocomma \beta}$ from
equation~(\ref{eq:phi-alpha-fundamental-def}), as well as the identity $\left(
1 + 2 \mymultiply x \right) = (1 + x)^2 +\mathcal{O} (x^2)$ with $x = \kappa
\mymultiply \frac{\mathd \nlPoisson}{\mathd \varepsilon} (\varepsilon)
\mymultiply T_2$, and we have identified the strain in simple traction
$\tmmathbf{E}_{\text{tr}}$ from
equation~(\ref{eq:beam-simple-traction-E-Sigma}).

By the constitutive law, the stress writes $\tmmathbf{\Sigma}^{(\varepsilon,
\kappa)} = \Sigma_{\text{tr}} \left( \varepsilon + \kappa \mymultiply
\nlPoisson (\varepsilon) \mymultiply T_2 \right) \mymultiply \tmmathbf{e}_3
\otimes \tmmathbf{e}_3 +\mathcal{O} (\kappa^2)$, see
equation~(\ref{eq:beam-simple-traction-E-Sigma}).

We proceed to check that $\tmmathbf{Y}^{(\varepsilon, \kappa)}$ satisfies the
variational problem~(\ref{eq:app-red-variational-pb-homogeneous}).

By design from equation~(\ref{eq:phi-alpha-fundamental-def}), the quantities
$\varphi_{\alpha} (\tmmathbf{T})$ and $T_1 \mymultiply \varphi_2
(\tmmathbf{T}) - T_2 \mymultiply \varphi_1 (\tmmathbf{T})$ average out to zero
on the cross-section: the kinematic constraints on the first two lines of
(\ref{eq:app-red-variational-pb-homogeneous}) are verified up to terms of
order $\kappa^2$.

Given that the stress is uniaxial to order $\kappa$, we only need the
projection along $\tmmathbf{e}_3 \otimes \tmmathbf{e}_3$ of the virtual change
of strain $\widehat{\tilde{\tmmathbf{E}}}^{\tmmathbf{h}} (\tmmathbf{T}) =
\frac{\mathd \tilde{\tmmathbf{E}}}{\mathd \tmmathbf{Y}} (\tmmathbf{T},
\tmmathbf{h}, \tmmathbf{Y}^{\tmmathbf{h}}) \cdot \hat{\tmmathbf{Y}}$ appearing
in the principle of virtual
work~(\ref{eq:app-red-variational-pb-homogeneous}). Using
equation~(\ref{eq:homogeneous-strain}), we find
$\widehat{\tilde{\tmmathbf{E}}}^{(\varepsilon, \kappa)} (\tmmathbf{T})
\doublecontract (\tmmathbf{e}_3 \otimes \tmmathbf{e}_3) = (1 + \varepsilon)
\mymultiply \kappa \mymultiply \hat{Y}_2 (\tmmathbf{T}) +\mathcal{O}
(\kappa^2)$, hence $\tmmathbf{\Sigma} (\tmmathbf{T},
\tmmathbf{E}^{\tmmathbf{h}} (\tmmathbf{T})) \doublecontract
\widehat{\tilde{\tmmathbf{E}}}^{\tmmathbf{h}} (\tmmathbf{T}) = (1 +
\varepsilon) \mymultiply \kappa \mymultiply \Sigma_{\text{tr}} \left(
\varepsilon + \kappa \mymultiply \nlPoisson (\varepsilon) \mymultiply T_2
\right) \mymultiply \hat{Y}_2 (\tmmathbf{T}) +\mathcal{O} (\kappa^2) = (1 +
\varepsilon) \mymultiply \kappa \mymultiply \Sigma_{\text{tr}} (\varepsilon)
\mymultiply \hat{Y}_2 (\tmmathbf{T}) +\mathcal{O} (\kappa^2)$. Therefore, the
variational problem~(\ref{eq:app-red-variational-pb-homogeneous}) writes
\begin{equation}
  \forall \hat{Y}_i (\tmmathbf{T}) \qquad \iint_{\Omega} \left[ (1 +
  \varepsilon) \mymultiply \kappa \mymultiply \Sigma_{\text{tr}} (\varepsilon)
  \mymultiply \hat{Y}_2 (\tmmathbf{T}) + F_i^{(\varepsilon, \kappa)}
  \mymultiply \hat{Y}_i (\tmmathbf{T}) + Q^{(\varepsilon, \kappa)} \mymultiply
  \left( T_1 \mymultiply \hat{Y}_2 (\tmmathbf{T}) - T_2 \mymultiply \hat{Y}_1
  (\tmmathbf{T}) \right) +\mathcal{O} (\kappa^2) \right] \mymultiply \mathd A
  = 0, \label{eq:beam-homogeneous-pvw}
\end{equation}
where $F_i^{(\varepsilon, \kappa)}$ and $Q^{(\varepsilon, \kappa)}$ are
Lagrange multipliers. This variational problem is indeed satisfied with the
choice $F_i^{(\varepsilon, \kappa)} = - \delta_{i \nocomma 2} \mymultiply (1 +
\varepsilon) \mymultiply \kappa \mymultiply \Sigma_{\text{tr}} (\varepsilon)$
and $Q^{(\varepsilon, \kappa)} = 0$, as can be checked.

There remains to calculate the strain energy $W_{\text{hom}} (\varepsilon,
\kappa) = \iint_{\Omega} w (\tmmathbf{E}^{(\varepsilon, \kappa)}
(\tmmathbf{T})) \mymultiply \mathd A$ associated with the homogeneous solution
$\tmmathbf{Y}^{(\varepsilon, \kappa)}$, see equation~(\ref{eq:Wh-def}). Let
$\tmmathbf{O} (\tmmathbf{T}) =\tmmathbf{E}^{(\varepsilon, \kappa)}
(\tmmathbf{T}) -\tmmathbf{E}_{\text{tr}} \left( \varepsilon + \kappa
\mymultiply \nlPoisson (\varepsilon) \mymultiply T_2 \right)$ denote the
deviation of the strain from the simple traction estimate: $\tmmathbf{O}
(\tmmathbf{T}) =\mathcal{O} (\kappa^2)$ from
equation~(\ref{eq:beam-E-expansion}). One can then expand $W_{\text{hom}}
(\varepsilon, \kappa)$ as
\[ \begin{array}{lll}
     W_{\text{hom}} (\varepsilon, \kappa) & = & \iint_{\Omega} w \left(
     \tmmathbf{E}_{\text{tr}} \left( \varepsilon + \kappa \mymultiply
     \nlPoisson (\varepsilon) \mymultiply T_2 \right) +\tmmathbf{O}
     (\tmmathbf{T}) \right) \mymultiply \mathd A\\
     & = & \iint_{\Omega} \left[ w \left( \tmmathbf{E}_{\text{tr}} \left(
     \varepsilon + \kappa \mymultiply \nlPoisson (\varepsilon) \mymultiply T_2
     \right) \right) +\tmmathbf{\Sigma}_{\text{tr}} \left( \varepsilon +
     \kappa \mymultiply \nlPoisson (\varepsilon) \mymultiply T_2 \right)
     \doublecontract \tmmathbf{O} (\tmmathbf{T}) +\mathcal{O} (\kappa^4)
     \right] \mymultiply \mathd A
   \end{array} \]
after identifying the stress $\tmmathbf{\Sigma}_{\text{tr}} (e) = \frac{\mathd
w}{\mathd \tmmathbf{E}} \left( \tmmathbf{E}_{\text{tr}} (e) \right)$ in simple
traction, see equations~(\ref{eq:gr-effect-homogeneous-qties}) and
(\ref{eq:beam-simple-traction-E-Sigma}). Now, the second term in the integrand
can be evaluated as follows by using
equations~(\ref{eq:beam-simple-traction-E-Sigma}), (\ref{eq:beam-Ehom-exact}),
(\ref{eq:beam-simple-traction-E-Sigma}) again, (\ref{eq:beam-homogeneous-Y})
and~(\ref{eq:phi-alpha-fundamental-def}) sequentially,
\[ \begin{array}{lll}
     \iint_{\Omega} \tmmathbf{\Sigma}_{\text{tr}} \left( \varepsilon +
     \kappa \mymultiply \nlPoisson (\varepsilon) \mymultiply T_2 \right)
     \doublecontract \tmmathbf{O} (\tmmathbf{T}) \mymultiply \mathd A & = &
     \iint_{\Omega} \Sigma_{\text{tr}} \left( \varepsilon + \kappa
     \mymultiply \nlPoisson (\varepsilon) \mymultiply T_2 \right) \mymultiply
     \left\{ \tmmathbf{e}_3 \otimes \tmmathbf{e}_3 \doublecontract \left(
     \tmmathbf{E}^{(\varepsilon, \kappa)} (\tmmathbf{T})
     -\tmmathbf{E}_{\text{tr}} \left( \varepsilon + \kappa \mymultiply
     \nlPoisson (\varepsilon) \mymultiply T_2 \right) \right) \right\}
     \mymultiply \mathd A\\
     & = & \iint_{\Omega} \left( \Sigma_{\text{tr}} (\varepsilon)
     +\mathcal{O} (\kappa) \right) \mymultiply \left\{ \frac{\left( 1 +
     \varepsilon + \kappa \mymultiply Y_2^{(\varepsilon, \kappa)}
     (\tmmathbf{T}) \right)^2 - 1}{2} - \frac{\left( 1 + \varepsilon + \kappa
     \mymultiply \nlPoisson (\varepsilon) \mymultiply T_2 \right)^2 - 1}{2}
     \right\} \mymultiply \mathd A\\
     & = & \iint_{\Omega} \left( \Sigma_{\text{tr}} (\varepsilon)
     +\mathcal{O} (\kappa) \right) \mymultiply \left\{ (1 + \varepsilon)
     \mymultiply \left( \kappa \mymultiply Y_2^{(\varepsilon, \kappa)}
     (\tmmathbf{T}) - \kappa \mymultiply \nlPoisson (\varepsilon) \mymultiply
     T_2 +\mathcal{O} (\kappa^2) \right) \right\} \mymultiply \mathd A\\
     & = & \iint_{\Omega} (1 + \varepsilon) \mymultiply \left(
     \Sigma_{\text{tr}} (\varepsilon) +\mathcal{O} (\kappa) \right)
     \mymultiply \kappa^2 \mymultiply \left( \nlPoisson (\varepsilon)
     \mymultiply \frac{\mathd \nlPoisson}{\mathd \varepsilon} (\varepsilon)
     \mymultiply \varphi_2 (\tmmathbf{T}) +\mathcal{O} (\kappa) \right)
     \mymultiply \mathd A\\
     & = & (1 + \varepsilon) \mymultiply \Sigma_{\text{tr}} (\varepsilon)
     \mymultiply \nlPoisson (\varepsilon) \mymultiply \frac{\mathd
     \nlPoisson}{\mathd \varepsilon} (\varepsilon) \mymultiply \kappa^2
     \mymultiply \iint_{\Omega} \varphi_2 (\tmmathbf{T}) \mymultiply \mathd
     A +\mathcal{O} (\kappa^3)\\
     & = & \mathcal{O} (\kappa^3)
   \end{array} \]
Returning to the expression of $W_{\text{hom}} (\varepsilon, \kappa)$, we are
therefore left with
\[ \begin{array}{lll}
     W_{\text{hom}} (\varepsilon, \kappa) & = & \iint_{\Omega} \left[ w
     \left( \tmmathbf{E}_{\text{tr}} \left( \varepsilon + \kappa \mymultiply
     \nlPoisson (\varepsilon) \mymultiply T_2 \right) \right) +\mathcal{O}
     (\kappa^3) \right] \mymultiply \mathd A\\
     & = & \iint_{\Omega} \left[ w_{\text{tr}} (\varepsilon) +
     \frac{\mathd w_{\text{tr}}}{\mathd \varepsilon} (\varepsilon) \mymultiply
     \kappa \mymultiply p (\varepsilon) \mymultiply T_2 + \frac{\kappa^2}{2}
     \mymultiply \frac{\mathd^2 w_{\text{tr}}}{\mathd \varepsilon^2}
     (\varepsilon) \mymultiply p^2 (\varepsilon) \mymultiply T_2^2
     +\mathcal{O} (\kappa^3) \right] \mymultiply \mathd A,
   \end{array} \]
where $w_{\text{tr}} (\varepsilon) = w \left( \tmmathbf{E}_{\text{tr}}
(\varepsilon) \right)$ is the strain energy density for simple traction.

The integral of $T_2$ is zero in a disk. Identifying the Young modulus
$Y_{\text{t}} (\varepsilon) = \frac{\mathd^2 w_{\text{tr}}}{\mathd
\varepsilon^2} (\varepsilon)$ from~(\ref{eq:beam-Yt}) and the initial
geometric moment of inertia $I_1^0 = \iint_{\Omega} T_2^2 \mymultiply
\mathd A$, we find
\[ W_{\text{hom}} (\varepsilon, \kappa) = A \mymultiply w_{\text{tr}}
   (\varepsilon) + \frac{\kappa^2}{2} \mymultiply Y_{\text{t}} (\varepsilon)
   \mymultiply p^2 (\varepsilon) \mymultiply I_1^0 +\mathcal{O} (\kappa^3) .
\]
The expression announced in equation~(\ref{eq:beam-Whom}) follows by observing
that $W_{\text{hom}} (\varepsilon, \kappa)$ is an even function of $\kappa$ by
symmetry, implying that the remainder $\mathcal{O} (\kappa^3)$ in equation
above is actually of order $\mathcal{O} (\kappa^4)$.

\subsection{Analysis of the gradient
effect}\label{ssec:beam-app-gradient-effect}

This section derives the gradient effect, following closely the outline given
in section~\ref{sec:gradient-effect}.

From equation~(\ref{eq:gr-effect-homogeneous-qties}), we find the
transformation gradient in the homogeneous configuration as $F^{(\varepsilon,
\kappa)}_{3 \nocomma 3} (\tmmathbf{T}) = 1 + \varepsilon + \kappa \mymultiply
\nlPoisson (\varepsilon) \mymultiply T_2 +\mathcal{O} (\kappa^2)$. Using the
identity $\Sigma_{\text{tr}} (e) = \frac{\mathd w_{\text{tr}}}{\mathd e} (e) /
(1 + e)$ from equation~(\ref{eq:beam-Sigma-tr-wtr-prime}), the operator
$\tmmathbf{C}_{\tmmathbf{h}}^{(1)}$ defined in~(\ref{eq:A-C1h}) is found as
\begin{equation}
  \begin{array}{lll}
    \tmmathbf{C}_{(\varepsilon, \kappa)}^{(1)} \cdot \tmmathbf{Z}^{\dag} & = &
    \iint_{\Omega} \left( 1 + \varepsilon + \kappa \mymultiply \nlPoisson
    (\varepsilon) \mymultiply T_2 \right) \mymultiply \Sigma_{\text{tr}}
    \left( \varepsilon + \kappa \mymultiply \nlPoisson (\varepsilon)
    \mymultiply T_2 \right) \mymultiply Z_3^{\dag} (\tmmathbf{T}) \mymultiply
    \mathd A +\mathcal{O} (\kappa^2)\\
    & = & \iint_{\Omega} w_{\text{tr}}' \left( \varepsilon + \kappa
    \mymultiply \nlPoisson (\varepsilon) \mymultiply T_2 \right) \mymultiply
    Z_3^{\dag} (\tmmathbf{T}) \mymultiply \mathd A +\mathcal{O} (\kappa^2)\\
    & = & w_{\text{tr}}' (\varepsilon) \mymultiply \iint_{\Omega}
    \mymultiply Z_3^{\dag} (\tmmathbf{T}) \mymultiply \mathd A + \kappa
    \mymultiply Y_{\text{t}} (\varepsilon) \mymultiply p (\varepsilon)
    \mymultiply \iint_{\Omega} T_2 \mymultiply Z_3^{\dag} (\tmmathbf{T})
    \mymultiply \mathd A +\mathcal{O} (\kappa^2)\\
    & = & \kappa \mymultiply Y_{\text{t}} (\varepsilon) \mymultiply p
    (\varepsilon) \mymultiply \iint_{\Omega} T_2 \mymultiply Z_3^{\dag}
    (\tmmathbf{T}) \mymultiply \mathd A +\mathcal{O} (\kappa^2)
  \end{array} \label{eq:beam-C1h}
\end{equation}
where $Y_{\text{t}} (\varepsilon)$ is the tangent Young modulus from
equation~(\ref{eq:beam-Yt}). The operator $\tmmathbf{C}_{\tmmathbf{h}}^{(1)}$
is applied exclusively to cross-sectional functions $\tmmathbf{Z}^{\dag}$
satisfying the constraints $\tmmathbf{q} (\tmmathbf{0}) =\tmmathbf{0}$ which
implies in particular $\iint_{\Omega} \mymultiply Z_3^{\dag} (\tmmathbf{T})
\mymultiply \mathd A = 0$, hence the last equality above.

In the remainder of this derivation, we limit attention to $\kappa = 0$.

For $\kappa = 0$, the transformation
gradient~(\ref{eq:gr-effect-homogeneous-qties}) is $F^{(\varepsilon, 0)}_{i
\nocomma j} (\tmmathbf{T}) = (1 + \varepsilon) \mymultiply \delta_{i \nocomma
3} \mymultiply \delta_{j \nocomma 3} + \nlPoisson (\varepsilon) \mymultiply
\delta_{i \nocomma \alpha} \mymultiply \delta_{j \nocomma \alpha}$. The
linearized strain $\mathcal{E}$ in equation~(\ref{eq:Lh}) writes
\begin{equation}
  \mathcal{E}^{(\varepsilon, 0)} (\tmmathbf{T}, \tmmathbf{h}^{\dag},
  \tmmathbf{Z}) = \left( \tmmathbf{h}^{\dag} \cdot (\nabla Y^{(\varepsilon,
  0)}_{\alpha} (\tmmathbf{T}))_{\kappa = 0} \mymultiply \nlPoisson
  (\varepsilon) + (1 + \varepsilon) \mymultiply \partial_{\alpha} Z_3
  (\tmmathbf{T}) \right) \mymultiply \tmmathbf{e}_{\alpha} \odot
  \tmmathbf{e}_3 + \nlPoisson (\varepsilon) \mymultiply \partial_{\alpha}
  Z_{\beta} (\tmmathbf{T}) \mymultiply \tmmathbf{e}_{\alpha} \odot
  \tmmathbf{e}_{\beta} \label{eq:beam-details-E-cal}
\end{equation}
where $\tmmathbf{h}^{\dag} = (\varepsilon^{\dag}, \kappa^{\dag})$ denotes the
strain gradient. Recalling that the $\nabla$ operator stands for a derivative
with respect to the macroscopic strain $\tmmathbf{h}= (\varepsilon, \kappa)$,
see~equation~(\ref{eq:nabla-notation}),
\[ \begin{array}{rcl}
     \tmmathbf{h}^{\dag} \cdot (\nabla Y^{(\varepsilon, 0)}_{\alpha}
     (\tmmathbf{T}))_{\kappa = 0} & = & \left( \left( \varepsilon^{\dag}
     \mymultiply \frac{\mathd}{\mathd \varepsilon} + \kappa^{\dag} \mymultiply
     \frac{\mathd}{\mathd \kappa} \right) \left( \nlPoisson (\varepsilon)
     \mymultiply \left( T_{\alpha} + \kappa \mymultiply \frac{\mathd
     \nlPoisson}{\mathd \varepsilon} (\varepsilon) \mymultiply
     \varphi_{\alpha} (\tmmathbf{T}) \right) +\mathcal{O} (\kappa^2) \right)
     \right)_{\kappa = 0}\\
     & = & \varepsilon^{\dag} \mymultiply \frac{\mathd \nlPoisson}{\mathd
     \varepsilon} (\varepsilon) \mymultiply T_{\alpha} + \kappa^{\dag}
     \mymultiply \nlPoisson (\varepsilon) \mymultiply \frac{\mathd
     \nlPoisson}{\mathd \varepsilon} (\varepsilon) \mymultiply
     \varphi_{\alpha} (\tmmathbf{T})
   \end{array} \]
and so
\begin{equation}
  \begin{array}{l}
    \mathcal{E}^{(\varepsilon, 0)} (\tmmathbf{T}, \tmmathbf{h}^{\dag},
    \tmmathbf{Z}) =\\
    \nobracket \nobracket \hspace{3em} \nlPoisson (\varepsilon) \mymultiply
    \frac{\mathd \nlPoisson}{\mathd \varepsilon} (\varepsilon) \mymultiply
    \left( \mymultiply T_{\alpha} \mymultiply \varepsilon^{\dag} + \nlPoisson
    (\varepsilon) \mymultiply \varphi_{\alpha} (\tmmathbf{T}) \mymultiply
    \kappa^{\dag} \right) \mymultiply \tmmathbf{e}_{\alpha} \odot
    \tmmathbf{e}_3 + \left( (1 + \varepsilon) \mymultiply \partial_{\alpha}
    Z_3 (\tmmathbf{T}) \mymultiply \tmmathbf{e}_3 + \nlPoisson (\varepsilon)
    \mymultiply \partial_{\alpha} Z_{\beta} (\tmmathbf{T}) \mymultiply
    \tmmathbf{e}_{\beta} \right) \odot \tmmathbf{e}_{\alpha} .
  \end{array} \label{eq:L-eps-zero-buckling}
\end{equation}

For $\kappa = 0$, the operator $\nabla \tmmathbf{C}^{(1)}_{(\varepsilon,
\kappa)}$ introduced in equation~(\ref{eq:minus-grad-C1h}) writes, with the
help of equation~(\ref{eq:beam-C1h}),
\begin{equation}
  \begin{array}{rcl}
    - (\varepsilon^{\dag}, \kappa^{\dag}) \cdot \nabla
    \tmmathbf{C}_{(\varepsilon, 0)}^{(1)} \cdot \tmmathbf{Z} & = & -
    \iint_{\Omega} \left( \frac{\partial \left( \kappa \mymultiply
    Y_{\text{t}} (\varepsilon) \mymultiply p (\varepsilon) \mymultiply T_2
    \right)}{\partial \varepsilon} \mymultiply \varepsilon^{\dag} +
    \frac{\partial \left( \kappa \mymultiply Y_{\text{t}} (\varepsilon)
    \mymultiply p (\varepsilon) \mymultiply T_2 \right)}{\partial \kappa}
    \mymultiply \kappa^{\dag} \right)_{\kappa = 0} \mymultiply Z_3
    (\tmmathbf{T}) \mymultiply \mathd A\\
    & = & - \iint_{\Omega} \left( 0 + Y_{\text{t}} (\varepsilon)
    \mymultiply p (\varepsilon) \mymultiply T_2 \mymultiply \kappa^{\dag}
    \right) \mymultiply Z_3 (\tmmathbf{T}) \mymultiply \mathd A\\
    & = & - \kappa^{\dag} \mymultiply Y_{\text{t}} (\varepsilon) \mymultiply
    \nlPoisson (\varepsilon) \mymultiply \iint_{\Omega} T_2 \mymultiply Z_3
    (\tmmathbf{T}) \mymultiply \mathd A.
  \end{array} \label{eq:beam-nabla-C1}
\end{equation}

We proceed to calculate the bilinear operator $\mathcal{B}^{\tmmathbf{h}}
(\tmmathbf{h}^{\dag}, \tmmathbf{Z})$ from equation~(\ref{eq:B-cal}):
\[ \begin{array}{l}
     \mathcal{B}^{(\varepsilon, 0)} (\tmmathbf{h}^{\dag}, \tmmathbf{Z}) =
     \iint_{\Omega} \frac{1}{2} \mymultiply \mathcal{E}^{(\varepsilon, 0)}
     (\tmmathbf{T}, \tmmathbf{h}^{\dag}, \tmmathbf{Z}) \doublecontract
     \tmmathbf{K}^{(\varepsilon, 0)} (\tmmathbf{T}) \doublecontract
     \mathcal{E}^{(\varepsilon, 0)} (\tmmathbf{T}, \tmmathbf{h}^{\dag},
     \tmmathbf{Z}) \mymultiply \mathd A\\
     \nobracket \nobracket \hspace{8em} \ldots + \iint_{\Omega} \left(
     \frac{1}{2} \mymultiply \sum_{\alpha} (\tmmathbf{h}^{\dag} \cdot (\nabla
     Y^{(\varepsilon, 0)}_{\alpha} (\tmmathbf{T}))_{\kappa = 0})^2 \mymultiply
     \Sigma_{\text{tr}} (\varepsilon) \right) \mymultiply \mathd A
     -\tmmathbf{h}^{\dag} \cdot \nabla \tmmathbf{C}_{(\varepsilon, 0)}^{(1)}
     \cdot \tmmathbf{Z},
   \end{array} \]
where $\nabla \tmmathbf{C}_{(\varepsilon, 0)}^{(1)}$ has just been obtained in
equation~(\ref{eq:beam-nabla-C1}).

Using the transverse isotropy of $\tmmathbf{K}^{(\varepsilon, 0)}
(\tmmathbf{T}) =\tmmathbf{K}_{\text{tr}} (\varepsilon)$ from
equation~(\ref{eq:beam-transverse-isotropic-K}), and the expression of
$\mathcal{E}^{(\varepsilon, 0)} (\tmmathbf{T}, \tmmathbf{h}^{\dag},
\tmmathbf{Z})$ in equation~(\ref{eq:beam-details-E-cal}), one can calculate
the elastic stiffness term appearing in $\mathcal{B}^{(\varepsilon, 0)}
(\tmmathbf{h}^{\dag}, \tmmathbf{Z})$ as
\[ \frac{1}{2} \mymultiply \mathcal{E}^{(\varepsilon, 0)} (\tmmathbf{T},
   \tmmathbf{h}^{\dag}, \tmmathbf{Z}) \doublecontract
   \tmmathbf{K}^{(\varepsilon, 0)} (\tmmathbf{T}) \doublecontract
   \mathcal{E}^{(\varepsilon, 0)} (\tmmathbf{T}, \tmmathbf{h}^{\dag},
   \tmmathbf{Z}) = \frac{1}{2} \mymultiply G_{\text{t}} (\varepsilon)
   \mymultiply \left( \sum_{\alpha} 2^2 \mymultiply
   (\mathcal{E}^{(\varepsilon, 0)}_{\alpha \nocomma 3})^2 \right) +
   \frac{1}{2} \mymultiply \mathcal{E}_{\alpha \nocomma \beta}^{(\varepsilon,
   0)} \mymultiply K_{\alpha \nocomma \beta \nocomma \gamma \nocomma
   \delta}^{(\varepsilon, 0)} (\tmmathbf{T}) \mymultiply \mathcal{E}_{\gamma
   \nocomma \delta}^{(\varepsilon, 0)}, \]
where $\mathcal{E}^{(\varepsilon, 0)}_{\alpha \nocomma 3}
=\mathcal{E}^{(\varepsilon, 0)} (\tmmathbf{T}, \tmmathbf{h}^{\dag},
\tmmathbf{Z}) \doublecontract (\tmmathbf{e}_{\alpha} \otimes \tmmathbf{e}_3) =
\frac{1}{2} \mymultiply \left( \tmmathbf{h}^{\dag} \cdot (\nabla
Y^{(\varepsilon, 0)}_{\alpha} (\tmmathbf{T}))_{\kappa = 0} \mymultiply
\nlPoisson (\varepsilon) + (1 + \varepsilon) \mymultiply \partial_{\alpha} Z_3
(\tmmathbf{T}) \right)$ from equation~(\ref{eq:beam-details-E-cal}). By
equation~(\ref{eq:L-eps-zero-buckling}), $\mathcal{E}^{(\varepsilon,
0)}_{\alpha \nocomma 3}$ depends on the longitudinal corrective displacement
$Z_3$ while $\mathcal{E}_{\alpha \nocomma \beta}^{(\varepsilon, 0)}$ depends
on the transverse one. This shows that the quantities $Z_{\alpha}
(\tmmathbf{T})$ are uncoupled from $Z_3 (\tmmathbf{T})$ in
$\mathcal{B}^{(\varepsilon, 0)} (\tmmathbf{h}^{\dag}, \tmmathbf{Z})$. The
source terms $\varepsilon^{\dag}$ and $\kappa^{\dag}$ in \
$\mathcal{B}^{(\varepsilon, 0)} (\tmmathbf{h}^{\dag}, \tmmathbf{Z})$ being
coupled to $Z_3 (\tmmathbf{T})$, see equation~(\ref{eq:beam-nabla-C1}), we
conclude that the optimal transverse displacement is zero,
\[ Z_{\text{opt}, \alpha}^{(\varepsilon, 0)} ((\varepsilon^{\dag},
   \kappa^{\dag}), \tmmathbf{T}) = 0. \]

We are left with
\[ \begin{array}{l}
     \mathcal{B}^{(\varepsilon, 0)} (\tmmathbf{h}^{\dag}, \tmmathbf{Z}) =
     \frac{1}{2} \mymultiply \iint_{\Omega} G_{\text{t}} (\varepsilon)
     \mymultiply \sum_{\alpha} \left( \nlPoisson (\varepsilon) \mymultiply
     \frac{\mathd \nlPoisson}{\mathd \varepsilon} (\varepsilon) \mymultiply
     T_{\alpha} \mymultiply \varepsilon^{\dag} + \nlPoisson^2 (\varepsilon)
     \mymultiply \frac{\mathd \nlPoisson}{\mathd \varepsilon} (\varepsilon)
     \mymultiply \varphi_{\alpha} (\tmmathbf{T}) \mymultiply \kappa^{\dag} +
     (1 + \varepsilon) \mymultiply \partial_{\alpha} Z_3 (\tmmathbf{T})
     \right)^2 \mymultiply \mathd A\\
     \nobracket \nobracket \hspace{4em} \ldots + \iint_{\Omega} \left(
     \frac{1}{2} \mymultiply \sum_{\alpha} \left( \frac{\mathd
     \nlPoisson}{\mathd \varepsilon} (\varepsilon) \mymultiply T_{\alpha}
     \mymultiply \varepsilon^{\dag} + \nlPoisson (\varepsilon) \mymultiply
     \frac{\mathd \nlPoisson}{\mathd \varepsilon} (\varepsilon) \mymultiply
     \varphi_{\alpha} (\tmmathbf{T}) \mymultiply \kappa^{\dag} \right)^2
     \mymultiply \Sigma_{\text{tr}} (\varepsilon) \right) \mymultiply \mathd A
     - \kappa^{\dag} \mymultiply \nlPoisson (\varepsilon) \mymultiply
     Y_{\text{t}} (\varepsilon) \mymultiply \iint_{\Omega} T_2 \mymultiply
     Z_3 (\tmmathbf{T}) \mymultiply \mathd A.
   \end{array} \]

Since we are only interested in capturing the gradient effect associated with
bending, we focus attention to the case where the argument of
$\mathcal{B}^{(\varepsilon, 0)}$ has $\varepsilon^{\dag} = 0$, {\tmem{i.e.}},
we set $\tmmathbf{h}^{\dag} = (0, \kappa^{\dag})$. Expanding the square in
equation above, we have
\[ \begin{array}{l}
     \mathcal{B}^{(\varepsilon, 0)} ((0, \kappa^{\dag}), \tmmathbf{Z}) =
     \frac{(\kappa^{\dag})^2}{2} \mymultiply \left( \nlPoisson (\varepsilon)
     \mymultiply \frac{\mathd \nlPoisson}{\mathd \varepsilon} (\varepsilon)
     \right)^2 \mymultiply \Sigma_{\text{tr}} (\varepsilon) M \ldots\\
     \nobracket \nobracket \hspace{3em} \ldots + \frac{1}{2} \mymultiply
     G_{\text{t}} (\varepsilon) \mymultiply \iint_{\Omega} \left(
     \nlPoisson^2 (\varepsilon) \mymultiply \frac{\mathd \nlPoisson}{\mathd
     \varepsilon} (\varepsilon) \mymultiply \varphi_{\alpha} (\tmmathbf{T})
     \mymultiply \kappa^{\dag} + (1 + \varepsilon) \mymultiply
     \partial_{\alpha} Z_3 (\tmmathbf{T}) \right)^2 \mymultiply \mathd A -
     \kappa^{\dag} \nlPoisson (\varepsilon) \mymultiply Y (\varepsilon)
     \iint_{\Omega} T_2 \mymultiply Z_3 (\tmmathbf{T}) \mymultiply \mathd A
   \end{array} \]
where $M$ is the constant defined in
equation~(\ref{eq:beam-1d-reduc-geom-constants-general-cross-sect}).

The optimal correction $Z_3 = Z_{\text{opt}, 3}^{(\varepsilon, 0)} ((0,
\kappa^{\dag}), \tmmathbf{T})$ associated with a bending gradient
$\kappa^{\dag}$ makes stationary $\mathcal{B}^{(\varepsilon, 0)} ((0,
\kappa^{\dag}), \tmmathbf{Z})$ among all functions $Z_3$ that have a zero
average on the cross-section, see equation~(\ref{eq:Z-variational-pb}): $Z_3
(\tmmathbf{T})$ is such that, for any virtual displacement $\hat{Z}_3
(\tmmathbf{T})$,
\begin{multline}
    \iint_{\Omega} G_{\text{t}} (\varepsilon) \mymultiply (1 + \varepsilon)
    \mymultiply \left( \nlPoisson^2 (\varepsilon) \mymultiply \frac{\mathd
    \nlPoisson}{\mathd \varepsilon} (\varepsilon) \mymultiply \varphi_{\alpha}
    (\tmmathbf{T}) \mymultiply \kappa^{\dag} + (1 + \varepsilon) \mymultiply
    \partial_{\alpha} Z_3 (\tmmathbf{T}) \right) \mymultiply \partial_{\alpha}
    \hat{Z}_3 (\tmmathbf{T}) \mymultiply \mathd A\\
    \ldots - \kappa^{\dag} \nlPoisson (\varepsilon) \mymultiply Y
    (\varepsilon) \mymultiply \iint_{\Omega} T_2 \mymultiply \hat{Z}_3
    (\tmmathbf{T}) \mymultiply \mathd A + \iint_{\Omega} F_3^{\dag}
    (\varepsilon) \mymultiply \hat{Z}_3 (\tmmathbf{T}) \overset{}{}
    \mymultiply \mathd A = 0,
	\label{eq:beam-Z3-variational-problem}
\end{multline}
where $F_3^{\dag} (\varepsilon)$ is a Lagrange multiplier enforcing the
constraint $\iint_{\Omega} Z_3 \mymultiply \mathd A = 0$. Taking a constant
virtual field $\hat{Z}_3 (\tmmathbf{T}) = 1$ and using $\iint_{\Omega} T_2
\mymultiply \mathd A = A \mymultiply \langle T_2 \rangle = 0$ by
equation~(\ref{eq:Euler-buckling-cross-section-symmetry}), we find $F_3^{\dag}
(\varepsilon) = 0$.

With $c_{\Gamma} (\varepsilon)$ as the material constant defined in
equation~(\ref{eq:beam-1d-reduction-coefficients}) the equation above can be
rewritten as
\[ \forall \hat{Z}_3 \qquad \iint_{\Omega} \left[
   \partial_{\alpha} Z_3 (\tmmathbf{T}) \mymultiply \partial_{\alpha}
   \hat{Z}_3 (\tmmathbf{T}) + \kappa^{\dag} \mymultiply \frac{\nlPoisson^2
   (\varepsilon) \mymultiply \frac{\mathd \nlPoisson}{\mathd \varepsilon}
   (\varepsilon)}{1 + \varepsilon} \left( \varphi_{\alpha} \left( \tmmathbf{T}
   \right) \mymultiply \partial_{\alpha} \hat{Z}_3 (\tmmathbf{T}) + c_{\Gamma}
   (\varepsilon) \times 2 \mymultiply T_2 \mymultiply \hat{Z}_3 (\tmmathbf{T})
   \right) \right] \mathd A = 0\textrm{.} \]
By linearity, the solution $Z_3 (\tmmathbf{T})$ of this variational problem is
the function $Z_{\text{opt}, 3}^{(\varepsilon, 0)} ((0, \kappa^{\dag}),
\tmmathbf{T})$ given in
equation~(\ref{eq:beam-optima-corrective-displacement}) in terms of the
cross-sectional functions $\Theta$ and $\Gamma$ introduced in
equation~(\ref{eq:beam-1d-reduc-pb-Theta-Gamma}).

Inserting this solution $Z_{\text{opt}, 3}^{(\varepsilon, 0)} ((0,
\kappa^{\dag}), \tmmathbf{T})$ into the expression of
$\mathcal{B}^{(\varepsilon, 0)} ((0, \kappa^{\dag}), \tmmathbf{Z})$ above, and
expanding the $\frac{1}{2} \mymultiply G_{\text{t}} (\varepsilon) \mymultiply
\iint_{\Omega} (\ldots)^2 \mymultiply \mathd A$ term, we find the operator
$\tmmathbf{B}$ defined in equation~(\ref{eq:B-C}) as
\[ \begin{array}{lll}
     \frac{1}{2} \mymultiply (0, \kappa^{\dag}) \cdot \tmmathbf{B}
     (\varepsilon, 0) \cdot (0, \kappa^{\dag}) & = &
     \mathcal{B}^{(\varepsilon, 0)} \left( (0, \kappa^{\dag}),
     \tmmathbf{Z}_{\text{opt}}^{(\varepsilon, 0)} ((0, \kappa^{\dag}))
     \right)\\
     & = & \frac{(\kappa^{\dag})^2}{2} \mymultiply \left( \nlPoisson
     (\varepsilon) \mymultiply \frac{\mathd \nlPoisson}{\mathd \varepsilon}
     (\varepsilon) \right)^2 \mymultiply \left( \Sigma_{\text{tr}}
     (\varepsilon) + \nlPoisson^2 (\varepsilon) \mymultiply G_{\text{t}}
     (\varepsilon) \right) \mymultiply M - \kappa^{\dag} \nlPoisson
     (\varepsilon) \mymultiply Y (\varepsilon) \iint_{\Omega} T_2
     \mymultiply Z_{\text{opt}, 3}^{(\varepsilon, 0)} (\tmmathbf{T})
     \mymultiply \mathd A \ldots\\
     &  & \nobracket \nobracket + G_{\text{t}} (\varepsilon) \mymultiply (1 +
     \varepsilon) \mymultiply \iint_{\Omega} \left( \nlPoisson^2
     (\varepsilon) \mymultiply \frac{\mathd \nlPoisson}{\mathd \varepsilon}
     (\varepsilon) \mymultiply \varphi_{\alpha} (\tmmathbf{T}) \mymultiply
     \kappa^{\dag} + \frac{1 + \varepsilon}{2} \mymultiply \partial_{\alpha}
     Z_{\text{opt}, 3}^{(\varepsilon, 0)} (\tmmathbf{T}) \right) \mymultiply
     \partial_{\alpha} Z_{\text{opt}, 3}^{(\varepsilon, 0)} (\tmmathbf{T})
     \mymultiply \mathd A
   \end{array} \]

This expression can be simplified using the following identity, obtained by
choosing the virtual motion in~(\ref{eq:beam-Z3-variational-problem}) to be
the actual solution, $\hat{Z}_3 = Z_{\text{opt}, 3}^{(\varepsilon, 0)}$,
\begin{multline}
     \iint_{\Omega} G_{\text{t}} (\varepsilon) \mymultiply (1 +
     \varepsilon) \mymultiply \left( \nlPoisson^2 (\varepsilon) \mymultiply
     \frac{\mathd \nlPoisson}{\mathd \varepsilon} (\varepsilon) \mymultiply
     \varphi_{\alpha} (\tmmathbf{T}) \mymultiply \kappa^{\dag} + (1 +
     \varepsilon) \mymultiply \partial_{\alpha} Z_{\text{opt},
     3}^{(\varepsilon, 0)} (\tmmathbf{T}) \right) \mymultiply
     \partial_{\alpha} Z_{\text{opt}, 3}^{(\varepsilon, 0)} (\tmmathbf{T})
     \mymultiply \mathd A \hspace{6em}\\
     \ldots - \kappa^{\dag} \nlPoisson (\varepsilon) \mymultiply Y
     (\varepsilon) \mymultiply \iint_{\Omega} T_2 \mymultiply
     Z_{\text{opt}, 3}^{(\varepsilon, 0)} (\tmmathbf{T}) \overset{}{}
     \mymultiply \mathd A = 0.
 \end{multline}
This yields
\[ \begin{array}{lll}
     \frac{1}{2} \mymultiply (0, \kappa^{\dag}) \cdot \tmmathbf{B}
     (\varepsilon, 0) \cdot (0, \kappa^{\dag}) & = &
     \frac{(\kappa^{\dag})^2}{2} \mymultiply \left( \nlPoisson (\varepsilon)
     \mymultiply \frac{\mathd \nlPoisson}{\mathd \varepsilon} (\varepsilon)
     \right)^2 \mymultiply \left( \Sigma_{\text{tr}} (\varepsilon) +
     \nlPoisson^2 (\varepsilon) \mymultiply G_{\text{t}} (\varepsilon) \right)
     \mymultiply M \ldots\\
     &  & \nobracket \nobracket \hspace{6em} \nobracket \nobracket -
     \frac{1}{2} \mymultiply G_{\text{t}} (\varepsilon) \mymultiply (1 +
     \varepsilon)^2 \mymultiply \iint_{\Omega} \sum^2_{\alpha = 1} \left(
     \mymultiply \partial_{\alpha} Z_{\text{opt}, 3}^{(\varepsilon, 0)}
     (\tmmathbf{T}) \right)^2 \mymultiply \mathd A\\
     & = & \frac{(\kappa^{\dag})^2}{2} \mymultiply \left( \nlPoisson
     (\varepsilon) \mymultiply \frac{\mathd \nlPoisson}{\mathd \varepsilon}
     (\varepsilon) \right)^2 \mymultiply \left[ \left( \Sigma_{\text{tr}}
     (\varepsilon) + \nlPoisson^2 (\varepsilon) \mymultiply G_{\text{t}}
     (\varepsilon) \right) \mymultiply M \ldots \right.\\
     &  & \nobracket \nobracket \hspace{6em} \nobracket \nobracket \left. -
     G_{\text{t}} (\varepsilon) \mymultiply \nlPoisson^2 (\varepsilon)
     \mymultiply \iint_{\Omega} \sum^2_{\alpha = 1} \left(
     \partial_{\alpha} \Theta (\tmmathbf{T}) + c_{\Gamma} (\varepsilon)
     \mymultiply \partial_{\alpha} \Gamma (\tmmathbf{T}) \right)^2 \mymultiply
     \mathd A \right]
	 \textrm{.}
   \end{array} \]
Using the constants from
equation~(\ref{eq:beam-1d-reduc-geom-constants-general-cross-sect}) and
eliminating $\Sigma_{\text{tr}} (\varepsilon)$ using
equation~(\ref{eq:beam-Sigma-tr-wtr-prime}), one obtains an following identity
concerning the operator $\tmmathbf{B} (\varepsilon, 0)$ introduced in
equation~(\ref{eq:B-C}):
\[ \frac{1}{2} \mymultiply (0, \kappa^{\dag}) \cdot \tmmathbf{B} (\varepsilon,
   0) \cdot (0, \kappa^{\dag}) = \frac{(\kappa^{\dag})^2}{2} \mymultiply B_{1
   \nocomma 1} (\varepsilon, 0) \]
where
\begin{equation}
  B_{1 \nocomma 1} (\varepsilon, 0) = \left( \nlPoisson (\varepsilon)
  \mymultiply \frac{\mathd \nlPoisson}{\mathd \varepsilon} (\varepsilon)
  \right)^2 \mymultiply \left( \frac{1}{1 + \varepsilon} \mymultiply
  \frac{\mathd w_{\text{tr}}}{\mathd \varepsilon} (\varepsilon) \mymultiply M
  + \nlPoisson^2 (\varepsilon) \mymultiply G_{\text{t}} (\varepsilon)
  \mymultiply \left( M - J_{\Theta \nocomma \Theta} - 2 \mymultiply c_{\Gamma}
  (\varepsilon) \mymultiply J_{\Theta \nocomma \Gamma} - c_{\Gamma}^2
  (\varepsilon) \mymultiply J_{\Gamma \nocomma \Gamma} \right) \right) .
  \label{eq:beam-B11}
\end{equation}

In addition, the operator $\tmmathbf{C} (\varepsilon, \kappa)$ defined in
equation~(\ref{eq:B-C}) as $\tmmathbf{C} (\varepsilon, \kappa) \cdot
(\varepsilon^{\dag}, \kappa^{\dag}) =\tmmathbf{C}_{(\varepsilon,
\kappa)}^{(1)} \cdot \tmmathbf{Z}_{\text{opt}}^{(\varepsilon, \kappa)}
(\varepsilon^{\dag}, \kappa^{\dag})$ is found from
equations~(\ref{eq:beam-C1h})
and~(\ref{eq:beam-optima-corrective-displacement}) as
\[ \begin{array}{lcl}
     \tmmathbf{C} (\varepsilon, \kappa) \cdot (\varepsilon^{\dag},
     \kappa^{\dag}) & = & \kappa \mymultiply Y_{\text{t}} (\varepsilon)
     \mymultiply p (\varepsilon) \mymultiply \iint_{\Omega} T_2 \mymultiply
     Z_3^{(\varepsilon, 0)} ((\varepsilon^{\dag}, \kappa^{\dag}),
     \tmmathbf{T}) \mymultiply \mathd A +\mathcal{O} (\kappa^2)\\
     & = & \kappa \mymultiply Y_{\text{t}} (\varepsilon) \mymultiply p
     (\varepsilon) \mymultiply \iint_{\Omega} T_2 \mymultiply \left(
     \varepsilon^{\dag} \times (\ldots) + \kappa^{\dag} \mymultiply
     \frac{\nlPoisson^2 (\varepsilon) \mymultiply \frac{\mathd
     \nlPoisson}{\mathd \varepsilon} (\varepsilon)}{1 + \varepsilon}
     \mymultiply \left( \Theta (\tmmathbf{T}) + c_{\Gamma} (\varepsilon)
     \mymultiply \Gamma (\tmmathbf{T}) \right) \right) \mymultiply \mathd A
     +\mathcal{O} (\kappa^2)\\
     & = & \varepsilon^{\dag} \mymultiply \kappa \times (\ldots) +
     \kappa^{\dag} \mymultiply \kappa \mymultiply Y_{\text{t}} (\varepsilon)
     \mymultiply \frac{\nlPoisson^3 (\varepsilon) \mymultiply \frac{\mathd
     \nlPoisson}{\mathd \varepsilon} (\varepsilon)}{1 + \varepsilon}
     \mymultiply \iint_{\Omega} T_2 \mymultiply \left( \Theta
     (\tmmathbf{T}) + c_{\Gamma} (\varepsilon) \mymultiply \Gamma
     (\tmmathbf{T}) \right) \mymultiply \mathd A +\mathcal{O} (\kappa^2)\\
     & = & \varepsilon^{\dag} \mymultiply C_0 (\varepsilon, \kappa) +
     \kappa^{\dag} \mymultiply C_1 (\varepsilon, \kappa)
   \end{array} \]
where the ellipsis stands for an expression that does not need to be
calculated,
\begin{equation}
  C_0 (\varepsilon, \kappa) = \kappa \times (\ldots) +\mathcal{O} (\kappa^2)
  =\mathcal{O} (\kappa) \label{eq:beam-app-C0}
\end{equation}
and
\begin{equation}
  C_1 (\varepsilon, \kappa) = \kappa \mymultiply Y_{\text{t}} (\varepsilon)
  \mymultiply \frac{\nlPoisson^3 (\varepsilon) \mymultiply \frac{\mathd
  \nlPoisson}{\mathd \varepsilon} (\varepsilon)}{1 + \varepsilon} \mymultiply
  \left( \iint_{\Omega} T_2 \mymultiply \Theta (\tmmathbf{T}) \mymultiply
  \mathd A + c_{\Gamma} (\varepsilon) \mymultiply \iint_{\Omega} T_2
  \mymultiply \Gamma (\tmmathbf{T}) \mymultiply \mathd A \right) +\mathcal{O}
  (\kappa^2) .
\end{equation}
To evaluate the integrals in the right-hand side, we set $\hat{Z}_3
(\tmmathbf{T}) = g (\tmmathbf{T})$ in the variational problem for $\Gamma$ in
equation~(\ref{eq:beam-1d-reduc-pb-Theta-Gamma}),
\begin{equation}
  \begin{array}{lll}
    \iint_{\Omega} T_2 \mymultiply g (\tmmathbf{T}) \mymultiply \mathd A &
    = & - \frac{1}{2} \mymultiply \iint_{\Omega} \partial_{\alpha} \Gamma
    (\tmmathbf{T}) \mymultiply \partial_{\alpha} g (\tmmathbf{T}) \mymultiply
    \mathd A\\
    & = & \left\{\begin{array}{ll}
      - J_{\Theta \nocomma \Gamma} / 2 & \text{if $g = \Theta$}\\
      - J_{\Gamma \nocomma \Gamma} / 2 & \text{if $g = \Gamma$} .
    \end{array}\right.
  \end{array}
\end{equation}
This yields, after using~(\ref{eq:beam-1d-reduction-coefficients}),
\begin{equation}
  \begin{array}{lll}
    C_1 (\varepsilon, \kappa) & = & \kappa \mymultiply Y_{\text{t}}
    (\varepsilon) \mymultiply \frac{\nlPoisson^3 (\varepsilon) \mymultiply
    \left( - \frac{\mathd \nlPoisson}{\mathd \varepsilon} (\varepsilon)
    \right)}{2 \mymultiply (1 + \varepsilon)} \mymultiply \left( J_{\Theta
    \nocomma \Gamma} + J_{\Gamma \nocomma \Gamma} \mymultiply c_{\Gamma}
    (\varepsilon) \right) +\mathcal{O} (\kappa^2)\\
    & = & \kappa \mymultiply G_{\text{t}} (\varepsilon) \mymultiply \left(
    \nlPoisson^2 (\varepsilon) \mymultiply \frac{\mathd \nlPoisson}{\mathd
    \varepsilon} (\varepsilon) \right)^2 \mymultiply \left( J_{\Theta \nocomma
    \Gamma} + J_{\Gamma \nocomma \Gamma} \mymultiply c_{\Gamma} (\varepsilon)
    \right) \mymultiply c_{\Gamma} (\varepsilon) +\mathcal{O} (\kappa^2) .
  \end{array} \label{eq:beam-C1}
\end{equation}
The elastic modulus $D_{1 \nocomma 1}$ associated with the gradient effect in
bending and defined in equation~(\ref{eq:D-of-h}) is found by combining
equations~(\ref{eq:beam-B11}) and (\ref{eq:beam-C1}) as
\[ \begin{array}{lll}
     D_{1 \nocomma 1} (\varepsilon, 0) & = & B_{1 \nocomma 1} (\varepsilon, 0)
     + 2 \mymultiply \frac{\partial C_1}{\partial \kappa} (\varepsilon, 0)\\
     & = & \left( \nlPoisson (\varepsilon) \mymultiply \frac{\mathd
     \nlPoisson}{\mathd \varepsilon} (\varepsilon) \right)^2 \mymultiply
     \left( \frac{1}{1 + \varepsilon} \mymultiply \frac{\mathd
     w_{\text{tr}}}{\mathd \varepsilon} (\varepsilon) \mymultiply M +
     \nlPoisson^2 (\varepsilon) \mymultiply G_{\text{t}} (\varepsilon)
     \mymultiply \left( M - J_{\Theta \nocomma \Theta} + J_{\Gamma \nocomma
     \Gamma} \mymultiply c_{\Gamma}^2 (\varepsilon) \right) \right)
   \end{array} \]
as announced in the third equation of
equation~(\ref{eq:beam-1d-reduction-result-generic}).

By a symmetry argument similar to that presented in
section~\ref{sssec:beam-symmetries}, one can show that $C_0 (\varepsilon,
\kappa)$ is even with respect to $\kappa$, so that
equation~(\ref{eq:beam-app-C0}) implies the stronger estimate
\begin{equation}
  C_0 (\varepsilon, \kappa) =\mathcal{O} (\kappa^2) . \label{eq:beam-C0}
\end{equation}
The modified strains are then found from equation~(\ref{eq:hi-tilde})
and~(\ref{eq:beam-Whom}) as $\tilde{\varepsilon} = \varepsilon + \xi_0
(\varepsilon, \kappa) \mymultiply \varepsilon''$ and $\tilde{\kappa} = \kappa
+ \xi_1 (\varepsilon, \kappa) \mymultiply \kappa''$, where
\[ \begin{array}{lll}
     \xi_0 (\varepsilon, 0) & = & \lim_{\kappa \rightarrow 0}  \frac{C_0
     (\varepsilon, \kappa)}{\frac{\partial W_{\text{hom}}}{\partial
     \varepsilon} (\varepsilon, \kappa)}\\
     & = & \lim_{\kappa \rightarrow 0}  \frac{\mathcal{O}
     (\kappa^2)}{\mathcal{O} (\kappa)}\\
     & = & 0
   \end{array} \]
and, recalling equations~(\ref{eq:beam-C1})
and~(\ref{eq:beam-1d-reduction-coefficients}),
\[ \begin{array}{lll}
     \xi_1 (\varepsilon, 0) & = & \lim_{\kappa \rightarrow 0}  \frac{C_1
     (\varepsilon, \kappa)}{\frac{\partial W_{\text{hom}}}{\partial \kappa}
     (\varepsilon, \kappa)}\\
     & = & \left( \frac{\kappa \mymultiply G_{\text{t}} (\varepsilon)
     \mymultiply \left( \nlPoisson^2 (\varepsilon) \mymultiply \frac{\mathd
     \nlPoisson}{\mathd \varepsilon} (\varepsilon) \right)^2 \mymultiply
     \left( J_{\Theta \nocomma \Gamma} + J_{\Gamma \nocomma \Gamma}
     \mymultiply c_{\Gamma} (\varepsilon) \right) \mymultiply c_{\Gamma}
     (\varepsilon) +\mathcal{O} (\kappa^2)}{\kappa \mymultiply Y_{\text{t}}
     (\varepsilon) \mymultiply \nlPoisson^2 (\varepsilon) \mymultiply I_1^0
     +\mathcal{O} (\kappa^3)} \right)_{\kappa = 0}\\
     & = & \frac{G_{\text{t}} (\varepsilon) \mymultiply \left( \nlPoisson^2
     (\varepsilon) \mymultiply \frac{\mathd \nlPoisson}{\mathd \varepsilon}
     (\varepsilon) \right)^2 \mymultiply \left( J_{\Theta \nocomma \Gamma} +
     J_{\Gamma \nocomma \Gamma} \mymultiply c_{\Gamma} (\varepsilon) \right)
     \mymultiply c_{\Gamma} (\varepsilon)}{Y_{\text{t}} (\varepsilon)
     \mymultiply \nlPoisson^2 (\varepsilon) \mymultiply I_1^0}\\
     & = & \nlPoisson (\varepsilon) \mymultiply \left( - \frac{\mathd
     \nlPoisson}{\mathd \varepsilon} (\varepsilon) \right) \mymultiply
     \frac{\left( J_{\Theta \nocomma \Gamma} + J_{\Gamma \nocomma \Gamma}
     \mymultiply c_{\Gamma} (\varepsilon) \right)}{2 \mymultiply (1 +
     \varepsilon) \mymultiply I_1^0}
   \end{array} \]
This completes the proof of the results announced in
equation~(\ref{eq:beam-1d-reduction-result-generic}).

 \bigskip
\bibliographystyle{elsarticle-harv}

\end{document}